\documentclass[aps, prd, amsmath, floats, floatfix, superscriptaddress, nofootinbib, showpacs,twocolumn]{revtex4-2} %

\usepackage{aas_macros}
\usepackage{amssymb}
\usepackage{amsmath}
\usepackage{verbatim}
\usepackage{mathrsfs}
\usepackage{amsfonts}
\usepackage{latexsym}
\usepackage{epsfig}
\usepackage{epstopdf}
\usepackage{float}
\usepackage[usenames, dvipsnames]{color}
\usepackage{mathtools}
\usepackage{cases}
\usepackage{hyperref}
\usepackage{soul}

\begin{document}


\definecolor{pink}{rgb}{0.9,0,0.6} 

\newcommand{\dario}[1]{\textcolor{red}{{\bf Dar\'io: #1}}}
\newcommand{\aldo}[1]{\textcolor{blue}{{  #1}}}
\newcommand{\OS}[1]{{\color{cyan}~\textsf{[(Olivier) #1]}}}
\newcommand{\paola}[1]{{\color{pink}~\textsf{[(Paola) #1]}}}
\newcommand{\CG}[1]{{\color{green}~\textsf{[(Carlos) #1]}}}


\title{Accretion of a Vlasov gas onto a black hole from a sphere of finite radius and the role of angular momentum} 

\author{Aldo Gamboa} 
\email{aldojavier@ciencias.unam.mx} 
\affiliation{Instituto de Ciencias Nucleares, Universidad Nacional
  Aut\'onoma de M\'exico, Circuito Exterior C.U., A.P. 70-543,
  M\'exico D.F. 04510, M\'exico}

\author{Carlos Gabarrete}
\email{carlos.gabarrete@umich.mx}
\affiliation{Instituto de F\'isica y Matem\'aticas, Universidad Michoacana de San Nicol\'as de Hidalgo,
Edificio C-3, Ciudad Universitaria, 58040 Morelia, Michoac\'an, M\'exico}
  
\author{Paola Dom\'inguez-Fern\'andez} 
\email[]{paoladominguez@unist.ac.kr} 
\affiliation{Department of Physics, School of Natural Sciences, Ulsan National Institute of Science and Technology, Ulsan 44919, Republic of Korea}

\author{Dar\'{\i}o N\'u\~nez}
\email[]{nunez@nucleares.unam.mx}
\affiliation{Instituto de Ciencias Nucleares, Universidad Nacional
  Aut\'onoma de M\'exico, Circuito Exterior C.U., A.P. 70-543,
  M\'exico D.F. 04510, M\'exico}

\author{Olivier Sarbach}
\email{olivier.sarbach@umich.mx}
\affiliation{Instituto de F\'isica y Matem\'aticas, Universidad Michoacana de San Nicol\'as de Hidalgo,
Edificio C-3, Ciudad Universitaria, 58040 Morelia, Michoac\'an, M\'exico}


\date{\today}


\begin{abstract}
The accretion of a spherically symmetric, collisionless kinetic gas cloud onto a Schwarzschild black hole is analyzed. Whereas previous studies have treated this problem by specifying boundary conditions at infinity, here the properties of the gas are given at a sphere of finite radius. The corresponding steady-state solutions are computed using four different models with an increasing level of sophistication, starting with the purely radial infall of Newtonian particles and culminating with a fully general relativistic calculation in which individual particles have angular momentum. The resulting mass accretion rates are analyzed and compared with previous models, including the standard Bondi model for a hydrodynamic flow. We apply our models to the supermassive black holes Sgr~A* and M87*, and we discuss how their low luminosity could be partially explained by a kinetic description involving angular momentum. Furthermore, we get results consistent with previous model-dependent bounds for the accretion rate imposed by rotation measures of the polarized light coming from Sgr~A* and with estimations of the accretion rate of M87* from the Event Horizon Telescope collaboration. Our methods and results could serve as a first approximation for more realistic black hole accretion models in various astrophysical scenarios in which the accreted material is expected to be nearly collisionless. 
\end{abstract}

 


\maketitle


\section{Introduction}
\label{sec:introduction}

Accretion of matter is one of the most important processes in astrophysical systems due to its fundamental role in the formation and evolution of galaxies, stars and compact objects. The fact that different types of matter (e.g. kinetic gases,  fluids or scalar fields) have distinctive features in their corresponding dynamics, makes essential to take into account the nature of the infalling matter for a physically correct description of the accretion process (the features of some of these types of matter can be seen e.g. in  \citet{Dominguez-Fernandez:2017nxx} where the dynamics of a collisionless kinetic gas in a dark matter halo is studied; in \citet{2002apa..book.....F} for accretion studies based on fluid dynamics; or in \citet{Burt:2011pv} where the peculiar distribution of a scalar field surrounding a black hole (BH) is described).

The first studies on the phenomenon of accretion were developed in \citet{1939PCPS...35..405H, 10.1093/mnras/104.5.273} for a star moving at a steady speed through an infinite pressureless gas cloud. Later on, \citet{1952MNRAS.112..195B} studied the hydrodynamical steady spherical accretion of a gas at rest at infinity onto a Newtonian star. In these models, effects such as viscosity, turbulence, self-gravity or magnetic fields are neglected. 
Further studies have been undertaken for different  scenarios in which matter accretion was shown to be astrophysically relevant, for example in X-ray binaries (e.g. \citet[][]{1997CAS....26.....L}), in gamma-ray bursts (e.g. \citet[][]{1999ApJ...518..356P}), in protoplanetary disks  (e.g. \citet[][]{2011ARA&A..49...67W}) or in active galactic nuclei (e.g. \citet[][]{1999agnc.book.....K}). A great part of these scenarios involve BHs, because they naturally appear in the life cycle of massive stars (\citet{1965PhRvL..14...57P, 1999CQGra..16A...3C}) and in the core of medium-to-large galaxies (\citet{1995ARA&A..33..581K, Kormendy:2013dxa}). The radiation emanated from these powerful sources originates from a region close to the BH's event horizon, and therefore the corresponding accretion requires a fully general-relativistic modeling. 

Substantial theoretical and numerical work has been done on the fluid or hydrodynamical approximation of the accreting flow onto BHs. In this context, the first general relativistic extension of the Bondi model was given by \citet[]{1972Ap&SS..15..153M} who studied the steady spherical accretion flow of simple polytropic gases onto a Schwarzschild BH. Additional generalizations of these hydrodynamic solutions have been worked out over the years (see \citet{Aguayo-Ortiz:2021jzv} and references therein for recent work providing a review of the Bondi and Michel solutions and their generalization to rotating BHs).
In particular, recent numerical works make use of general relativistic magnetohydrodynamic (GRMHD) simulations to study different accretion models (see e.g. \citet{Porth_2017}). The fluid approximation in these models considers that the effective mean free path for the particles is sufficiently short with respect to the length over which macroscopic quantities, such as the particle number density, the bulk velocity or the temperature vary in a significant way, so that thermal equilibrium is attained locally. However, there are some cases in which the hydrodynamical approximation does not correspond to the nature of the infalling matter. This is the situation, e.g. for underluminous sources (with respect to the Eddington luminosity) such as Sgr~A*, the supermassive black hole (SMBH) in the center of our galaxy (\citet[][]{2003ApJ...586L.127G,2009ApJ...692.1075G, Falcke_2013}), and M87*, the SMBH in the  galactic center of Messier 87 (\citet[][]{2019ApJ...875L...1E, EHTCV}), whose accreting plasmas are in a low-density, high-temperature regime which makes them effectively collisionless (\citet{Mahadevan_1997, 1998AJ....115.1801H, 2003ApJ...591..891B, EHTCV}). 

A further important example is the accretion of dark matter (an exotic entity which is expected to be collisionless), which has been suggested to play a prominent role in the formation of SMBHs (see e.g. \citet{Read:2002wb, Choquette:2018lvq, 2021MNRAS.502.4227A}). The low collisionality of these kind of flows makes necessary to take into account the kinetic approximation in order to correctly describe the dynamics of the accreting matter.

Analytically, the problem of accretion of kinetic gases onto BHs is studied using the formalism of the relativistic Boltzmann equation (see e.g. \citet[][]{CercignaniKremer-Book}. In particular, the collisionless approximation (also known as a \textit{Vlasov gas}) in which the component particles do not interact directly with each other, has been studied both in the Newtonian  (e.g. \citet[][]{1971reas.book.....Z, 1983bhwd.book.....S} and in the general relativistic regimes (e.g. \citet[][]{pRoS17a, pRoS17b, pMaO21a, pMaO21b}. Numerically, models of collisionless (or weakly collisional) plasmas which include some kinetic effects into the equations of GRMHD flows, have been developed (\citet[][]{Sharma:2005ez, Chandra:2015iza, Foucart:2017axc}). Nonetheless, modelling a fully 3D kinetic simulation has been a complex subject due to the high computational effort of calculating the evolution of the 6D distribution functions of ions and electrons in the accreting plasma (see e.g. \citet{Kunz_2016} where a local shearing-box model of a collisionless accretion disk is used, and references therein).\footnote{For recent analytic work analyzing the dynamics of a collisionless gas  in the equatorial plane of a (rotating) Kerr black hole and the phase-space mixing phenomenon, see~\cite{pRoS18}.}

The analytical hydrodynamic and kinetic models with spherical symmetry mentioned so far, assume that the boundary conditions determining the properties of the gas (temperature and density) are specified at infinity. However, in practice such properties are measured at a finite distance from the black hole. Therefore, analytical modeling should take into account the finite nature of the accretion phenomenon in order to produce more realistic models. In the hydrodynamic case, such finite models have recently been proposed in the context of the `choked' accretion mechanism (see \citet{Aguayo-Ortiz:2019fap,Tejeda:2019fwr,Aguayo-Ortiz:2020qro} and references therein) in which the gas is injected from a sphere of finite radius, named
the `injection sphere', with a slight equatorial to polar density contrast, resulting in an inflow-outflow configuration.

In this article we present a series of illustrative and simplified analytic finite models which aim to solve the BH accretion problem from a kinetic and relativistic standpoint. To this purpose, we analyze the steady, spherical accretion flow of a collisionless kinetic gas with negligible self-gravity from an injection sphere of finite radius $R$. We study the case of purely radial infall, in which none of the particles have angular momentum, as well as the case where individual particles have arbitrary angular momentum but the gas as a whole (averaged over the momentum space) moves in the pure radial direction. In the latter case, we obtain a general formula for the mass accretion rate which reduces to previous known results for $R\to \infty$ (see \citet[][]{1983bhwd.book.....S, pRoS17a}), while for fixed $R$ one obtains new solutions.

Despite the simplicity of our models, we get reasonable results  when applied to the flows onto Sgr~A* and M87*. A smaller mass accretion rate for these BHs is predicted by our kinetic approach, which may contribute to the understanding of their low luminosity (\citet[][]{2003ApJ...591..891B, 2003ApJ...582..133D}) and the presence of polarized light at 230~GHz coming from regions near to the BH horizon (\citet{2000ApJ...534L.173A, 2014ApJ...783L..33K,ehtc7,EHTCVIII}). The presence of this polarization would not be possible for an accretion rate similar to the predicted from the Bondi model, because larger mass accretion rates would depolarize the light through extreme Faraday rotation gradients (see e.g. \citet[][]{2000ApJ...545..842Q, 2018MNRAS.478.1875J}). The conventional solutions to these problems involve radiatively inefficient accretion flow (RIAF) models (\citet{1995ApJ...452..710N, 1999ApJ...516..399Q, 2003ApJ...598..301Y}). In these models the low luminosity is explained by an inefficiency in the process of energy exchange between protons and electrons causing the advection of most of the viscously released energy  into the BH's horizon. On the other hand, the presence of linearly polarized light at 230~GHz is explained by allowing a loss of mass in the inner regions of the flow through convection and/or outflows, reducing effectively the mass accretion rate via a dependence of the accretion rate with radius (see \citet{2014ARA&A..52..529Y} for a comprehensive review of accretion flow models).

In this work, we derive the kinetic gas mass accretion formula analogous to the spherical Bondi fluid model. In the examples provided, we show that the accretion rate in a fluid model can be similar to the corresponding kinetic gas one, closing in this way the gap that existed in previous analysis of both models. This issue is discussed for instance in \citet{1983bhwd.book.....S} where there is a huge difference between the fluid and kinetic accretion rates. We also derive the generalizations of the accretion rate predicted in \citet{1971reas.book.....Z} for a finite radius and show, in the examples presented, that the value of the accretion rate strongly depends on that radius and the environment. The expressions obtained in our manuscript should be ideally suited to be applied also in dark matter studies in which case the dynamic is expected to be described by a totally non-interacting matter. Furthermore, the obtained equations could be useful for studying the accretion of hot, low-density matter which is trapped inside the gravitational potential of a Schwarzschild BH. The developed formalism and our ideal solutions could serve as a starting point for more complex scenarios, such as the non-spherical accretion onto a Kerr BH, and/or the addition of magnetic fields. 

This manuscript is organized as follows: in Section~\ref{sec:dynamics}, we present an overview of the formalism of the general relativistic Vlasov equation and the definitions of the physical quantities relevant for the accretion flow; in Section~\ref{sec:radial_infall},  we treat the purely radial spherical infall of particles, both in the non-relativistic and relativistic limits, and we apply the resulting equations to particles obeying mono-energetic and Maxwell-J\"uttner distribution functions; in Section~\ref{sec:finite}, we analyze the spherical accretion in which the assumption of zero angular momentum for the individual particles is relaxed and we also apply the results to mono-energetic and Maxwell-J\"uttner distribution functions; in Section~\ref{sec:summary}, we summarize our results in a concise form suitable for its immediate application; in Section~\ref{sec:applications}, we apply our results to the accretion flows onto Sgr A* and M87*; finally, in Section~\ref{sec:conclusions} we give final comments and suggestions for future research. An additional model which considers a distribution of particles with fixed angular momentum which is useful for the interpretation of some of our results is given in an appendix. Throughout this work, we use the signature convention $(-,+,+,+)$ for the space-time metric.


\section{Review of the general relativistic Vlasov equation} \label{sec:dynamics}

The study of a collisionless kinetic gas interacting with a central object is based on the \textit{one-particle distribution function}. In Newtonian theory, the distribution function $f$ is a function of time and coordinates $(\mathbf{x}, \mathbf{p})$ of the six-dimensional phase-space, such that $f(t, \mathbf{x}, \mathbf{p}) \,  d^3x \, d^3p$ represents the expected number of particles in the phase-space volume element $d^3x \, d^3p$ at time $t$. In general relativity, the distribution function can be defined, \textit{a priori}, on the eight-dimensional cotangent bundle  $T^*\mathcal{M}$ associated with the curved space-time manifold $(\mathcal{M}, g)$, that is, the set consisting of pairs $(x,p)$ where $x\in \mathcal{M}$ is a space-time event and $p$ is a momentum co-vector at $x$. Thus, locally the distribution function can be regarded as a function of the coordinates $(x^\mu,p_\mu)$, $\mu=0,1,2,3$, parametrizing the cotangent bundle.\footnote{Alternatively, one can work on the tangent bundle $T\mathcal{M}$ with local coordinates $(x^\mu,p^\mu)$. Both formulations are equivalent since the space-time metric provides a natural way to identify $T\mathcal{M}$ with $T^*\mathcal{M}$.}

For a relativistic, collisionless gas, the distribution function $f$ is required to solve the \textit{Vlasov} (or \textit{collisionless Boltzmann}) \textit{equation} which can be conveniently written as:
\begin{equation}
    \{ \mathcal{H}, f \} \equiv \frac{\partial \mathcal{H}}{\partial p_\mu}\frac{\partial f}{\partial x^\mu} - \frac{\partial \mathcal{H}}{\partial x^\mu}\frac{\partial f}{\partial p_\mu} = 0,
    \label{Eq:RelVlasov}
\end{equation}
where here $\mathcal{H}$ denotes the free particle Hamiltonian
\begin{equation}
    \mathcal{H}(x,p) := \frac{1}{2} g^{\mu\nu}(x) p_\mu p_\nu,
\end{equation}
with $g^{\mu\nu}(x)$ the components of the inverse metric at $x$. It follows immediately from Eq.~(\ref{Eq:RelVlasov}), that any distribution function $f$ which is only a function of integrals of motion satisfies the Vlasov equation.

Note that the Hamiltonian itself is an integral of motion; the corresponding conserved quantity is $-(mc)^2/2$ with $m$ the rest mass of the particle. For the following, we consider a collisionless gas of identical particles of positive mass $m > 0$, such that $f$ can be restricted on the future mass shell, the seven-dimensional submanifold of $T^*\mathcal{M}$ consisting of those points $(x^\mu,p_\mu)$ for which
\begin{equation}
    p^\mu p_\mu = -(mc)^2,
    \label{Eq:MassShell}
\end{equation}
and $p^\mu$ is future-directed. The future mass shell can be parametrized in terms of the coordinates $(x^\mu,p_i)$, with $i=1,2,3$, where the time component of the momentum does not appear as an independent coordinate since it can be reconstructed from the mass-shell constraint~(\ref{Eq:MassShell}). For further details on the geometry of the relativistic phase-space we refer the reader to \citet{2009PhyA..388.1079D,Sarbach:2013uba,rAcGoS2021}.

Among the space-time observables, the central quantity in our work describing  the most relevant physical properties of the solution is the particle current density vector field, defined as (see for instance equation (12.35) in \citet{CercignaniKremer-Book}):
\begin{equation}
    J^\mu(x) := c\int\limits_{P_x^+(m)} p^\mu f(x,p) \, \mbox{dvol}_x(p),
\label{Eq:ParticleDensity}    
\end{equation}
where $P_x^+(m)$ is the future mass hyperboloid consisting of those future-directed timelike vectors $p^\mu$ for which \eqref{Eq:MassShell} is satisfied and $\mbox{dvol}_x(p)$ is the Lorentz-invariant volume element on $P_x^+(m)$, defined as:
\begin{equation} \label{eq:vol_elem}
    \mbox{dvol}_x(p) := \frac{1}{\sqrt{-g}} \frac{d^3p_{*}}{p^0},
\end{equation}
where $\sqrt{-g}$ is the square root of the metric's determinant and $p_* = (p_{i})$ (with $i=1,2,3$) refer to the covariant spatial components of the linear momentum (\citet[][]{2009PhyA..388.1079D}).\footnote{The covariant and contravariant momentum volume elements can be related through $\frac{1}{\sqrt{-g}} \frac{d^3p_{*}}{p^0} = \sqrt{-g} \frac{d^3p}{|p_0|}$.}  

The corresponding invariant particle number density and mean four-velocity at $x$ are given by:
\begin{align}
    n(x) &\coloneqq \frac{1}{c}\,\sqrt{-J^\mu(x) J_\mu(x)},    \label{eq:inv_dens} \\
    u^\mu(x) &\coloneqq \frac{J^\mu(x)}{n(x)}. \label{eq:curr_dens_vel}
\end{align}

For the following, we assume that the gravitational potential is dominated by the BH such that the self-gravity of the gas can be neglected. Omitting the rotation of the BH for simplicity, we thus consider a spherically symmetric static background described by the metric: 
\begin{equation}
    ds^2 = g_{ab}(r)\,dx^a dx^b
 + r^2\left( d\vartheta^2 + \sin^2\vartheta d\varphi^2 \right),
\label{eq:el_se}
\end{equation}
with $(x^a) = (ct,r)$, where $c$ denotes the speed of light in vacuum, $t$ is the time coordinate, $r$ is the areal radius and $(\vartheta,\varphi)$ denote the usual angular coordinates on the two-sphere. The integrals of motion in this case consist of the rest mass $m = \sqrt{-2\mathcal{H}}$, the energy $E$ and the angular momentum vector $\mathbf{L}$ associated with the spherical symmetry.

It can be shown (e.g. \citet{CercignaniKremer-Book}) that for a distribution function $f$ satisfying the Vlasov equation, $J^\mu$ automatically satisfies the continuity equation $\nabla_\mu J^\mu = 0$, which allows us to define the conserved (rest) mass accretion rate for the metric in Eq.~(\ref{eq:el_se}):
\begin{equation}\label{Eq:Mdotn}
    \dot{M} \coloneqq 4\pi r^2 m \,J^r(x).
\end{equation}
Note that this definition is coordinate-independent, since it is defined in terms of the areal radius $r$ and the contravariant $r$-component of the current density vector field, which can be written as $J^r = dr(J) = J^\mu \nabla_\mu r$.

Finally, it is straightforward to show that in the non-relativistic limit ($|u^i|\ll c$ and $p^0 = mc$), the well-known expressions for the particle number density, the mean radial velocity and the mass accretion rate are recovered:
\begin{align}
&n(x) = \int f(x,p)\,d^3 p_*, \label{eq:inv_dens_clas} \\
 &u^r(x)  = \frac{1}{n(x)}\int \frac{p^r}{m}\,f(x,p)\,d^3 p_*, \label{eq:vel_clas} \\
&\dot M = 4\pi  r^2 m \, n(x) \, u^r(x) = 4\pi r^2 \int p^r \, f(x,p)\,d^3 p_*. \label{eq:acc_rate_clas}
\end{align}

In the results presented in this article, the distribution function $f$ is assumed to depend on $(x,p)$ only through the integrals of motion, $E$ and $\mathbf{L}$. Due to dispersion and mixing, it is in fact expected that any gas configuration relaxes in time to one described by such a distribution function (\citet{pRoS17a,pRoS20}), provided the boundary conditions specified at the injection sphere are compatible with it.
In addition, we focus on purely spherical accretion for which the distribution function depends only on the energy $E$ and the total angular momentum $L = |\mathbf{L}|$ of each particle. We shall use $F$ to denote the distribution function expressed in terms of $E$ and $L$.


\section{Purely radial infall from a finite radius}
\label{sec:radial_infall}

In this section, we focus on the spherically symmetric steady radial infall of a Vlasov gas into a central object, assuming that each individual particle has zero angular momentum. We assume that the particles are being accreted from an injection sphere at finite radius with specific density and energy or temperature which provide the boundary conditions for the problem. The distribution function describing this scenario depends only on the radial coordinate $r$ and its momentum $p_r$, and the corresponding observables only on $r$.  We treat both the non-relativistic and relativistic limits.  The definitions given in the previous section are specialized in order to describe adequately the radial accretion process. 


\subsection{Non-relativistic limit}
\label{sec:radial_nr}

In this limit, the particles are under the effect of a gravitational central potential $\Phi(r)$ generated by a mass $M$ (e.g. $\Phi(r) =-GMm/r$), and the injection sphere of the particles is at radius $R$, where we specify the particle number density. We ignore interactions with the surface of the central object since we are interested in a scenario analogous to a Schwarzschild BH, where there is no physical surface. For spherical coordinates and under the assumption that the particles have zero angular momentum, the volume element (\ref{eq:vol_elem}) in momentum space can be replaced by  (\citet{Dominguez-Fernandez:2017nxx}):\footnote{There is a $2 \pi$ difference with the result shown in \citet{Dominguez-Fernandez:2017nxx} due to a change of variable in the momentum space done in that work.}
\begin{equation} \label{eq:dpr}
    \frac{d^3 p_*}{\sqrt{-g}}  \rightarrow \frac{1}{r^2} \, dp_r, 
\end{equation}
Thus, from Eqs.~(\ref{eq:inv_dens_clas}-\ref{eq:acc_rate_clas}), we find that
\begin{align}
n(r) &= \frac{1}{r^2}\int_{-\infty}^{p_{\rm \textrm{m}}(r, R)}f(r,p_r)\,dp_r, \label{eq:inv_dens_rad_clas} \\
 u^r(r)  &=\frac{1}{r^2\, m \, n(r)}\int_{-\infty}^{p_\textrm{m}(r, R)} p_r\,f(r,p_r)\,dp_r, \label{eq:vel_rad_clas} \\
\dot M &= 4\pi  r^2 \,m \, n(r) \,  u^r(r)  = 4\pi \int_{-\infty}^{p_\textrm{m}(r, R)}p_r\, f(r,p_r)\,dp_r, \label{eq:acc_rate_clas_rad}
\end{align}
where $p^r = p_r$. Here, the upper integration limit $p_\textrm{m} (r, R) \coloneqq - \sqrt{2\,m[\Phi(R) - \Phi(r)]}$ incorporates the physical requirement that all the particles in the system are falling radially from a radius $R$ into the central mass, with $E = \Phi(R)$ being the minimum possible energy for the particles. Note that, if $R\rightarrow \infty$, then $n(r\rightarrow\infty) = 0 $ (which is consistent with the fact that the distribution function vanishes at infinite radius), and thus we cannot apply the boundary condition $n(R) = n_R$.

A specific scenario, which is directly related with the analyzed case by \citet{1983bhwd.book.....S}, is the radial infall of mono-energetic particles with energy $E_0 \geq \Phi(R)$. The distribution function in this case is:
\begin{equation}
    F(E) = f_0 \, \delta( E - E_0) = f_0 \, \delta\left( \frac{p_r^2}{2m} + \Phi(r) - E_0\right),
\end{equation}
where $f_0$ is a constant with units of [time]$^{-1}$ (because the radial distribution function, $f=f(r,p_r)$, has units of [length $\times$ momentum]$^{-1}$) and it is related with $n_R$. Next, we can use the properties of the Dirac delta distribution\footnote{Namely, the composition of the Dirac delta distribution with a smooth function $g(x)$, is given by $\delta(g(x)) = \sum_i \frac{\delta(x-x_i)}{|g'(x_i)|} $, where the sum goes over all the different roots $x_i$ of $g$. } to rewrite the distribution function as:
\begin{equation}\label{eq:monoenergetic_distribution}
    f(r, p_r) =  f_0 \, \sqrt{\frac{m}{2}} \, \frac{\delta\left( p_r + \sqrt{2m[E_0 -\Phi(r)]}\right)}{\sqrt{{E_0 - \Phi(r)}}}  ,
\end{equation}
where we have used the fact that all particles are falling and hence they can only have negative momentum. According to Eqs.~(\ref{eq:inv_dens_rad_clas}-\ref{eq:acc_rate_clas_rad}) and the boundary condition $ n_R$, the particle density, the average radial velocity and the accretion rate are, respectively:
\begin{align}
    n(r) &= \frac{f_0}{r^2} \, \sqrt{\frac{m}{ 2[E_0 - \Phi(r)]}}, \label{eq:inv_dens_rad_clas_mono} \\
   u^r(r) &= - \sqrt{\frac{2[E_0 - \Phi(r)]}{m}},  \label{eq:vel_rad_clas_mono} \\
    |\dot M| &=  4 \pi r^2 \, m \, n(r) |u^r(r)| = 4 \pi m f_0, \label{eq:acc_rate_rad_clas_mono}
\end{align}
where 
\begin{equation}
    f_0 = R^2 n_R \sqrt{\frac{2[E_0 - \Phi(R)]}{m} }, 
\end{equation}
which yields
\begin{equation}
\frac{|\dot{M}|}{m n_R} = 4\pi R^2 \sqrt{\frac{2\left[E_0 - \Phi(R) \right]}{m}},   
\label{Eq:Eq:AccretionRateModelMonoEner}
\end{equation}
valid for $r\leq R$. If $E_0 = 0$ and the gravitational potential is due to a central mass such that $\Phi(r) \sim 1/r$, Eq.~(\ref{eq:vel_rad_clas_mono}) is easily recognized as the free-fall velocity. Furthermore, we see from Eqs.~(\ref{eq:inv_dens_rad_clas_mono}) and  (\ref{eq:vel_rad_clas_mono}) that the particle number density and velocity are proportional to $r^{-3/2}$ and $r^{-1/2}$, respectively, which is the expected behaviour for the fluid limit (\citet{1983bhwd.book.....S}). If we set $E_0 = \Phi(R) + \frac{1}{2} m v_R^2$, where $v_R = u^r(R)$ is the speed of the particles at the injection radius with respect to the central object, then Eq.~(\ref{Eq:Eq:AccretionRateModelMonoEner}) can be written as the well-known expression:
\begin{equation}
\frac{|\dot{M}|}{m n_R v_R} = 4\pi R^2.
\label{Eq:Eq:AccretionRateModelMonoEnerBis}
\end{equation}
Another scenario, similar to the Bondi case, consists of a stationary cloud of particles with mass $m$ described by a Maxwell-Boltzmann distribution function (e.g. \citet[][]{binney2011galactic}), falling radially into a central object according to the gravitational potential $\Phi(r)$ generated by $M$. This has the form:
\begin{equation} \label{eq:f_E}
f(r, p_r) = A \,\exp\left[-\beta\,\left(\frac{p_r^2}{2\,m} + \Phi(r) \right)\right], 
\end{equation}
where as usual $\beta = 1/k_\textrm{B} T$, with $k_\textrm{B}$ the Boltzmann constant, $k_\textrm{B}=1.38\,\times\,10^{-23}\,\rm{m^2 \, kg \, K^{-1} \, s^{-2}}$, $T$ the temperature of the cloud and $A$ is a constant with units of [length $\times$ momentum]$^{-1}$. In this case, following Eqs.~(\ref{eq:inv_dens_rad_clas}-\ref{eq:acc_rate_clas_rad}) and the boundary condition $ n_R$, the particle number density, the average radial velocity and the accretion rate for $r\leq R$ are, respectively:
\begin{align}
 n(r) &= \frac{A}{r^2}\,\sqrt{\frac{m\pi}{2\,\beta}}\, e^{\textstyle - \beta \Phi(r)} \nonumber \\
 & \times \left[ 1-{\rm Erf}\left(\sqrt{\beta [\Phi(R) - \Phi(r)]}\right)\right],  \\
u^r(r) &= - \sqrt{\frac{2}{\beta m \pi} }  e^{ \textstyle - \beta[\Phi(R) - \Phi(r)]  }  \nonumber \\ 
&\times \left[ 1-{\rm Erf}\left(\sqrt{\beta [\Phi(R) - \Phi(r)]}\right)\right]^{-1} ,  \\
|\dot{M}| &= 4 \pi r^2 \, m \, n(r) |u^r(r)| =    4 \pi m \frac{A}{\beta} e^{\textstyle - \beta \Phi(R)},   
\end{align}
where Erf$(x)$ denotes the error function, and
\begin{equation}
    A = n_R \, R^2  \sqrt{\frac{2 \beta}{m \pi}} e^{\textstyle \, \beta \Phi(R)},
\end{equation}
corresponding to the injection sphere at a radius $R$, which yields
\begin{equation}
\frac{|\dot{M}|}{m n_R} = 4\pi R^2 \sqrt{\frac{2}{\pi m\beta}} = 4\pi R^2 \sqrt{\frac{2}{\pi}\frac{k_\textrm{B} T}{m}}  \, .
\label{Eq:AccretionRateModelExp}
\end{equation}
%


\subsection{Relativistic case}
\label{sec:radial_r}

In the relativistic case one considers a Vlasov gas on a Schwarzschild background, with metric components $-g_{00}=1/g_{rr}=\alpha(r)^2$ and $g_{0r} = 0$ in Eq.~(\ref{eq:el_se}), where
\begin{equation}
\alpha(r)^2 \coloneqq 1- \frac{r_\textrm{S}}{r},
\end{equation}
with $r_S$ the Schwarzschild radius defined by $r_\textrm{S}\coloneqq2GM/c^2$. The volume element in momentum space, Eq.~(\ref{eq:vol_elem}), takes the form:
\begin{equation} \label{eq:vol_elem_rad_clas}
    \mbox{dvol}_x(p)  =\frac{1}{r^2} \frac{dp_r}{p^0}, 
\end{equation}
with $p^0$ related to the relativistic energy $E$ through
\begin{equation}
    E = \alpha(r)^2 c\,p^0.
\end{equation} 
With Eqs.~(\ref{Eq:ParticleDensity}, \ref{eq:inv_dens}) we find:
\begin{align}
   J^0(r) &= \frac{c}{r^2}  \int_{-\infty}^{p_\textrm{m}(r,R)} f(r,p_r) \, d p_r \,,  \label{eq:num_dens_rel_1d} \\
    J^r(r)  &= \frac{ c}{r^2} \int_{-\infty}^{p_\textrm{m}(r,R)} f(r,p_r) \,  \frac{p^r}{p^0} \, d p_r , \label{eq:vel_rel_1d}\\
    n(r) &= \frac{1}{c} \sqrt{  [\alpha(r) J^0(r)]^2 - [J^r(r)/\alpha(r)]^2 }, \label{eq:inv_dens_rad}
\end{align}
where $p_{\textrm{m}}(r,R) := - \frac{m c}{ \alpha(r)^2} \sqrt{\alpha(R)^2 - \alpha(r)^2}$ (whose definition reduces to the one used in Eqs.~(\ref{eq:inv_dens_rad_clas}-\ref{eq:acc_rate_clas_rad}) in the non-relativistic limit) originates from the requirement that all the particles are infalling and have minimum possible energy equal to $E = \alpha(r) mc^2$ [see Eq.~(\ref{eq:energy_schwa})]. The boundary conditions are given by the particle number density at the injection sphere $ n_R$, and the energy or temperature as before.

As an example, we reconsider the Vlasov gas of mono-energetic particles of mass $m$, now with relativistic energy $E_0 \geq \alpha(R) m c^2$. The expected distribution function is:
\begin{equation}
    F(E) = f_0 \, \delta( E - E_0),
    \label{Eq:MonoEnergetic}
\end{equation}
where, again, $f_0$ is a constant with units of [time]$^{-1}$ related to $n_R$. From the general relation \eqref{Eq:MassShell}, we obtain:
\begin{equation} \label{eq:energy_schwa}
    E = \sqrt{ [ \alpha(r)^2 \, p_r \, c]^2 + \alpha(r)^2 \, m^2 c^4  },
\end{equation}
so that the distribution function is written as:
\begin{align}
    f(r, p_r) &= f_0 \, \delta \left( \sqrt{ [\alpha(r)^2 \, p_r \, c]^2 +\alpha(r)^2 \, m^2 c^4} - E_0 \right) \nonumber\\
    &= \frac{f_0 \, \delta \left( p_r + \frac{mc}{\alpha(r)^2}\sqrt{\left( \frac{E_0}{mc^2} \right)^2 -\alpha(r)^2} \, \right)}{ c \, \alpha(r)^2 \, \sqrt{1- \alpha(r)^2\left( \frac{m c^2}{E_0} \right)^2 \, }}   ,
\end{align}
where we have used the fact that all the particles have negative radial momentum. The invariant particle number density, the average radial velocity and the mass accretion rate computed from Eqs.~(\ref{Eq:Mdotn},  \ref{eq:num_dens_rel_1d}-\ref{eq:inv_dens_rad}) and the boundary condition $ n_R$, yield
\begin{align}
    n(r) &=  \frac{f_0}{r^2 c} \left[ \left( \frac{E_0}{m c^2}\right)^2 -\alpha(r)^2 \right]^{-1/2}, \label{eq:dens_rel_rad_mono}\\
    u^r(r) &=  - c  \left[  \left( \frac{E_0}{m c^2}\right)^2 - \alpha(r)^2 \, \right]^{1/2},  \label{eq:vel_rad_rel_mono} \\
   |\dot M| &= 4\pi m f_0, \label{eq:acc_rate_rad_rel_mono} 
\end{align}
valid for $r\leq R$, and with $f_0$ given by:
\begin{equation}
       f_0 = n_R \, c R^2 \sqrt{ \left( \frac{E_0}{m c^2}\right)^2 -\alpha(R)^2 } ,
\end{equation}
which yields
\begin{equation}
    \frac{|\dot{M}|}{m c n_R} = 4\pi R^2\sqrt{\left(\frac{E_0}{m c^2}\right)^2 - \alpha(R)^2} .
    \label{Eq:AccretionRateRelativisticMonoEner}
\end{equation}
Using Eq.~(\ref{eq:vel_rad_rel_mono}), we get the familiar result
\begin{equation} \label{eq:mono_rel_rad}
    \frac{|\dot{M}|}{m n_R u^r(R)} = 4\pi R^2,
\end{equation}
as follows directly from integrating the continuity equation for radially infalling dust in which case $\rho = m n$.
Nevertheless, for our purposes it is convenient to express the accretion rate in terms of the  3-velocity $v_R$ of the gas particles calculated by a static observer at the shell $r=R$, because the injection sphere is static with respect to the black hole. The relation between $v_R$ and $u^r(R)$ is given by (see e.g. \citet{2002GReGr..34.2075C})
\begin{equation}
    u^r(R) = \alpha(R) \, v_R \, \gamma,
\end{equation}
where $\gamma := (1 - v_R^2/c^2)^{-1/2}$ is the Lorentz factor associated with $v_R$, which implies that 
\begin{equation}
 \frac{|\dot{M}|}{m n_R v_{R}} = 4 \pi R^2 \alpha(R) \gamma. \label{eq:mono_rel_rad_stationary}  
\end{equation}
In the non-relativistic limit, with $v_R \ll c$ and $r_\textrm{S} \ll R$, the previous equations reduce to  Eqs.~(\ref{eq:inv_dens_rad_clas_mono}-\ref{Eq:Eq:AccretionRateModelMonoEnerBis}), as expected.

We now consider a distribution function of the Maxwell-J\"uttner-type (\citet{https://doi.org/10.1002/andp.19113390503}), 
\begin{equation} \label{eq:max_bolt}
    F(E) = A \, e^{- \beta E},
\end{equation}
where $\beta = 1/ k_\textrm{B} T$, $A$ is a constant with units of [length $\times$ momentum]$^{-1}$, the energy is given by Eq.~(\ref{eq:energy_schwa}), and $T$ is the temperature of the gas at the injection sphere.\footnote{Strictly speaking, the distribution function described by \eqref{eq:max_bolt} does not describe a configuration in thermodynamical equilibrium since in this section we restrict all the particles to have zero angular momentum.}
The resulting expressions from Eqs.~(\ref{eq:num_dens_rel_1d}-\ref{eq:inv_dens_rad}) have no analytical closed form. Nevertheless, we can make a change of integration variable from $p_r$ to the relativistic energy $E$ through Eq.~(\ref{eq:energy_schwa}). In this way, for the Schwarzschild metric we get:
\begin{align}
    J^{0}(r) &= \frac{1}{\alpha(r)^2 r^2} \int_{\alpha(R) m c^2}^{\infty} F(E) \, \frac{E}{\sqrt{E^2 - \alpha(r)^2 m^2 c^4}} \, dE  , \\
   J^{r}(r) &= \frac{1}{r^2} \int_{\alpha(R) m c^2}^{\infty} F(E) \, dE  . 
\end{align}
This set of equations can also be applied to the distribution function in Eq.~(\ref{Eq:MonoEnergetic}) and the resulting expressions are again Eqs.~(\ref{eq:dens_rel_rad_mono}-\ref{Eq:AccretionRateRelativisticMonoEner}). For the distribution function in Eq.~(\ref{eq:max_bolt}), we obtain:
\begin{align}
    n_R &= \frac{A m c}{R^2} \, \frac{ \sqrt{\textbf{K}_1(z)^2 e^{2z} - z^{-2} }}{ e^{z}},  \\
    u^r(R) &= - \frac{1}{\beta m c} \frac{1}{\sqrt{\textbf{K}_1(z)^2 e^{2z} -z^{-2}}}, \\ 
    |\dot M | &= 4 \pi m \frac{A}{\beta} e^{-z },
\label{Eq:MdotMaxBoltz}
\end{align}
where $z \coloneqq m c^2 \alpha(R)\beta $ and $\textbf{K}_1(z)$ is the modified Bessel function of the second kind and first order (see e.g. \citet{abramowitz2012handbook}). Note that the previous expressions are evaluated at $r=R$; this was necessary to get an analytical closed form. Eliminating $A$, we get an expression for the accretion rate:
\begin{equation}\label{eq:MJ-rad-rel}  
  \frac{|\dot M  |}{m c n_R} = \frac{4 \pi R^2 \alpha(R)}{\sqrt{\left[ \mathbf{K}_1(z) z e^{z} \right]^2 - 1 }}.
\end{equation}
Finally, considering the non-relativistic limit $ mc^2 \gg k_\textrm{B} T$, so that $z \gg 1$, we obtain: 
\begin{equation}
\frac{|\dot M  |}{m c n_R}  \approx  4\pi R^2 \alpha(R) \sqrt{\frac{2}{\pi z}} = 4\pi R^2 \sqrt{\frac{2\alpha(R)}{\pi}\frac{k_\textrm{B} T}{m c^2}} ,\label{Eq:AccretionRateRelativisticModelExp}
\end{equation}
which reduces to the expression in Eq.~(\ref{Eq:AccretionRateModelExp}) when $R \gg r_\textrm{S}$.


\section{Spherical accretion with angular momentum from a finite radius}
\label{sec:finite}

In this section, we  generalise the calculations of the previous section to the case in which individual gas particles are allowed to have angular momentum; however, we assume that the averaged quantities describing the gas (i.e. the space-time observables) are still spherical. For simplicity, we shall assume a uniform distribution in the total angular momentum $L$, and take the same mono-energetic or Maxwell-J\"uttner-like distribution in the energy as considered in the previous section. The analysis in this section is performed directly in the relativistic case with the Schwarzschild BH with mass $M$ as an accretor.

We assume that the injection sphere is located at a radius $R > r_{\text{ISCO}} = 6GM/c^2$ larger than the radius of the innermost stable circular orbit (ISCO) (it will become clear in a moment why the restriction $R > r_{\text{ISCO}}$ is required). As in the previous section, we impose the particle number density $n_R$ on the injection sphere, and we compute the solution satisfying this boundary condition and the corresponding accretion rate.

\begin{figure}
	\includegraphics[width=\columnwidth]{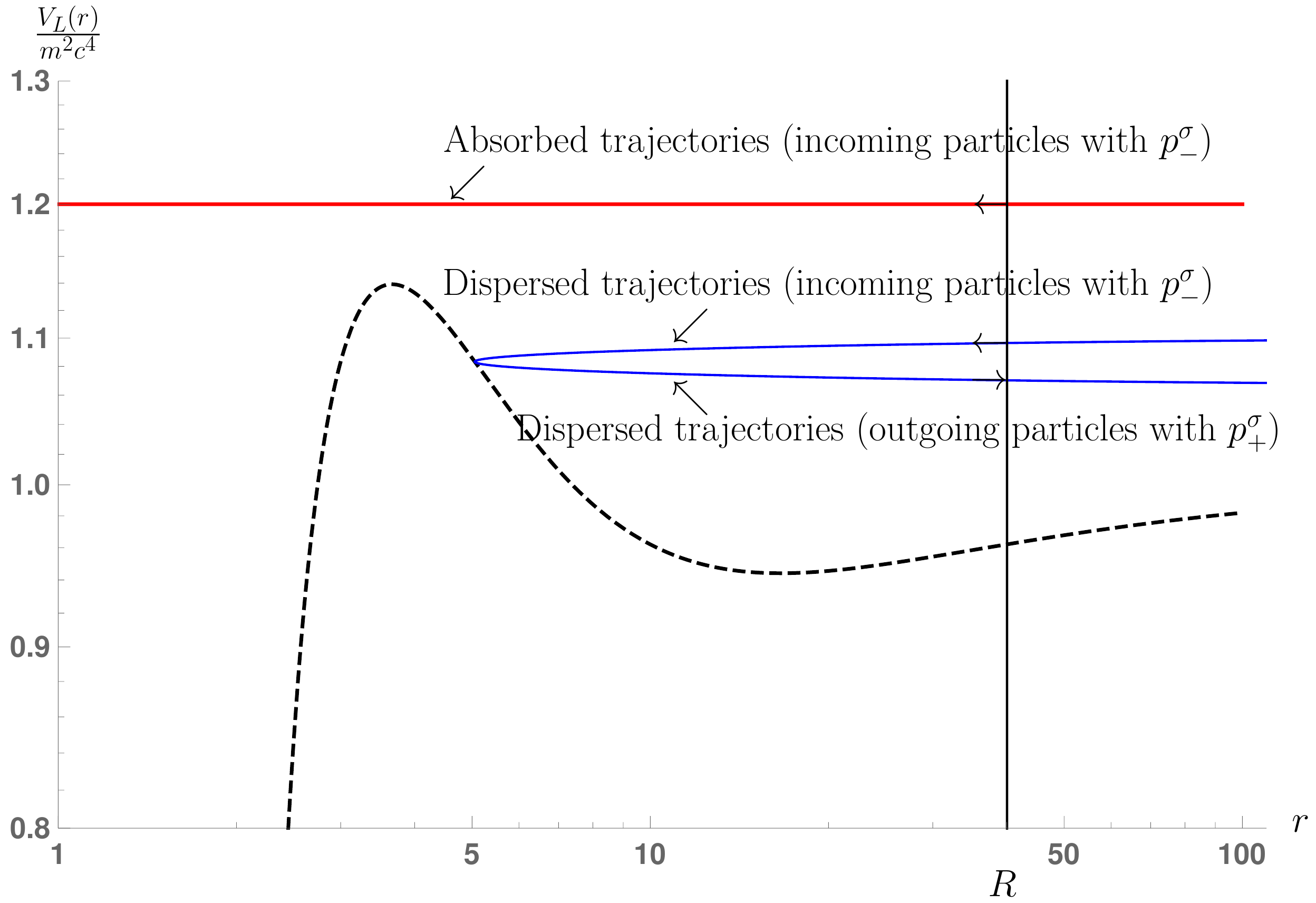}
    \caption{Plot of the Schwarzschild effective potential $V_L(r)$ vs $r$ (with $r$ in units of $r_\textrm{S}$) and $L = 4.5 G M m/c$. We have identified for the absorbed trajectories the incoming moments by $p^{\sigma}_{-}$, while for the scattered trajectories the incoming moments are $p^{\sigma}_{-}$ and the outgoing moments are $p^{\sigma}_{+}$. Here, the $\pm$ sign in $p^\sigma_{\pm}$ refers to the same sign appearing in the $p^1$ component of Eq.~(\ref{Eq:VolumeForm}). For more details on the effective potential see~\citet{MTW-Book, Straumann-Book} and Appendix A of~\citet{pRoS17a}.}
    \label{Fig:EffectiveSchPotential}
\end{figure}

For the following, it is convenient to express the momentum in terms of orthonormal components $(p^0,p^1,p^2,p^3)$ such that
\begin{equation}
    p^\mu\frac{\partial}{\partial x^\mu} = p^0\frac{1}{\alpha(r)}\frac{\partial}{\partial (ct)}
    + p^1\alpha(r)\frac{\partial}{\partial r}
    + p^2\frac{1}{r}\frac{\partial}{\partial \vartheta}
    + p^3\frac{1}{r\sin\vartheta}\frac{\partial}{\partial \varphi}
\end{equation}
where as before, $\displaystyle \alpha(r)^2= 1-\frac{r_\textrm{S}}{r}$. In terms of these orthonormal components the volume element~(\ref{eq:vol_elem}) reads
\begin{equation}
    \mbox{dvol}_x(p) = \frac{dp^1 dp^2 dp^3}{\sqrt{m^2c^2 + (p^1)^2 + (p^2)^2 + (p^3)^2}}.
\end{equation}
Expressed in terms of the integrals of motion $(E,L)$ and the angle $\chi$ defined by $L_z/L = \sin\vartheta\sin\chi$, one obtains 
\begin{gather}
    (p^\sigma_\pm) = \left(\frac{E}{c\,\alpha(r)}, \pm\frac{\sqrt{E^2 - V_L(r)}}{c\,\alpha(r)}, \frac{L\cos\chi}{r}, \frac{L\sin\chi}{r} \right), \label{Eq:VolumeForm} \\
      \mbox{dvol}_x(p) = \frac{dE\;LdL\;d\chi}{r^2\sqrt{E^2 - V_L(r)}},\label{Eq:VolumeForm2}
\end{gather}
where the $\pm$ sign in $p^1$ determines whether the particle is infalling or outgoing and $V_L(r)$ is the effective potential describing the radial motion, defined as
\begin{equation}
    V_L(r):= c^2\alpha(r)^2\left(m^2 c^2 + \frac{L^2}{r^2} \right).
\end{equation}
The behavior of the effective potential $V_L$ is well-known but critical for what follows, so we briefly review its main features (see e.g. Appendix A in \citet{pRoS17a}  for more details). For $L\leq L_{\text{ISCO}} \coloneqq \sqrt{12} G M m/c$ the function $V_L$ is monotonously increasing, which means that any infalling particle released from $r = R$ whose total angular momentum lies in this range inevitably falls into the black hole within a finite amount of its proper time. For $L > L_{\text{ISCO}}$ the function $V_L$ has a local maximum inside the interval $(3GM/c^2,6GM/c^2)$, which is due to the presence of the centrifugal term and which gives rise to a potential well with corresponding minimum lying in the interval $(6GM/c^2,\infty)$. Whether or not an infalling particle released from $r = R$ with $L > L_{\text{ISCO}}$ falls into the black hole depends on its energy (see Fig.~\ref{Fig:EffectiveSchPotential}). If $E^2$ is larger than the maximum of the potential, the particle is absorbed by the black hole; otherwise it bounces off the centrifugal barrier and is reflected towards $r = R$.

Therefore, given $L\geq 0$, the relevant energy range for describing the aforementioned accretion scenario is $\sqrt{V_L(R)}\leq E < \infty$, with particles being absorbed or reflected depending on whether or not $E^2$ is larger than the centrifugal barrier of $V_L$. (We do not consider particles with energies lower than $\sqrt{V_L(R)}$ since they correspond to either bound trajectories whose turning points $r_i$ satisfy $r_\textrm{S} < r_1 < r_2 < R$, and hence do not affect the value of $n_R$ nor the accretion rate, or to particles emanating at a radius $r < R$ which are absorbed by the black hole in finite proper time).

When computing the current density~(\ref{Eq:ParticleDensity}) with the volume form expressed in terms of $E$, $L$ and $\chi$ as in Eq.~(\ref{Eq:VolumeForm2}), the appropriate limits of integration for each variable have to be taken into account. The range for $\chi$ is obviously $(0,2\pi)$, while $0\leq L < \infty$ and $\sqrt{V_L(R)}\leq E < \infty$, as we have just described, where values of $E^2$ exceeding the local maximum of $V_L$ give rise only to incoming particles with momentum $p_-^\sigma$ and values of $E^2$ less than this maximum giving rise to both incoming and outgoing particles with momenta $p_\pm^\sigma$.

Since the distribution functions considered in this work depend only on $E$, one can perform the integrals over $L$ explicitly, as shown below. For this, it is necessary to fix the energy level first and determine the correct limits for $L$ as a function of $E$. Thus, the energy range is now $\sqrt{V_{L=0}(R)} = m c^2 \alpha(R)\leq E < \infty$, while the value of the total angular momentum $L$ is limited by the requirement that $V_L(R)\leq E^2$, which translates into an upper bound $L\leq L_{\text{max}}(E,R)$ for $L$. Furthermore, there is a critical value $L = L_{\text{c}}(E)$, corresponding to the value of $L$ for which the effective potential has a local maximum equal to $E^2$, such that particles with $L < L_{\text{c}}(E)$ are absorbed and particles with $L > L_{\text{c}}(E)$ are reflected.

By analyzing the behavior of the two limits $L_{\text{max}}(E,R)$ and $L_{\text{c}}(E)$ as functions of $E$, one finds that there is a critical energy $E = E_{\text{c}}(R)$ for which they are equal, $L_{\textrm{max}}(E_\textrm{c}(R), R) = L_{\textrm{c}}(E_{\textrm{c}})$. Furthermore, we have $L_{\text{max}}(E,R) < L_{\text{c}}(E)$ for $mc^2\alpha(R)\leq E < E_{\text{c}}(R)$, and $L_{\text{max}}(E,R) > L_{\text{c}}(E)$ when $E > E_{\text{c}}(R)$. This leads to the following characterizations in the parameter space $(E,L)$:
\begin{enumerate}
\item \emph{Absorbed particles}
$$
\begin{cases}
\quad m c^2\alpha(R) \leq E < E_\text{c}(R) \quad \textrm{and} \quad  0\leq L \leq L_{\text{max}}(E,R),\\
\quad E_\text{c}(R) < E < \infty \quad \textrm{and} \quad
    0\leq L < L_{\text{c}}(E).
\end{cases}
$$

\item \emph{Scattered particles}
\begin{equation}
    E_\text{c}(R) \leq E < \infty \quad \textrm{and} \quad
    L_{\text{c}}(E) < L < L_{\text{max}}(E,R).
\end{equation}
\end{enumerate}
The explicit expressions for $E_{\text{c}}$, $L_{\text{max}}$ and $L_{\text{c}}$ are derived in Gabarrete and Sarbach (paper in preparation) and \citet{pRoS17a} and they are:
\begin{align}
     &E_{\text{c}}(R) = m c^2\frac{R + r_\textrm{S}}{\sqrt{R(R+3r_\textrm{S})}}, \label{eq:E_c} \\
     &L_{\text{max}}(E,R) =  mc R\sqrt{\frac{E^2}{m^2 c^4\alpha(R)^2} - 1}, \label{Eq:Lmax}\\
     &L_{\text{c}}(E) =\frac{4\sqrt{2}G M m^3 c^3}{\sqrt{36m^2c^4 E^2 - 8m^4c^8 - 27E^4 + E(9E^2 - 8m^2c^4)^{3/2}}}, \label{eq:Lc}
\end{align}
where Eq.~\eqref{eq:Lc} is defined for $E \geq E_{\text{ISCO}} = 2\sqrt{2} m c^2/3$, which is the energy corresponding to the ISCO.\footnote{As $E$ increases from $E_{\text{ISCO}}$ to $\infty$, $L_\text{c}(E)$ increases from $L_{\text{ISCO}}$ to $\infty$, with $L_\text{c}(E) = 4 G M m/c$ for $E = m c^2$.} 

After these comments, it is straightforward to compute the mass accretion rate $\dot{M}$ and the particle number density $n_R$ at the injection sphere. Using Eqs.~(\ref{Eq:ParticleDensity}, \ref{Eq:VolumeForm}), we obtain
\begin{align}
        \left.  J^\sigma_{\text{abs}} \right|_{r=R} &= \, c\int\limits_{mc^2\alpha(R)}^{E_{\text{c}}(R)} \int\limits_{0}^{L_{\text{max}}(E,R)} \int\limits_{0}^{2\pi} \frac{p^\sigma_- F(E)dE LdL d\chi}{R^2 \sqrt{E^2 - V_{L}(R)}}
    \nonumber \\ 
    &+ c \int\limits_{E_{\text{c}}(R)}^{+\infty} \int\limits_{0}^{ L_{\text{c}}(E) } \int\limits_{0}^{2\pi} \frac{p^\sigma_- F(E)dE LdL d\chi}{R^2 \sqrt{E^2 - V_{L}(R)}}, 
    \label{Eq:Jabs}
\end{align}
\begin{equation}
    \left. J^\sigma_{\text{sca}} \right|_{r=R} = c\sum_{\pm} \int\limits_{E_{\text{c}}(R)}^{+\infty} \int\limits_{L_{\text{c}}(E)}^{L_{\text{max}}(E,R)} \int\limits_{0}^{2\pi} \frac{p^\sigma_\pm F(E)dE LdL d\chi}{R^2 \sqrt{E^2 - V_{L}(R)}},
    \label{Eq:Jsca}
\end{equation}
with $p^\sigma_\pm$ given by Eq.~(\ref{Eq:VolumeForm}). We note from these expressions, that only the absorbed trajectories contribute to the mass accretion rate $\dot{M}$, since the terms $p^1_+$ and $p^1_-$ in Eq.~(\ref{Eq:Jsca}) cancel each other out. In contrast to this, all the trajectories (absorbed and scattered) contribute to the particle number density $n_R$. The non-vanishing orthonormal components yield:
\begin{align}
         \left. J^{0}_{\text{abs}} \right|_{r=R} =& 
    \left\{ \int\limits_{m c^2\alpha(R)}^{+\infty} E \sqrt{E^2 - m^2 c^4 \alpha(R)^2} F(E)dE \right. \nonumber \\
    &- \left. \int\limits_{E_{\text{c}}(R)}^{+\infty} E\sqrt{E^2 - V_{\text{c}}(E,R)} F(E)dE \right\}\frac{2\pi}{\alpha(R)^3 c^2}, \label{Eq:ComponentsParticleCurrentAbs0}   
\end{align}
\begin{align}
           \left. J^{1}_{\text{abs}} \right|_{r=R} =& - \left\{ \int\limits_{m c^2\alpha(R)}^{E_{\text{c}}(R)} L_{\text{max}}(E,R)^2 F(E) dE  \right. \nonumber \\
    &+ \left. \int\limits_{E_{\text{c}}(R)}^{+\infty} L_{\text{c}}(E)^2 F(E) dE \right\}\frac{\pi}{R^2 \alpha(R)},
    \label{Eq:ComponentsParticleCurrentAbs1} 
\end{align}
\begin{equation}
       \left. J^{0}_{\text{sca}} \right|_{r=R} = \frac{4\pi}{\alpha(R)^3 c^2} \int\limits_{E_{\text{c}}(R)}^{+\infty} E \sqrt{E^2 - V_{\text{c}}(E,R)} F(E) dE,
    \label{Eq:ComponentsParticleCurrentSca0} 
\end{equation}
where we have introduced the shorthand notation $\displaystyle V_{\text{c}}(E,R):= V_{L_{\text{c}}(E)}(R)$. Note that by definition, $E_{\text{c}}(R)^2\geq V_{\text{c}}(E,R)$ for all $R\geq r_{\text{ISCO}}$ and $V_{L_{\text{c}}(E)}(R) \to m^2 c^4$ for $R\rightarrow\infty$; hence only the scattered particles yield a non-vanishing contribution to $\left. J^\alpha \right|_{r=R}$ when $R\to \infty$. Using Eqs.~(\ref{Eq:Mdotn}, \ref{Eq:ComponentsParticleCurrentAbs0}, \ref{Eq:ComponentsParticleCurrentAbs1}, \ref{Eq:ComponentsParticleCurrentSca0}), and $J^r = \alpha J^1_{\text{abs}}$, one obtains the mass accretion rate
\begin{align}
      \dot{M} &:= 4\pi m R^2 \left. J^{r} \right|_{r=R}  = - 4\pi^2 m \left\{ \int\limits_{E_{\text{c}}(R)}^{+\infty} L_{\text{c}}(E)^2 F(E) dE  \right.\nonumber\\ 
      &+ \left. \int\limits_{m c^2\alpha(R)}^{E_{\text{c}}(R)} L_{\text{max}}(E,R)^2 F(E) dE 
    \right\},
    \label{Eq:MassAccretionRate}  
\end{align}
and the particle number density at $r = R$,
\begin{equation}
    n_R = \frac{1}{c}\sqrt{\left[ J_{\text{abs}}^0(R) + J_{\text{sca}}^0(R) \right]^2 - \left[ J_{\text{abs}}^{1}(R)\right]^2}.
    \label{Eq:ParticlesDensity2}
\end{equation}
In the following, we further analyze these results for the mono-energetic and Maxwell-J\"uttner-type distributions in the energy.


\subsection{Mono-energetic model}
\label{sec:L_mono}

For the mono-energetic model $F(E) = f_0\,\delta(E - E_0)$, one obtains
\begin{flalign*}
\frac{|\dot{M}|}{m c n_R } = 4\pi R^2\alpha(R) &&
\end{flalign*}
\vspace{-0.5cm}
\begin{subequations}
\begin{numcases}{  \times}
   \displaystyle 
   \sqrt{\frac{\gamma^2 - 1}{3\gamma^2 + 1}} \quad \textrm{for} \quad 1 < \gamma < \gamma_{\text{c}}(R), \label{Eq:RelMassAccretionRateBis_a} \\
     \displaystyle \frac{h(R,\gamma)}{
\left[4\gamma^2\left(\sqrt{\gamma^2 - 1} + \sqrt{\gamma^2 - 1 - h(R,\gamma)}\right)^2  - h(R,\gamma)^2 \right]^{1/2} } \nonumber\\
\quad \textrm{for} \quad \gamma > \gamma_{\text{c}}(R), \label{Eq:RelMassAccretionRateBis_b}
\end{numcases}
\end{subequations}
where we recall that $\gamma = (1 - v_R^2/c^2)^{-1/2}$ is the Lorentz factor associated with the 3-velocity of the gas particles measured by a static observer at the injection sphere (see Section~\ref{sec:radial_r}), such that the energy $E_0$ is given by\footnote{The relation between the energy $E_0$ and the speed $v_R$ in Eq.~\eqref{eq:E0_L} can be computed using the formula $\displaystyle \frac{|\vec{v}_R|^2}{c^2} = \frac{|\vec{p}|^2}{(p^0)^2}$, with $|\vec{p}|^2 \equiv (p^1)^2 + (p^2)^2 + (p^3)^2  $, where $p^i$ are the orthonormal components defined in Eq.~\eqref{Eq:VolumeForm}. \label{footnote}}
\begin{equation}\label{eq:E0_L}
    E_0 = m c^2\alpha(R)\gamma .
\end{equation}
Further, $\gamma_{\text{c}}(R) \coloneqq E_{\text{c}}(R)/(mc^2\alpha(R))$ and $h$ denotes the function
\begin{align}
h(R,\gamma) & \coloneqq \left[ \frac{L_\textrm{c}(E_0)}{mcR} \right]^2
\nonumber\\
 & = \frac{8r_\textrm{S}^2}{R^2} \frac{1}{36\,\alpha^2\gamma^2 - 8  - 27\alpha^4\gamma^4 + \alpha\gamma [9\alpha^2\gamma^2 - 8]^{3/2}}.
\label{Eq:f1Def}
\end{align}
The formulae~(\ref{Eq:RelMassAccretionRateBis_a}, \ref{Eq:RelMassAccretionRateBis_b})  generalize the Bondi-type formula that can be found, for instance in \citet[Chapter 14, Section 2]{1983bhwd.book.....S}, to the accretion of a mono-energetic gas of arbitrary energy $E_0 > mc^2\alpha(R)$ accreting from a sphere of finite radius $R > r_{\text{ISCO}}$. 

Using the fact that for $E = E_{\textrm{c}}(R)$ one has $L_{\textrm{c}}(E) = L_{\textrm{max}}(E,R)$, it is simple to verify that $|\dot{M}|$ is continuous at the transition point $\gamma = \gamma_{\textrm{c}}(R)$, where it has the value
\begin{equation}
\frac{|\dot{M}|}{m c n_R} = \frac{4\pi r_\textrm{S}R \alpha(R)  }{\sqrt{1+\frac{2r_\textrm{S}}{R}}}.
\label{Eq:RelMassAccretionRate_max}
\end{equation}
In fact, for fixed $R$, $|\dot M|$ is a monotonically increasing function of $\gamma$  in the interval $1 < \gamma < \gamma_{\text{c}}(R)$, while it decreases monotonically for $\gamma > \gamma_{\text{c}}(R)$. Thus, Eq.~\eqref{Eq:RelMassAccretionRate_max} is the maximum accretion rate for the mono-energetic model with angular momentum. In the limit $R\to \infty$ it follows that $E_{\text{c}}(R) \rightarrow mc^2$ [see Eq.~\eqref{eq:E_c}] such that $\gamma_{\text{c}}(R) \to 1$ and  Eq.~(\ref{Eq:RelMassAccretionRateBis_b}) reduces to
\begin{align} 
\frac{|\dot{M}|}{m c n_\infty}
 &= \frac{\pi L_\textrm{c}^2(m c^2\gamma_\infty)}{m^2 c^2 \gamma_\infty \sqrt{\gamma_\infty^2 - 1}}
\nonumber\\
 & = \frac{16\pi G^2 M^2}{c^3 v_\infty}\left[
  1 + \frac{v_\infty^2}{c^2} - \frac{v_\infty^4}{c^4} + {\cal O}\left( \frac{v_\infty^6}{c^6}\right) \right],
\label{eq:acc_rate_infty}
\end{align}
where $n_\infty \coloneqq  \lim\limits_{R\rightarrow\infty} n_R$,
$v_\infty \coloneqq  \lim\limits_{R\rightarrow\infty} v_R$, and $\gamma_\infty \coloneqq  \lim\limits_{R\rightarrow\infty} \gamma_R$. The leading-order term in $v_\infty/c$ agrees with Eq.~(14.2.20) in \citet{1983bhwd.book.....S}.

Comparing Eq.~\eqref{Eq:RelMassAccretionRateBis_a} with the corresponding expression for the mass accretion rate in the absence of angular momentum [see Eq.~\eqref{Eq:AccretionRateRelativisticMonoEner}], the difference relies in the factor $(3\gamma^2+1)^{-1/2}\leq 1$ which implies that for $\gamma < \gamma_{\text{c}}(R)$ the accretion rate is smaller when the angular momentum is considered. This is expected since the tangential movement of particles with angular momentum reduces the net infall of particles. Note that in the non-relativistic limit $\gamma\to 1$ and fixed $R$ one obtains half the value given in Eq.~(\ref{eq:mono_rel_rad_stationary}) computed for the purely radial infall. As further analyzed in Appendix~\ref{ap:fixedLmodels}, this is due to the fact that when angular momentum is present, the three-velocity contains non-trivial angular components.

A simplified form of Eqs.~(\ref{Eq:RelMassAccretionRateBis_a}, \ref{Eq:RelMassAccretionRateBis_b}) can be obtained in the limit when the injection sphere is far from the horizon: $R\gg r_\textrm{S}$ and for non-relativistic energies, such that $v_R \ll c$. For this, one notices that
\begin{equation}
\gamma_{\text{c}}(R)-1 = 2\left(\frac{r_\textrm{S}}{R}\right)^2 + \mathcal{O}\left(\frac{r_\textrm{S}}{R}\right)^3,
\label{Eq:KExp}
\end{equation}
and that the denominator of the second factor on the right-hand side of \eqref{Eq:f1Def} converges to $2$ when $\alpha(R)\to 1$ and $\gamma\to 1$. Using this, one finds to leading order,
\begin{subequations}
\begin{numcases}{ \frac{|\dot{M}|}{m c n_R} = 4\pi R^2\times}
   \displaystyle 
   \frac{v_R}{2c} \quad \textrm{for} \quad  \frac{v_R}{2c} < \frac{r_\textrm{S}}{R},
   \label{Eq:RelMassAccretionRateApprox_a}\\
     \displaystyle \frac{2c}{v_R}\left( \frac{r_\textrm{S}}{R} \right)^2\frac{1}{1 + \sqrt{1 - \left( \frac{2c}{v_R} \frac{r_\textrm{S}}{R} \right)^2}} 
     \nonumber\\ \quad \textrm{for} \quad \frac{v_R}{2c} > \frac{r_\textrm{S}}{R}, 
     \label{Eq:RelMassAccretionRateApprox_b}
\end{numcases} 
\end{subequations}
which is valid for $R \gg r_\textrm{S}$ and $v_R\ll c$.

\begin{figure}
	\includegraphics[width=\columnwidth]{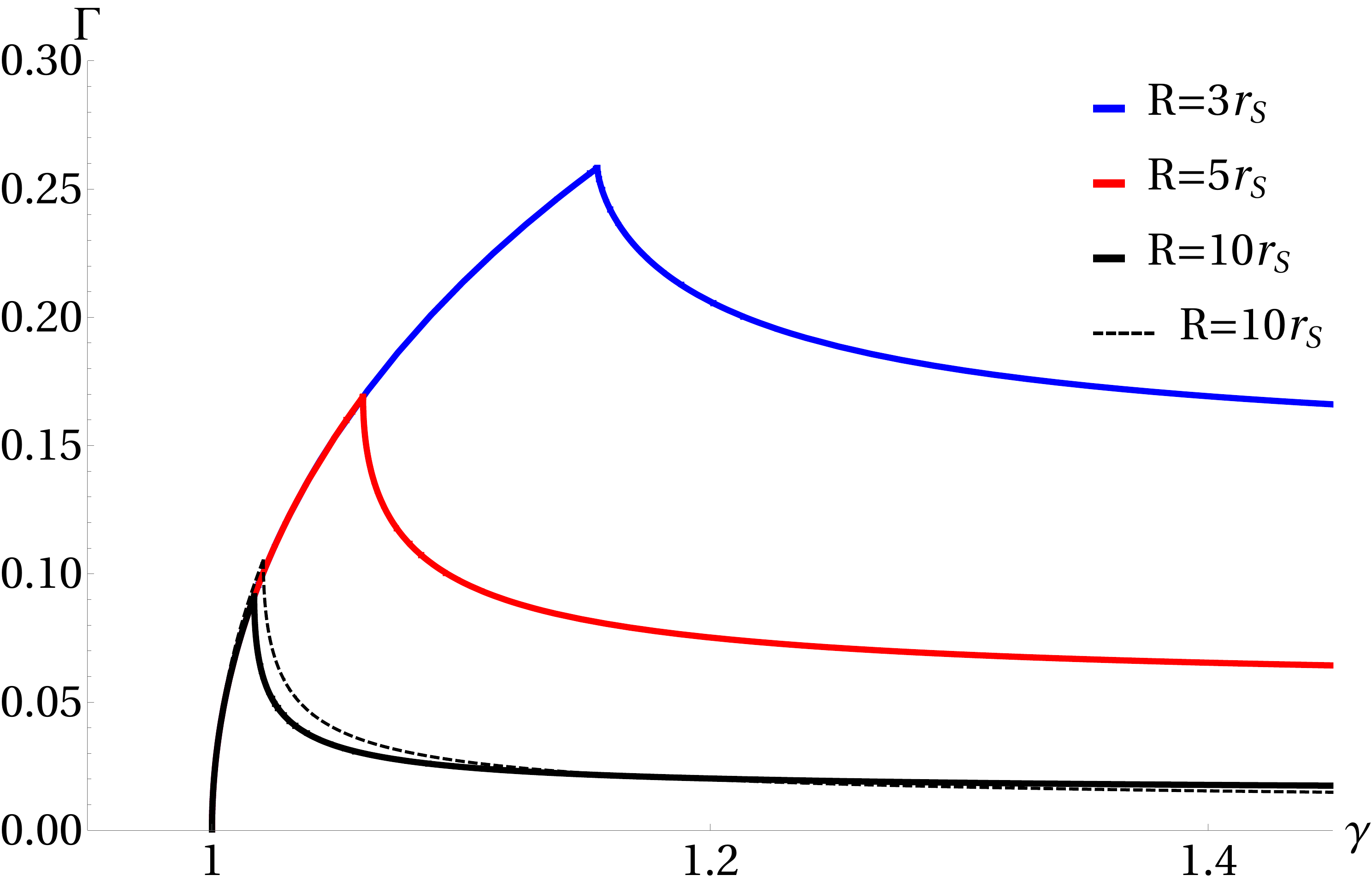}
    \caption{The dimensionless quantity $\Gamma = |\dot{M}|/(4\pi R^2\alpha(R) m c n_R)$ vs the Lorentz factor $\gamma = (1-v_R^2/c^2)^{-1/2}$ for some fixed values of the injection sphere's radius $R$.
    The solid lines are computed from Eqs.~(\ref{Eq:RelMassAccretionRateBis_a}, \ref{Eq:RelMassAccretionRateBis_b}) for different values of $R$. The black dashed line shows the same quantity $\Gamma$ for the case $R=10r_\textrm{S}$, using the approximation from Eqs.~(\ref{Eq:RelMassAccretionRateApprox_a}, \ref{Eq:RelMassAccretionRateApprox_b}) which is valid for $R\gg r_\textrm{S}$ and non-relativistic velocities $v_R \ll c$.}
    \label{Fig:MassAccretionRateMonoEnergetic}
\end{figure}

In Fig.~\ref{Fig:MassAccretionRateMonoEnergetic} we show the behaviour of the dimensionless quantity $\Gamma = |\dot{M}|/(4\pi R^2\alpha(R) m c n_R)$ as a function of $\gamma$ for different values of $R$. As can be observed from this figure, $|\dot{M}|$ increases with $\gamma$ for small velocities $v_R$, the quantity $\Gamma$ being independent of $R$, as follows from Eq.~(\ref{Eq:RelMassAccretionRateBis_a}). Hence, in this regime the qualitative behaviour of the accretion rate as a function of $v_R$ is similar to the case of purely radial infall (the only difference consisting of the factor $(3\gamma^2+1)^{-1/2}$, as explained above). However, as soon as $\gamma$ reaches the critical value $\gamma_{\textrm{c}}(R)$, $|\dot{M}|$ starts decreasing, converging to a finite ($R$-dependent value) in the limit $\gamma\to \infty$. This can be understood as follows: when $\gamma < \gamma_{\textrm{c}}(R)$, all the particles have their energy below the critical value $E_{\textrm{c}}(R)$ and thus all of them are absorbed by the black hole. This leads to an accretion rate which increases with $v_R$. However, when $\gamma > \gamma_{\textrm{c}}(R)$, the particles have their energy lying above $E_{\textrm{c}}(R)$ and hence a fraction of them (namely, those with angular momentum larger than $L_{\textrm{c}}(E)$) are scattered off the effective potential, leading to a smaller accretion rate. As $v_R$ increases this fraction becomes larger which leads to a smaller mass accretion rate (see~\citet{pRoS17b} for a more extended discussion regarding this effect for a similar model with $R\to \infty$).

\subsection{Maxwell-J\"uttner-type distribution function}
\label{sec:L_jutt}

Next, we analyze the Maxwell-J\"uttner-type distribution~(\ref{eq:max_bolt}) which was also considered in \citet{pRoS17a, pRoS17b}.\footnote{Again, one should be careful with associating $T$ with temperature. Although in this section the gas particles are not restricted to zero angular momentum, the gas is still not in strict  thermodynamic equilibrium at finite $R$ because we are not considering hypothetical incoming particles emanating from the white hole. See the discussion in Section 4 of \citet{pRoS17b}.} To understand this limit, it is convenient to perform the variable substitutions $E = m c^2\alpha(R)(1 + x/z)$ and $E = E_{\text{c}}(R)(1 + y/z)$ in the integrals Eqs.~(\ref{Eq:ComponentsParticleCurrentAbs0}, \ref{Eq:ComponentsParticleCurrentAbs1}, \ref{Eq:ComponentsParticleCurrentSca0}), where we set $z(R,T):=z=m c^2\beta\alpha(R)$. This yields
\begin{align}
J^0 &\coloneqq \left.\left( J^0_{\text{abs}} + J^0_{\text{sca}}\right)\right|_{r=R} = \frac{2\pi A m^3 c^4}{z^{3/2}} e^{-z} I_1(R,z),
\label{Eq:J0}\\
J^1 &\coloneqq \left. J^1_{\text{abs}} \right|_{r=R}
 = -\frac{\pi A m^3 c^4}{z^2} e^{-z} I_2(R,z),
\label{Eq:J1}
\end{align}
with the integrals $I_1(R,z)$ and $I_2(R,z)$ given by
\begin{align}
   I_1(R,z) &= \int\limits_0^{\infty}\left(1+\frac{x}{z}\right)\sqrt{2x + \frac{x^2}{z}} e^{-x} dx 
\nonumber\\
 &+ \gamma_{\text{c}}(R)^3
 e^{-\Lambda(R,z)} 
 \int\limits_0^\infty \left(1+\frac{y}{z}\right) e^{-\gamma_{\text{c}}(R) y} \nonumber \\
 & \times \sqrt{z\left[ 1 - \frac{V_{\text{c}}\left[ E_{\text{c}}(R)\left(1+\frac{y}{z}\right),R\right] }{E_{\text{c}}(R)^2} \right] + 2y + \frac{y^2}{z}}  dy,
\label{Eq:I1} 
\end{align}
\begin{align}
    I_2(R,z) &=  \int\limits_0^{\Lambda(R,z)} \left(2x + \frac{x^2}{z} \right) e^{-x} dx
 + \frac{\gamma_{\text{c}}(R)}{R^2}z e^{-\Lambda(R,z)} \nonumber \\
 & \times  \int\limits_0^\infty \frac{L_{\text{c}}\left[  E_{\text{c}}(R)\left(1+\frac{y}{z}\right) \right]^2}{m^2 c^2} e^{-\gamma_{\text{c}}(R) y} dy,
\label{Eq:I2}
\end{align}
where we recall the shorthand notation $\gamma_\text{c}(R) \coloneqq E_{\text{c}}(R)/(mc^2\alpha(R))$ and where we have set $\Lambda(R,z) \coloneqq (\gamma_\text{c}(R)-1)z$. From Eqs.~(\ref{Eq:J0}, \ref{Eq:J1}), one obtains the following expression for the mass accretion rate:
\begin{equation}
\frac{|\dot{M}|}{m c n_R} = \frac{4\pi R^2\alpha(R)}{\sqrt{4z \left[\frac{I_1(R,z)}{I_2(R,z)}\right]^2 - 1}}.
\label{Eq:MassAccretionJüttner}
\end{equation}
Eq.~(\ref{Eq:MassAccretionJüttner}), together with the integrals defined in Eqs.~(\ref{Eq:I1}, \ref{Eq:I2}), provides an exact expression for the mass accretion rate as a function of the injection radius $R$ and the temperature $T$. Unfortunately, the integrals involved are rather complicated, and for this reason it is advantageous to obtain simplified expressions for certain limits. One such expression can be obtained assuming that the gas temperature is low, such that $k_\textrm{B} T \ll m c^2$, and that $R \gg r_\textrm{S}$ is much larger than the Schwarzschild radius of the accreting black hole. In order to discuss this limit, we first note that
\begin{equation}
  1 - \frac{V_{\text{c}}\left[ E_{\text{c}}(R) \left( 1 + \frac{y}{z} \right) , R \right]}{E_{\text{c}}^2(R)} =  - \frac{16r_\textrm{S}^2}{R^2 - r_\textrm{S}^2} \frac{y}{z} + \mathcal{O}\left(\frac{y^2}{z^2}\right),
\end{equation}
and hence for $z\gg 1$ one obtains
\begin{align}
I_1(R,z) &\approx \int\limits_0^\infty \sqrt{2x} e^{-x} dx \nonumber \\
 &+ \gamma_{\text{c}}(R)^3
 e^{-\Lambda(R,z)} 
 \int\limits_0^\infty \sqrt{2y
 \frac{1-\left(3r_\textrm{S}/R\right)^2}{1-\left(r_\textrm{S}/R\right)^2}}
e^{-\gamma_{\text{c}}(R) y} dy.
\end{align}
Now the integrals can be evaluated explicitly which yields, for $z\gg 1$,
\begin{equation}
I_1(R,z)\approx \sqrt{\frac{\pi}{2}}\left[ 1 + \gamma_{\text{c}}(R)^{3/2}
 e^{-\Lambda(R,z)} \sqrt{
 \frac{1-\left(3r_\textrm{S}/R\right)^2}{1-\left(r_\textrm{S}/R\right)^2}} \right].
\label{Eq:I1Bis}
\end{equation}
Similarly,
\begin{align}
I_2(R,z) &\approx 2 - 2[1+\Lambda(R,z)] e^{-\Lambda(R,z)} \nonumber \\
 &+ \frac{4r_\textrm{S}^2}{R^2[\gamma_{\text{c}}(R)-1]}\frac{\Lambda(R,z) e^{-\Lambda(R,z)}}{\alpha(R)^2\left(1+3r_\textrm{S}/R\right)},
\label{Eq:I2Bis}
\end{align}
for $z\gg 1$, where we have used that $\displaystyle L_{\text{c}}\left[E_{\text{c}}(R)\right]^2/(m c)^2 = 4r_\textrm{S}^2/[\alpha(R)^2\left(1+3r_\textrm{S}/R\right)]$. The expressions~(\ref{Eq:I1Bis}, \ref{Eq:I2Bis}) are valid when $z$ is much larger than one, independent of the value of $R$. When $R\gg r_\textrm{S}$ one can use the expansion~(\ref{Eq:KExp}) to show that $\Lambda(R,z) \approx 2(r_\textrm{S}/R)^2 z$. Therefore, $\Lambda(R,z)$ depends on the ratio between the two large quantities $z$ and $R^2$, implying that it varies over the whole range $(0,\infty)$. Assuming that $R\gg r_\textrm{S}$ in Eqs.~(\ref{Eq:I1Bis}, \ref{Eq:I2Bis}) leads to a further simplification,
\begin{align}
   I_1(R,z) &\approx \sqrt{\frac{\pi}{2}}\left[ 1 + e^{-\Lambda(R,z)} \right],  \\
   I_2(R,z) &\approx 2\left[ 1 - e^{-\Lambda(R,z)} \right],
\end{align}
and introduced into Eq.~(\ref{Eq:MassAccretionJüttner}) one obtains the simple expression
\begin{equation}
\frac{|\dot{M}|}{m c n_R} \approx
 R^2 \alpha(R) \tanh\left( \frac{r_\textrm{S}^2}{R^2}z \right) \sqrt{\frac{32\pi}{z}},
\label{Eq:MassAccretionJüttnerApprox}
\end{equation}
which is valid for arbitrary values of $R \gg r_\textrm{S}$ and $z = m c^2/(k_\textrm{B} T) \gg 1$. In the limit $R\to \infty$ one obtains, setting $\displaystyle z_\infty := \lim\limits_{R\to\infty} z$,
\begin{equation}
\frac{|\dot{M}|}{m c n_\infty} \approx \sqrt{32\pi z_\infty} r_\textrm{S}^2,
\label{eq:Juttner1}
\end{equation}
which agrees with Eq.~(87) in \citet{pRoS17a}.

\begin{figure}
	\includegraphics[width=\columnwidth]{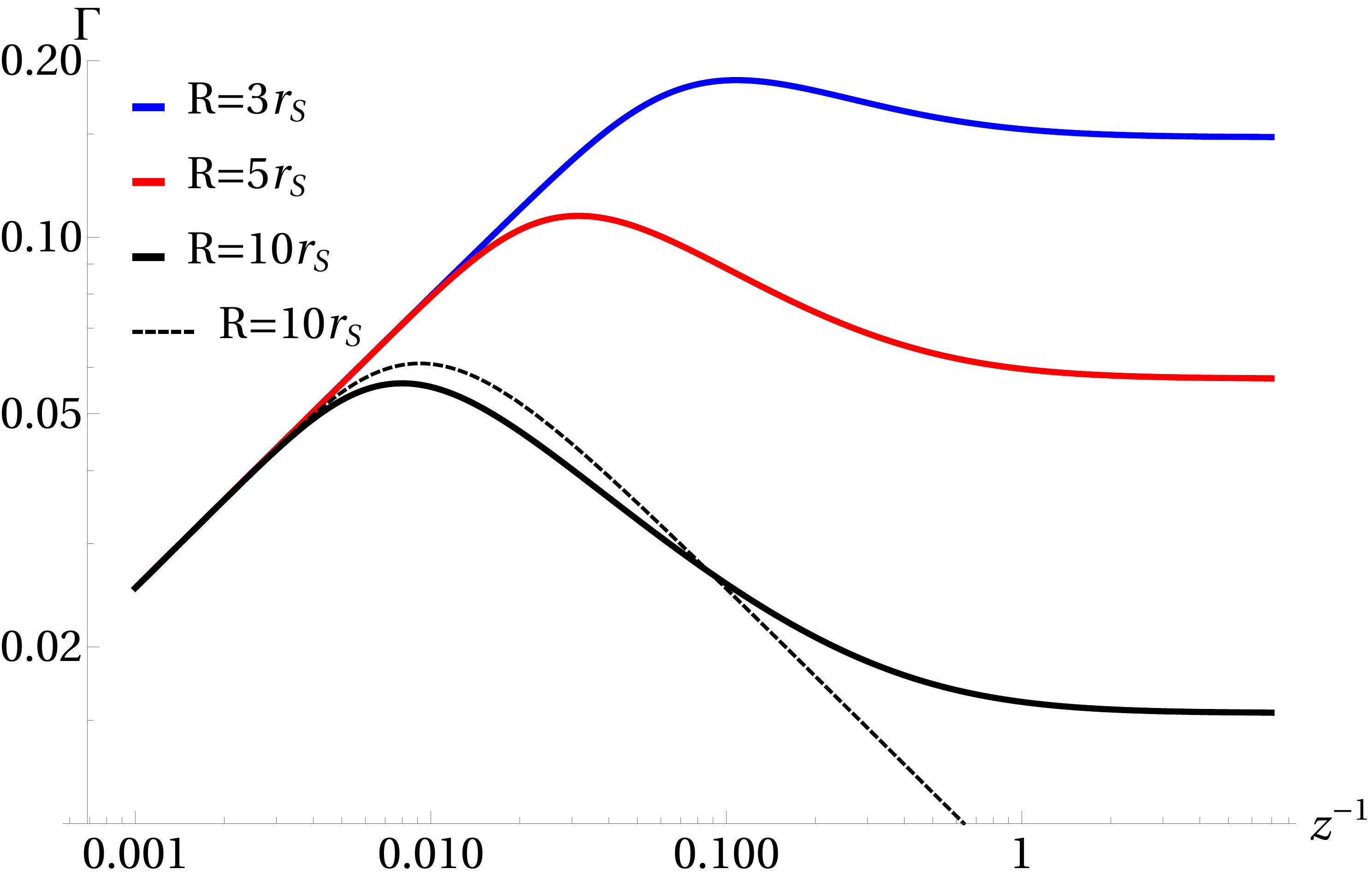}
    \caption{The dimensionless quantity $\Gamma = |\dot{M}|/(4\pi R^2\alpha(R) m c n_R)$ vs $z^{-1} = k_B T/(mc^2\alpha(R))$ for some fixed values of the radius $R$ of the injection sphere. The solid lines are computed from Eq.~(\ref{Eq:MassAccretionJüttner}) for different values of $R$.
    The black dashed line shows the same quantity $\Gamma$ for the case $R=10r_\textrm{S}$, using the approximation from Eq.~(\ref{Eq:MassAccretionJüttnerApprox}) which is valid for $R\gg r_\textrm{S}$ and non-relativistic temperatures  $z^{-1} \ll 1$.}
    \label{Fig:MassAccretionRateJüttner}
\end{figure}

In Fig.~\ref{Fig:MassAccretionRateJüttner} we show the dimensionless quantity $\Gamma = |\dot{M}|/(4\pi R^2\alpha(R) m c n_R)$ as a function of the temperature $T$ for different values of $R$. The behaviour is very similar to the one of the mono-energetic model, except that the function is smooth at the maximum value of the accretion rate, which is due to the non-trivial velocity dispersion in the distribution function.

\section{Summary of analytic models}
\label{sec:summary}

In this section we provide a summary of the expressions we have obtained for the mass accretion rate in the different models presented in Sections \ref{sec:radial_infall} and \ref{sec:finite}. We restrict ourselves to the relativistic models which describe the spherical accretion of a collisionless kinetic gas onto a Schwarzschild BH. In all these models, the mass accretion rate can be written in the following general form:
\begin{equation}
|\dot{M}| = 4\pi R^2 \alpha(R) m c n_R \Gamma,
\end{equation}
where $R$ is the areal radius of the injection sphere, $\alpha(R) = \sqrt{1 - r_\textrm{S}/R}$, with $r_\textrm{S} = 2GM/c^2$ the Schwarzschild radius of the black hole, $m$ the mass of the particles and $n_R$ the particle density at the injection sphere. Here, $\Gamma$ is a model-dependent dimensionless factor which is defined as follows.
\begin{enumerate}
\item Purely radial mono-energetic model:
\begin{equation}
\Gamma = \frac{v_{R}}{c}
\frac{1}{\sqrt{1 - \frac{v_R^2}{c^2}}},
\end{equation}
where $v_R$ is the magnitude of the three-velocity measured by static observers at the injection sphere.

\item Purely radial Maxwell-J\"uttner model:
\begin{equation}
\Gamma = \frac{1}{\sqrt{\left[ \mathbf{K}_1(z) z e^{z} \right]^2 - 1 }},
\end{equation}
where $z = mc^2\alpha(R)/k_\textrm{B} T$ with $T$ the gas temperature at the injection sphere and where $\textbf{K}_1(z)$ is the modified Bessel function of the second kind of first order. In the low-temperature limit $z\gg 1$ this factor reduces to $\Gamma \approx \sqrt{2/(\pi z)}$.

\item Mono-energetic model with angular momentum: In this case, $\Gamma$ can be read off from Eqs.~(\ref{Eq:RelMassAccretionRateBis_a}, \ref{Eq:RelMassAccretionRateBis_b}). However, in the limit $R \gg r_\textrm{S}$ and $v_R\ll c$ this simplifies to
\begin{subequations}
\begin{numcases}{\Gamma \approx }
    \displaystyle
    \frac{v_R}{2c}, \quad \textrm{for} \quad \frac{v_R}{2c} < \frac{r_\textrm{S}}{R}, \\
    \displaystyle
    \frac{\frac{2c}{v_R}\left( \frac{r_\textrm{S}}{R} \right)^2}{1 + \sqrt{1 - \left( \frac{2c}{v_R}\frac{r_\textrm{S}}{R} \right)^2}}, \quad \textrm{for} \quad \frac{v_R}{2c} > \frac{r_\textrm{S}}{R}.
\end{numcases} 
\end{subequations}
\item Maxwell-J\"uttner model with angular momentum:
\begin{equation}
\Gamma = \frac{1}{\sqrt{4z \left[\frac{I_1(R,z)}{I_2(R,z)}\right]^2 - 1}},
\end{equation}
where the integrals $I_1$ and $I_2$ are defined in Eqs.~(\ref{Eq:I1}, \ref{Eq:I2}). In the limit $R \gg r_\textrm{S}$ and $z \gg 1$ this simplifies to
\begin{equation}
\Gamma \approx \sqrt{\frac{2}{\pi z}}\tanh\left( \frac{r_\textrm{S}^2}{R^2}z \right).
\label{Eq:GammaMaxJutt}
\end{equation}
Note that in the limit $z\gg R^2/r_\textrm{S}^2$ one obtains precisely the same result as the low-temperature limit of the purely radial Maxwell-J\"uttner model. This shows that at very low temperatures the angular momentum is unimportant which is expected, since at low temperature most of the particles have low energy and hence must have low angular momentum as well.\footnote{See Eq.~(\ref{Eq:Lmax}): $L_{\text{max}}$ is small if $E$ is close to its minimum value $\alpha(R) m c^2$.} 
\end{enumerate}


\section{Applications}
\label{sec:applications}

In this section, we discuss a couple of astrophysical scenarios which allow us to point out some important quantitative differences between the mass accretion rates predicted by our and previous models.

\subsection{Accretion onto Sgr~A* from matter located near the Bondi radius}

In most scenarios, the Bondi accretion can be considered as a reliable first approximation.\footnote{Strictly speaking, the mass accretion rate for the case the central object is a black hole should be computed by means of a general relativistic calculation. However, when $\gamma_\textrm{a}\leq 5/3$ and for low temperatures (such that $k_\textrm{B} T_\infty/(m_\textrm{p}c^2)\ll1$), the Newtonian calculation yields a very good approximation to the general relativistic case even if the central black hole is rotating~\citep{Aguayo-Ortiz:2021jzv}. } For a polytropic transonic accretion flow, the  Bondi mass accretion rate $\dot M_\textrm{B}$ is defined through the following expression (\citet{1952MNRAS.112..195B, Korol_2016}):
\begin{equation}\label{eq:Bondi_1952}
    \frac{|\dot{M}_{\textrm{B}}|}{c_{\infty} \rho_\infty } = 4\pi\lambda r_\textrm{B}^2,
\end{equation}
where $c_{\infty}$ is the sound speed of the fluid at infinity, $\rho_\infty$ is the fluid density at infinity, and $\lambda$ is a function of the adiabatic index of the fluid $\gamma_\textrm{a}$ (e.g. \citet{1952MNRAS.112..195B}). For a monoatomic adiabatic process, one has $\gamma_\textrm{a}=5/3$ and $\lambda=1/4$. Note that we have written Eq.~(\ref{eq:Bondi_1952}) as a ratio in order to keep the same format that we followed in the previous sections. The Bondi radius $r_\textrm{B}$ in Eq.~(\ref{eq:Bondi_1952}) is defined as
\begin{equation}\label{eq:bondi}
    r_\textrm{B} = \frac{GM}{c_{\infty}^2}.
\end{equation}
Note that the adiabatic speed of sound of the fluid is defined as
\begin{equation}
c_s = \left( \frac{\partial P}{\partial\rho} \right)^{1/2}
= \left( \gamma_\textrm{a}\frac{P}{\rho} \right)^{1/2}
 = \left( \frac{\gamma_\textrm{a} k_\textrm{B} T}{\mu m_\textrm{p}} \right)^{1/2},
\end{equation}
where the ideal gas equation of state is assumed, $P=\frac{\rho k_\textrm{B} T}{\mu m_\textrm{p}}$, and $\mu = \langle m \rangle/m_\textrm{p}$ is the mean molecular weight which depends on the ionisation state of the gas. The relation between the particle number density $n$ and mass density is $\rho = n \langle m \rangle$, and the former has a contribution from the electrons and ions. Therefore it is common to consider the mean molecular weight separately. In the case of electrons, the mean molecular weight per electron is $\mu_\textrm{e} \approx 2/(1+X)$, where $X$ is the hydrogen mass fraction. For example, a fully ionised purely hydrogen gas has $\mu = 0.5$ and $\mu_\textrm{e} = 1$.

In practice, one substitutes the values of the particle number density and temperature obtained from X-ray observations at a finite radius from the SMBH in the Bondi accretion rate model given by Eq.~(\ref{eq:Bondi_1952}). In this way, considering the characteristic parameters for Sgr~A*: a mass of $\sim 4.3 \times 10^6 \, \mathrm{M}_{\odot}$  (\citet[][]{2003ApJ...586L.127G,2009ApJ...692.1075G}), a flow with temperature of $2.2 \times 10^7$\,K (1.9\,keV), a sound speed of 550\,km\,s$^{-1}$ and an electron number density of 160\,cm$^{-3}$ measured at $R\sim 0.06$\,pc  (e.g. \citet{2003ApJ...591..891B, 2013Natur.501..391E}), the Bondi accretion rate yields
\begin{align}
 |\dot{M}_\textrm{B}| &\sim 10^{-4} \, \mathrm{M}_{\odot} \mathrm{yr}^{-1} \left( \frac{\mathrm{M}}{4.3 \times 10^6 \mathrm{M}_{\odot}}\right)^2 \left( \frac{n_\textrm{e}}{160 \, \mathrm{cm}^{-3}}\right) \nonumber \\ &\times  \left( \frac{k_\textrm{B} T}{1.9 \, \mathrm{keV}} \right)^{-\frac{3}{2}},\label{eq:Bondi_values}  
\end{align}
where an adiabatic index $\gamma_\textrm{a}=5/3$, $\rho = \mu_\textrm{e} m_\textrm{p} n_\textrm{e}$ and $\mu = 1$ and $\mu_\textrm{e} = 1$ was assumed as in \citet{Falcke_2013}. In this case, it is implicit that one assumes the selected finite radius to be a good approximation for the values $n_{\infty}$ and $T_{\infty}$. Nevertheless, this assumption can lead to an overestimate of the actual mass accretion rate (see e.g. \citet{Korol_2016}. In the past years, accretion models based on numerical hydrodynamical simulations have estimated a mass accretion rate of order $\sim10^{-6} {M}_{\odot} \, \mathrm{yr}^{-1}$ at the Bondi radius scales (e.g. \citet{2006JPhCS..54..436C, 2008MNRAS.383..458C,cuadra2015}).

Another plausible way to estimate the mass accretion rate in these systems is through rotation measures (RM) derived from the observed polarized emission as has been previously done with Sgr~A*. The RM is proportional to the integrated electron density and the parallel component of the magnetic field to the line of sight.\footnote{RM = $0.81 \int n_\textrm{e} B_{\parallel} \, dl$ rad m$^{-2}$, with $n_\textrm{e}$ in units of cm$^{-3}$, $B$ in $\mu$G and $dl$ in pc.} Therefore, the interpretation of the RM relies on a radial model for both the density and the magnetic field. On the other hand, the radial dependency changes according to the assumed accretion model (e.g. Bondi or the various RIAF models). These model-dependent radial profiles are used in combination with certain assumptions on the magnetic field, such as equipartition between the magnetic and gravitational energy, in order to get upper and lower limits on the mass accretion rate in the vicinity of the BH horizon (\citet{2003ApJ...588..331B,2006ApJ...640..308M,2007ApJ...654L..57M}). 

In these polarimetry observations, a low mass accretion rate near Sgr~A*'s horizon is inferred from the RM of a few $\times 10^5$\,rad\,m$^{-2}$ at 230 GHz (\citet{2006ApJ...640..308M,2018ApJ...868..101B}). The inferred mass accretion rate lies in the range of $10^{-9}$--$10^{-7} \, \mathrm{M}_{\odot}$yr$^{-1}$, or even lower depending on the assumed inner and outer radii enclosing the polarized emission (see fig. 4 in \citet[][]{2006ApJ...640..308M}, and \citet[][]{2007ApJ...654L..57M}). The polarized emission is variable in time and as a consequence, there is variability on the RM. This RM variability can then be translated into a specific radius from which one takes the RM to originate, i.e. the observed RM comes from a wide range of radii. 
At very large radii, the RM inferred and extrapolated mass accretion rate is at least one order of magnitude smaller than the $\sim 10^{-5}$--$10^{-4} \, \mathrm{M}_{\odot}$yr$^{-1}$ rate estimated from X-ray measurements at a few $\times 10^5 \, r_\textrm{S}$ (see \citet{2003ApJ...588..331B}). At such radii and assuming a magnetic field strength of $\sim 1$ mG, the RM should be of the order of $\sim 10^3$ rad m$^{-2}$, which is two orders of magnitude smaller than what is obtained in observations. In principle this suggests that the observed RM is produced at distances much smaller than $10^5 \, r_\textrm{S}$. Nevertheless, \citet{2018ApJ...868..101B} more recently carried out a careful study on the RM variability of Sgr~A* where they found that the variability remains consistent with the average long-term RM value. Based on this long-term variability result, the authors suggest that the magnetic configuration at large radii could be stable and therefore,  they support the interpretation of the observed Faraday Rotation to arise mainly from $\sim 10^3$--$10^5 \,r_\textrm{S}$ (see more details in \citet{2018ApJ...868..101B}). This means that overall, RM observations favor models predicting low mass accretion rates in the vicinities of Sgr~A*'s horizon.

In the following, we estimate the mass accretion rate of Sgr A* for our different models. One variable needed for our estimates is the velocity of the particles at the injection radius, $v_R$.
Although this velocity is not known a priori, it is reasonable to suppose that it is of the order of the fluid sound speed or to assume that its value is of the order of the wind velocity from known massive stars (O-type stars and/or Wolf–Rayet (WR) stars) that are embedded within the dilute accretion flow towards Sgr~A*. These wind velocities are of the order of $\sim 450$--3000\,km\,s$^{-1}$ depending if they are O-type stars  (e.g. \citet{1990MNRAS.244..706A,1996A&A...305..171P,2004A&A...415..349R}), or WR stars (\citet{2006ApJ...643.1011P,2007A&A...468..233M}). The accretion flow of Sgr~A* from stellar winds has been extensively studied using hydrodynamical numerical simulations. For example, results from simulations of wind-fed accretion from WR stars have been shown to be consistent with observational constraints such as X-ray luminosities and RM 
(\citet{2006JPhCS..54..436C, 2008MNRAS.383..458C, 2018MNRAS.478.3544R, 2020ApJ...896L...6R,2020ApJ...888L...2C}). Knowing that these stellar winds could be the main contributors of the accretion towards Sgr~A*, we will assume an infall velocity that lies within the aforementioned velocity range. 

In our models, we use the same temperature for the cases described by a Maxwell-Boltzmann or Maxwell-J\"uttner distribution, whereas in the mono-energetic cases we convert this temperature to a velocity by choosing $v_R$ such that $m_\textrm{p} v_R^2$ matches $2k_\textrm{B} T$ in a first approximation. This yields $v_R=600$\,km\,s$^{-1}$, which lies within the velocity range of stellar winds. Furthermore, we also use this $v_R$ to compute the mass accretion rate for the Shapiro-Teukolsky model for mono-energetic particles (Eq.~(14.2.20) in \cite{1983bhwd.book.....S}) and the same temperature for the Rioseco-Sarbach model (Eq.~(87) in \cite{pRoS17a}). We show the comparisons in Table~\ref{table:applications}. The mass accretion rates are obtained at the fixed finite radius $R=0.06$\,pc for the different models studied in this work. Note that in this case we have considered the observational approach and assumed that the values of $n_{\infty}$ or $n_R$ correspond to the measured electron number density $n_\textrm{e}$ at $R=0.06$\,pc as is regularly done when applying the Bondi model. 

We see from Table~\ref{table:applications} that our models predict significantly different mass accretion rates depending on whether or not the infalling particles have angular momentum. The results from the purely radial infall in the kinetic description, both in the relativistic and non-relativistic cases, agree in order of magnitude with those of the hydrodynamical Bondi model.
In contrast to this, the models with angular momentum predict a significantly lower mass accretion rate (by about $\sim4$--5 orders of magnitude) than the Bondi formula, and have rates similar to the ones from the Zeldovich-Novikov model \cite{1971reas.book.....Z, 1983bhwd.book.....S}, where the infall is assumed to start from infinity, clearly indicating that the angular momentum is a decisive parameter in determining the magnitude of the mass accretion rate. 

To shed some light on these results, we first note that in the scenario considered in Table~\ref{table:applications}, the parameter $z \approx m c^2/k_\textrm{B} T \approx 0.5\times 10^6$ is still smaller than the ratio $\left(\frac{R}{r_\textrm{S}}\right)^2 \approx 2.1 \times 10^{10}$, and thus the results for Maxwell-J\"uttner model with or without angular momentum differ significantly.\footnote{The only way the role of the angular momentum could be neglected is to have $z \gg\left(\frac{R}{r_\textrm{S}}\right)^2$. This means that only at very low temperatures (of the order of the Cosmic Microwave background, $T \sim 2.73$\,K or lower), the models with and without angular momentum yield comparable mass accretion rates for the ratio between $R$ and $r_\textrm{S}$ considered in our example.} Next, we note that the ratio between the Bondi radius and the radius of the injection sphere is about $r_{\textrm{B}}/R \approx 1.02$. Furthermore, we observe that the models describing a pure radial infall can be written in the form (taking into account the aforementioned relation $2k_\textrm{B} T = m v_R^2$ which yields $v_R/c\approx \sqrt{2/z}$):
\begin{equation}
\frac{|\dot{M}|}{m n_R v_R}
 = 4\pi\lambda_{L=0} R^2,
\end{equation}
with the factor $\lambda_{L=0} = \alpha(R)\Gamma c/v_{\textrm{R}}$ being of order unity. This has the same form as the Bondi formula~(\ref{eq:Bondi_1952}) with the Bondi radius $r_\textrm{B}$ replaced with $R$ and the sound speed $c_\infty$ replaced with $v_R$. Since in our example $R\approx
r_\textrm{B}$ and $v_R$ is comparable with $c_\infty$, it follows that the mass accretion rates yield similar results. In contrast, the models with angular momentum in the limit $z\ll (R/r_\textrm{S})^2$ relevant for our example have
\begin{equation}
\frac{|\dot{M}|}{m n_R v_R} 
 = 4\pi \lambda_{L>0} r_\textrm{S}^2 z,
\end{equation}
with $\lambda_{L > 0} = R^2\alpha(R)\Gamma c/(v_{R} r_S^2 z)$ a numerical factor of order one. Accordingly, the mass accretion rate is suppressed by a factor of  $(r_\textrm{S}/r_\textrm{B})^2 z\sim 2.4\times 10^{-5}$ compared to the Bondi rate.

\begin{table*}
\centering
\caption{Mass accretion rate inferred for Sgr A* at $R = 0.06\,{\rm pc} $ for the models studied in this work and results from the literature. We assume $m=m_\textrm{p} =1.67\,\times\,10^{-27}\,{\rm kg}$, the characteristic values of $M = 4.3 \times 10^6$ M$_{\odot}$, $T=2.2\,\times\,10^7\,{\rm K}$ and $n_R=160\,{\rm cm^{-3}}$ as mentioned in the text. We consider an infall velocity of $v_R = 600$\,km\,s$^{-1}$ at the radius $R$ for the mono-energetic models. The Bondi model result is taken from Eq.~(\ref{eq:Bondi_values}). In the mono-energetic models we have used $E_{0} = \frac{1}{2} m_\textrm{p} v_R^2 - G M  m_\textrm{p}/R\,$ for the non-relativistic case and $E_{0} =  mc^2\alpha(R)\gamma $ for the relativistic case. Note that in the latter case with angular momentum it turns out that $\gamma > \gamma_\textrm{c}(R)$, and since both conditions $R \gg r_\textrm{S}$ and $v_\textrm{R}\ll c$ are met, one can use the corresponding approximation~\eqref{Eq:RelMassAccretionRateApprox_b}. `Non-rel' and `rel' stand for the non-relativistic and relativistic cases, respectively.}
\label{table:applications}
\begin{tabular}{cccccc}
    &&&& \\ \hline
      Accretion model & Approximation & Distribution function &$|\dot{M}|\,[\mathrm{M}_{\odot} \, \mathrm{yr}^{-1}]$ & Reference  
        \\ \hline \hline
       Bondi & {\rm non-rel} & (Perfect fluid)  & $\sim 10^{-4}$ & \citet{Falcke_2013} \\
       Zeldovich-Novikov & {\rm non-rel} &  Mono-energetic & $\sim 1.29 \times 10^{-9}$  & \citet[][]{1971reas.book.....Z} \\
       Rioseco-Sarbach &  {\rm rel}  & Maxwell-J\"uttner &   $\sim 1.48 \times 10^{-9}$    &\citet{pRoS17a}\\
       Radial infall & {\rm non-rel} & Mono-energetic  & $\sim 1.09 \times 10^{-4}$ & This work, Eq.~\eqref{Eq:Eq:AccretionRateModelMonoEnerBis} \\
       Radial infall & {\rm non-rel}  & Maxwell-Boltzmann  & $\sim 6.22 \times 10^{-5}$ & This work, Eq.~\eqref{Eq:AccretionRateModelExp} \\
       Radial infall & {\rm rel}  & Mono-energetic  & $\sim 1.09 \times 10^{-4}$ & This work, Eq.~\eqref{Eq:AccretionRateRelativisticMonoEner} \\
       Radial infall & {\rm rel}  & Maxwell-J\"uttner & $\sim 6.22 \times 10^{-5}$ & This work, Eq.~\eqref{Eq:AccretionRateRelativisticModelExp}\\
       Infall with angular momentum & {\rm rel} & Mono-energetic  & $\sim 1.23\,\times 10^{-9}$ & This work, Eq.~\eqref{Eq:RelMassAccretionRateApprox_b} \\
       Infall with angular momentum & {\rm rel} & Maxwell-J\"uttner & $\sim 1.38 \times 10^{-9}$ & This work, Eq.~\eqref{Eq:MassAccretionJüttnerApprox}  \\
      \hline
\end{tabular}
\end{table*}
As mentioned in the introduction, a long standing problem is that the measured luminosity of Sgr A* (and other underluminous sources such as M87*) is way lower than that expected from the Eddington luminosity. Since the luminosity of the accreting flow of BHs is proportional to the mass accretion rate, there have been mainly two proposed solutions to explain the observed low luminosity: 1) a Bondi accretion rate with a very low radiative efficiency or 2) a much lower mass accretion rate than the Bondi rate. In the literature, various RIAF models have been proposed to solve this problem by taking into account one or both of these solutions.

Comparing the results from Table~\ref{table:applications}, 
we conclude that part of the solution to the low luminosity problem of Sgr~A* could be that the mass accretion rate should be inferred from the accretion of a kinetic gas at a finite radius, taking into account the angular momentum of the individual particles. In this case, our mass accretion rate estimates for the models with angular momentum are of the order of the mass accretion rates bounds inferred from RM. Note, however, that these bounds are supposed to be for the vicinity of the BH horizon. Therefore, there is a significant difference in the results of the hydrodynamical and kinetic approaches for the wind accretion at $R\sim 0.06$\,pc, at least for our simplified models with angular momentum. A more complete theoretical kinetic treatment and future kinetic simulations of this accretion scenario could explain this difference, by confirming the important disparity between the accretion rates, or by endowing the kinetic flow with an accretion rate-reduction mechanism as a consequence of the more complex modelling of the problem. We note that our explanation of the low luminosity problem relies solely in the assumption that the luminosity is proportional to the accretion rate, not in the mechanism of radiation of a collisionless kinetic gas, which we do not study here.


\subsection{Accretion in the vicinity of Sgr~A* and M87*}

As a second example, we also cautiously apply our models to the vicinity of Sgr A*'s event horizon. The values of temperatures and densities near the BH are estimated from the results of GRMHD simulations for a two-temperature plasma of electrons and ions (e.g. \citet{2009ApJ...706..497M,2013A&A...559L...3M}). The accretion in these simulations proceeds through a geometrically thick, optically thin hydrodynamical flow, coming initially from a weakly magnetized torus in hydrodynamic equilibrium, orbiting a Kerr BH.  Despite the significant physical differences with the more realistic case studied in the GRMHD simulations, we apply our models of the Maxwell-J\"uttner distribution function as an illustrative first approximation for the kinetic scenario.\footnote{In this case, we do not consider the mono-energetic models due to their more idealized nature.} In particular, we take the values of density and temperature of a two-temperature radial inflow–outflow hydrodynamical model near Sgr A* with self-consistent feeding and conduction presented in \citet{2010ApJ...716..504S}; we assume a fully ionized plasma so that there is an equality between electron and proton densities, $n_\textrm{e} = n_\textrm{p}$, and we take  $R=5\, r_\textrm{S}$, $n_R = n_\textrm{e} = 2\times10^6 $\,cm$^{-3}$ and $T = T_\textrm{p} = 30 \, T_\textrm{e} = 1.2\times10^{12}$\,K.  Furthermore, since the proton mass is much greater than the electron mass, we assume $m = m_\textrm{p}$. The accretion rates calculated are shown in Table \ref{table:applications2}. We found consistency with the values of the RM constraints for the vicinity of Sgr A*. Moreover, the radial infall in our models produces a greater mass accretion rate by a factor of $\sim3$. Thus, the role of angular momentum is not significant close to the BH horizon. This is consistent with the well-known fact that particles fall almost radially as they approach to the ISCO (see e.g. \citet{Chandrasekhar83, Nunez05}). It is important to note that we took values of temperature and densities from hydrodynamical models as a first approximation, due to the lack of model-independent estimations of the conditions near the SMBHs. 
\begin{table*}
\centering
\caption{Mass accretion rate inferred for Sgr A* at $R = 5\, r_\textrm{S}$, for the Maxwell-J\"uttner models. We have used  $m=m_\textrm{p}$, $M = 4.3 \times 10^6$\,M$_{\odot}$, $T =1.2\times10^{12}$\,K and $n_R =  2\times10^6 $\,cm$^{-3}$ (see \citet{2010ApJ...716..504S}).}
\label{table:applications2}
\begin{tabular}{cccccc}
    &&&& \\ \hline
      Accretion model & Distribution function &$|\dot{M}|\,[\mathrm{M}_{\odot} \, \mathrm{yr}^{-1}]$ & Reference  
        \\ \hline \hline
       Radial infall   & Maxwell-J\"uttner & $\sim 1.92 \times 10^{-7}$ & This work, Eq.~\eqref{eq:MJ-rad-rel}\\
       Infall with angular momentum  & Maxwell-J\"uttner & $\sim 5.77 \times 10^{-8}$ & This work, Eq.~\eqref{Eq:MassAccretionJüttner} \\
      \hline
\end{tabular}
\end{table*}
\begin{table*}
\centering
\caption{Mass accretion rate inferred for M87* at $R = 5\, r_\textrm{S}$, for the Maxwell-J\"uttner models. We have used $m=m_\textrm{p}$, $M = 6.5 \times 10^9$\,M$_{\odot}$, $T = 6.25 \times 10^{10}$\,K and $n_R = 2.9 \times 10^4$\,cm$^{-3}$  (see \citet{EHTCVI,EHTCVIII}).}
\label{table:applications3}
\begin{tabular}{cccccc}
    &&&& \\ \hline
      Accretion model & Distribution function &$|\dot{M}|\,[\mathrm{M}_{\odot} \, \mathrm{yr}^{-1}]$ & Reference  
        \\ \hline \hline
       Radial infall   & Maxwell-J\"uttner & $\sim 1.526 \times 10^{-3}$ & This work, Eq.~\eqref{eq:MJ-rad-rel}\\
       Infall with angular momentum  & Maxwell-J\"uttner & $\sim 1.521 \times 10^{-3}$ & This work, Eq.~\eqref{Eq:MassAccretionJüttner}  \\
      \hline
\end{tabular}
\end{table*}

Finally, another interesting scenario to study with our models (with the same caveats as in the Sgr~A* case), is the vicinity of the galactic centre of M87*. Recently, the Event Horizon Telescope Collaboration (EHTC) has provided results for the mass accretion rate due to the plasma around this SMBH. They report an estimated average number density of $n_\textrm{e} \sim 2.9 \times 10^{4\textrm{--}7}$\,cm$^{-3}$, an electron temperature $T_\textrm{e} \sim $ (1--12) $\times 10^{10}$\,K, and an inferred mass accretion rate for M87* of (3 -- 20)$\,\times 10^{-4}$\,M$_{\odot}$\,yr$^{-1}$ from a simple one-zone emission model (\citet{EHTCV,ehtc7,EHTCVIII}). For the specific isothermal sphere model, they estimate the plasma number density $n_\textrm{e} \simeq 2.9\times 10^4$ cm$^{-3}$ and the electron temperature $T_\textrm{e} \simeq 6.25\times 10^{10}$\,K, for an emission radius assumed to be $r \simeq 5 r_\textrm{S}/2$. In Table~\ref{table:applications3} we present the mass accretion rates obtained from these values, assuming a fully ionized hydrogen plasma ($n_{\textrm{ions}} = n_\textrm{e}$) and assuming thermal equilibrium between the ions and electrons ($T_{\textrm{ions}} = T_\textrm{e}$) as a first approximation. Furthermore, we impose these values at radius $ R \sim 5r_\textrm{S}$, as in the example of Sgr~A*.\footnote{The reason for not choosing $R = 5 r_\textrm{S}/2$ is that this value is smaller than the ISCO radius of a Schwarzschild BH assumed in our model.} In this case, we got similar mass accretion rates for the models with and without angular momentum. Therefore, angular momentum does not play an important role, just like the case of the vicinity of Sgr~A*'s event horizon. Despite our crude approximation for the accretion flow of M87*, the inferred mass accretion rates are consistent with the reported bounds by the \citet{EHTCVIII}.


\section{Summary and conclusions}
\label{sec:conclusions}

Low collisionality is a general property expected in underluminous flows near BHs due to the conditions of high temperatures and low densities. Thus, a kinetic approach is needed in order to explain correctly the accretion process onto these BHs. Additionally, most previous analytic studies treat the problem of spherical mass accretion specifying boundary conditions at infinity, both in the fluid and the kinetic approximations, whereas in many situations of interest the gas is accreted from a region of finite radius. 

In this work, we presented several analytic models and their corresponding steady-state solutions for the mass accretion of a spherically symmetric, collisionless kinetic gas cloud onto a Schwarzschild BH. The novelty of this article consists in specifying the properties of the kinetic gas (its particle density $n_R$ and mean velocity or temperature) at an injection sphere of finite radius $R$. The models we have discussed include the  simple case of purely radial infall, in which all the particles have zero angular momentum (both in the Newtonian and relativistic regimes) and the case of a kinetic gas with a uniform distribution in the angular momentum, such that individual gas particles may rotate about the BH, yet the gas configuration as a whole is spherically symmetric. Regarding the energy distribution, we considered the mono-energetic case in which all particles have the same energy $E$ (or associated three-velocity $v_R$ at the injection sphere) as well as the Maxwell-Boltzmann (non-relativistic case) and Maxwell-Jüttner distribution with corresponding temperature $T$, assuming that the gas is accreted from a reservoir of particles in thermodynamic equilibrium. In each model, the mass accretion rate depends linearly on $n_R$ which is a direct consequence of our test field approximation (we have neglected the self gravity of the kinetic gas) while its dependency on $R$ and $v_R$ or $T$ is more intricate and is summarized in Section~\ref{sec:summary}.

We have checked that for fixed positive values of $n_R$, $v_R$ and $T$, our models with angular momentum have the property that their mass accretion rates converge to the corresponding expressions of previously known results (\citet{1971reas.book.....Z, 1983bhwd.book.....S,pRoS17a}) in the limit $R\to \infty$, while the models with purely radial infall have associated to them a mass accretion rate which diverges as $R\to \infty$. This means the latter do not possess  well-defined Bondi-type formulae which relate the mass accretion rate to the properties of the gas at infinity. This is analogous to the case of accretion of pure dust in spherical symmetry, in which the only steady-state solution has vanishing mass density at infinity (\citet{Chaverra:2015bya}). In fact, our mono-energetic model with purely radial infall does correspond precisely to this simple dust model, since its velocity dispersion is exactly zero.

However, for the more realistic case of an injection sphere of finite radius $R$ it turns out that the purely radial infall models may lead to similar accretion rates than those predicted from the well-known Bondi model for a hydrodynamic flow. In particular, we have shown that for boundary conditions corresponding to non-relativistic velocities or temperatures, the mass accretion rates in both models yield comparable results, provided $R$ is of the same order as the Bondi radius. Regarding our models with angular momentum, their mass accretion rate behaves qualitatively similarly to the purely radial infall models, and it increases with increasing values of $v_R$ or $T$ as long as they lie below a critical value. However, above this critical value the mass accretion rate reverses its behaviour and decreases with increasing $v_R$ or $T$ until it reaches a finite value. As we have explained, this reversal is due to the fact that as the particle's energy increases above a certain threshold, not all the particles are absorbed by the BH, and the fraction of absorbed particles becomes smaller as the energy increases, leading to a diminishing mass accretion rate.

Finally, we calculated the mass accretion rate onto the SMBHs Sgr~A* and M87* with our models, estimating the condition of the gas at different radii based on recent observations. Our results, which are summarized in Tables~\ref{table:applications}, \ref{table:applications2} and \ref{table:applications3}, are of the order of the model-dependent RM bounds for the mass accretion rate of Sgr~A* and the bounds estimated for M87* by the Event Horizon Telescope collaboration. We found that our kinetic models can overall predict lower mass accretion rates than the Bondi fluid model. The above suggests that a complete kinetic treatment to the accretion problem could explain some of the current questions associated with underluminous sources such as Sgr~A* or M87*.

There are several ingredients which, for simplicity, we did not take into account in our models. In particular, we restricted ourselves to spherical steady accretion onto a Schwarzschild BH, instead of the more realistic  non-spherical and unsteady accretion onto a (rotating) Kerr BH. Furthermore, we did not included the effects of radiative processes, magnetic fields, the consequences of outflows, convection currents or jets nor the effects due to net angular momentum of the gas. We intend to generalize our models to include some of these effects in future work.

Despite its simplicity, the presented models could serve as reference for more generic kinetic models, and they could be useful as a starting point to describe other physical scenarios where the assumptions of very low collisionality or quasi-spherical symmetry are approximately satisfied. This is the case, for example, in the BH accretion of dark matter, which is expected to be very weakly interactive, or in the accretion of low-luminosity active galactic nuclei whose corresponding flows are in a hot and low-density state. Thus, future generalizations of the presented formalism could be a key step in understanding accretion processes.



\acknowledgments
\vspace{-5mm}
We thank Laurent Loinard and Luis Felipe Rodr\'{\i}guez for comments on a previous version of this manuscript. CG and OS also thank Emilio Tejeda for fruitful discussions.
This work was partially supported by 
DGAPA-UNAM through grant IN110218, by a CIC grant to Universidad Michoacana and by the CONACyT Network Projects 
No. 376127 `Sombras, lentes y ondas gravitatorias generadas por objetos compactos astrof\'isicos', and No. 304001 `Estudio de campos escalares con aplicaciones en cosmolog\'ia y astrof\'isica'. AG and CG acknowledge  financial support from CONACyT graduate grants program. PDF acknowledge the support by the National Research Foundation (NRF) of Korea through grants 2016R1A5A1013277 and 2020R1A2C2102800.


\appendix

\section{Fixed-\texorpdfstring{$L$}{TEXT}-models}
\label{ap:fixedLmodels}

The main purpose of this appendix is to shed some light on the difference between the accretion rates predicted by the mono-energetic model in the pure radial case [see Eq.~(\ref{eq:mono_rel_rad_stationary})] and the same model in the presence of angular momentum [see Eq.~(\ref{Eq:RelMassAccretionRateBis_a})]. As we have discussed below Eq.~(\ref{eq:acc_rate_infty}), in the non-relativistic limit $v_R\ll c$ the latter case yields an accretion rate that is smaller (by a factor of $2$) compared to the purely radial case. In contrast, there is not such difference for the Maxwell-J\"uttner type model when the low temperature limit is taken [cf. the comments below Eq.~(\ref{Eq:GammaMaxJutt})]. 

In order to illustrate the role played by the angular momentum in this behaviour, we consider the following simple model:
\begin{equation}
F(E,L) = f(E)\frac{\delta(L - L_0)}{L_0},
\label{Eq:L0Model}
\end{equation}
in which all the particles have the same angular momentum $L_0 > 0$ and are subject to the energy distribution $f(E)$ which we specify later. Assuming that $L_0$ is small enough such that $L_0 < L_{\text{c}}(E)$ for all $E > mc^2\alpha(R)$ (which is guaranteed to be the case if $L_0 < L_{\text{ISCO}}$), one obtains from Eqs.~(\ref{Eq:Jabs}, \ref{Eq:Jsca}) the expressions
\begin{align}
\left. J^\sigma_{\text{abs}} \right|_{r=R} &= c\int\limits_{\sqrt{V_{L_0}(R)}}^\infty  \int\limits_{0}^{2\pi} \frac{p^\sigma_- f(E)dE d\chi}{R^2 \sqrt{E^2 - V_{L_0}(R)}},\\
\left. J^\sigma_{\text{sca}} \right|_{r=R} &= 0,
\end{align}
from which one immediately obtains
\begin{align}
\left. J^0_{\text{abs}} \right|_{r=R} 
&= \frac{2\pi}{\alpha(R) R^2}\int\limits_{\sqrt{V_{L_0}(R)}}^\infty \frac{f(E) E dE}{\sqrt{E^2 - V_{L_0}(R)}},
\label{Eq:J0absDelta}
\\
\left. J^1_{\text{abs}} \right|_{r=R} 
&= -\frac{2\pi}{\alpha(R) R^2}\int\limits_{\sqrt{V_{L_0}(R)}}^\infty f(E) dE,
\label{Eq:J1absDelta}
\end{align}
and $\left. J^2_{\text{abs}} \right|_{r=R} = \left. J^3_{\text{abs}} \right|_{r=R} = 0$, where we have introduced the shorthand notation $\left. V_{L_0}(r) := V_{L}(r) \right|_{L\rightarrow L_0}$. For the mono-energetic model with $f(E) = f_0\delta(E-E_0)$ and $E_0 = mc^2\alpha\gamma$ this yields (assuming $V_{L_0}(R) < E_0^2$ or, equivalently, $L_0 < L_{\text{max}}(E_0,R)$)
\begin{equation}
\frac{|\dot{M}|}{mc n_R} = 4\pi R^2\alpha(R)\sqrt{\frac{\gamma^2}{1+\kappa^2}\ - 1},
\label{Eq:MdotL0ModelMonoEnergetic}
\end{equation}
where we have defined $\kappa \coloneqq L_0/(R m c)$. In the limit $L_0\to 0$ one recovers the result from Eq.~(\ref{eq:mono_rel_rad_stationary}) which has been derived directly with the assumption that all the gas particles have vanishing angular momentum. If instead of $L_0\to 0$ one sets $L_0 = R |v_\perp|$ with $v_\perp$ the angular components of the velocity, one obtains in the limit $|v_R|\ll c$,
\begin{equation}
\frac{|\dot{M}|}{mc n_R} \approx 4\pi R^2\alpha(R)\frac{v_\textrm{rad}}{c},
\end{equation}
where $v_\textrm{rad} = \sqrt{v_R^2 - |v_\perp|^2}$ denotes the radial component of the three-velocity of the particles. For purely radial infall $v_\textrm{rad} = v_R$ and this result agrees precisely with Eq.~(\ref{Eq:Eq:AccretionRateModelMonoEnerBis}). However, when angular momentum is present, the accretion rate is suppressed by a factor of $v_\textrm{rad}/v_R$. This illustrates why the accretion rate is smaller for models with angular momentum when $n_R$ and $v_R$ are fixed at the injection sphere.

It is also interesting to apply the model described in Eq.~(\ref{Eq:L0Model}) to the Maxwell-J\"uttner-type distribution function. Inserting $f(E) = Ae^{-\beta E}$ into Eqs.~(\ref{Eq:J0absDelta}, \ref{Eq:J1absDelta}) yields the following non-vanish components of the current density
\begin{align}
\left. J^0_{\text{abs}} \right|_{r=R} 
&= \frac{2\pi A}{\alpha(R) R^2\beta} z_\kappa \mathbf{K}_1\left( z_\kappa \right),
\\
\left. J^1_{\text{abs}} \right|_{r=R} 
&= -\frac{2\pi A}{\alpha(R) R^2\beta} e^{-z_\kappa},
\label{Eq:J1absDeltaBis}
\end{align}
where we have set $z_\kappa \coloneqq \sqrt{1 + \kappa^2} z$ and $z = mc^2\alpha(R)\beta$, as defined below Eq.~(\ref{Eq:MdotMaxBoltz}). This in turn leads to the mass accretion rate 
\begin{equation}
     \frac{|\dot{M}|}{m c n_R} = \frac{4\pi \alpha(R) R^2}{\sqrt{\left[ \mathbf{K}_1\left( z_\kappa \right) z_\kappa e^{z_\kappa} \right]^2 - 1 }}, 
\label{Eq:MdotL0ModelMaxBoltz}
\end{equation}
which converges to the same expression as in Eq.~(\ref{eq:MJ-rad-rel}) in the limit $L_0\to 0$. Whereas the leading-order behaviour of the mass accretion rate given in Eq.~(\ref{Eq:MdotL0ModelMonoEnergetic}) for $v_R\ll c$ and $\kappa \ll 1$ depends on the relation between $v_R/c$ and $\kappa$, the limit of the right-hand side of Eq.~(\ref{Eq:MdotL0ModelMaxBoltz}) for $z\gg 1$ and $\kappa \ll 1$ always yields $4\alpha(R) R^2\sqrt{2\pi/z}$, regardless of the relation between $z$ and $\kappa$. This explains why in the Maxwell-J\"uttner case the accretion rate for the models with and without angular momentum agree with each other in the low temperature limit.

%
%
%

\vfill\null

\bibliography{referencias.bib}

\begin{thebibliography}{88}%
\makeatletter
\providecommand \@ifxundefined [1]{%
 \@ifx{#1\undefined}
}%
\providecommand \@ifnum [1]{%
 \ifnum #1\expandafter \@firstoftwo
 \else \expandafter \@secondoftwo
 \fi
}%
\providecommand \@ifx [1]{%
 \ifx #1\expandafter \@firstoftwo
 \else \expandafter \@secondoftwo
 \fi
}%
\providecommand \natexlab [1]{#1}%
\providecommand \enquote  [1]{``#1''}%
\providecommand \bibnamefont  [1]{#1}%
\providecommand \bibfnamefont [1]{#1}%
\providecommand \citenamefont [1]{#1}%
\providecommand \href@noop [0]{\@secondoftwo}%
\providecommand \href [0]{\begingroup \@sanitize@url \@href}%
\providecommand \@href[1]{\@@startlink{#1}\@@href}%
\providecommand \@@href[1]{\endgroup#1\@@endlink}%
\providecommand \@sanitize@url [0]{\catcode `\\12\catcode `\$12\catcode
  `\&12\catcode `\#12\catcode `\^12\catcode `\_12\catcode `\%12\relax}%
\providecommand \@@startlink[1]{}%
\providecommand \@@endlink[0]{}%
\providecommand \url  [0]{\begingroup\@sanitize@url \@url }%
\providecommand \@url [1]{\endgroup\@href {#1}{\urlprefix }}%
\providecommand \urlprefix  [0]{URL }%
\providecommand \Eprint [0]{\href }%
\providecommand \doibase [0]{https://doi.org/}%
\providecommand \selectlanguage [0]{\@gobble}%
\providecommand \bibinfo  [0]{\@secondoftwo}%
\providecommand \bibfield  [0]{\@secondoftwo}%
\providecommand \translation [1]{[#1]}%
\providecommand \BibitemOpen [0]{}%
\providecommand \bibitemStop [0]{}%
\providecommand \bibitemNoStop [0]{.\EOS\space}%
\providecommand \EOS [0]{\spacefactor3000\relax}%
\providecommand \BibitemShut  [1]{\csname bibitem#1\endcsname}%
\let\auto@bib@innerbib\@empty
\bibitem [{\citenamefont {{Dom{\'\i}nguez-Fern{\'a}ndez}}\ \emph
  {et~al.}(2017)\citenamefont {{Dom{\'\i}nguez-Fern{\'a}ndez}}, \citenamefont
  {{Jim{\'e}nez-V{\'a}zquez}}, \citenamefont {{Alcubierre}}, \citenamefont
  {{Montoya}},\ and\ \citenamefont
  {{N{\'u}{\~n}ez}}}]{Dominguez-Fernandez:2017nxx}%
  \BibitemOpen
  \bibfield  {author} {\bibinfo {author} {\bibfnamefont {P.}~\bibnamefont
  {{Dom{\'\i}nguez-Fern{\'a}ndez}}}, \bibinfo {author} {\bibfnamefont
  {E.}~\bibnamefont {{Jim{\'e}nez-V{\'a}zquez}}}, \bibinfo {author}
  {\bibfnamefont {M.}~\bibnamefont {{Alcubierre}}}, \bibinfo {author}
  {\bibfnamefont {E.}~\bibnamefont {{Montoya}}},\ and\ \bibinfo {author}
  {\bibfnamefont {D.}~\bibnamefont {{N{\'u}{\~n}ez}}},\ }\bibfield  {title}
  {\bibinfo {title} {{Description of the evolution of inhomogeneities on a dark
  matter halo with the Vlasov equation}},\ }\href
  {https://doi.org/10.1007/s10714-017-2286-8} {\bibfield  {journal} {\bibinfo
  {journal} {General Relativity and Gravitation}\ }\textbf {\bibinfo {volume}
  {49}},\ \bibinfo {eid} {123} (\bibinfo {year} {2017})},\ \Eprint
  {https://arxiv.org/abs/1703.03286} {arXiv:1703.03286 [gr-qc]} \BibitemShut
  {NoStop}%
\bibitem [{\citenamefont {{Frank}}\ \emph {et~al.}(2002)\citenamefont
  {{Frank}}, \citenamefont {{King}},\ and\ \citenamefont
  {{Raine}}}]{2002apa..book.....F}%
  \BibitemOpen
  \bibfield  {author} {\bibinfo {author} {\bibfnamefont {J.}~\bibnamefont
  {{Frank}}}, \bibinfo {author} {\bibfnamefont {A.}~\bibnamefont {{King}}},\
  and\ \bibinfo {author} {\bibfnamefont {D.~J.}\ \bibnamefont {{Raine}}},\
  }\href@noop {} {\emph {\bibinfo {title} {{Accretion Power in
  Astrophysics}}}},\ \bibinfo {edition} {3rd}\ ed.\ (\bibinfo  {publisher}
  {Cambridge University Press},\ \bibinfo {address} {Cambridge, UK},\ \bibinfo
  {year} {2002})\BibitemShut {NoStop}%
\bibitem [{\citenamefont {Barranco}\ \emph {et~al.}(2011)\citenamefont
  {Barranco}, \citenamefont {Bernal}, \citenamefont {Degollado}, \citenamefont
  {Diez-Tejedor}, \citenamefont {Megevand}, \citenamefont {Alcubierre},
  \citenamefont {Nunez},\ and\ \citenamefont {Sarbach}}]{Burt:2011pv}%
  \BibitemOpen
  \bibfield  {author} {\bibinfo {author} {\bibfnamefont {J.}~\bibnamefont
  {Barranco}}, \bibinfo {author} {\bibfnamefont {A.}~\bibnamefont {Bernal}},
  \bibinfo {author} {\bibfnamefont {J.~C.}\ \bibnamefont {Degollado}}, \bibinfo
  {author} {\bibfnamefont {A.}~\bibnamefont {Diez-Tejedor}}, \bibinfo {author}
  {\bibfnamefont {M.}~\bibnamefont {Megevand}}, \bibinfo {author}
  {\bibfnamefont {M.}~\bibnamefont {Alcubierre}}, \bibinfo {author}
  {\bibfnamefont {D.}~\bibnamefont {Nunez}},\ and\ \bibinfo {author}
  {\bibfnamefont {O.}~\bibnamefont {Sarbach}},\ }\bibfield  {title} {\bibinfo
  {title} {{Are black holes a serious threat to scalar field dark matter
  models?}},\ }\href {https://doi.org/10.1103/PhysRevD.84.083008} {\bibfield
  {journal} {\bibinfo  {journal} {Phys. Rev.}\ }\textbf {\bibinfo {volume}
  {D84}},\ \bibinfo {pages} {083008} (\bibinfo {year} {2011})},\ \Eprint
  {https://arxiv.org/abs/1108.0931} {arXiv:1108.0931 [gr-qc]} \BibitemShut
  {NoStop}%
\bibitem [{\citenamefont {{Hoyle}}\ and\ \citenamefont
  {{Lyttleton}}(1939)}]{1939PCPS...35..405H}%
  \BibitemOpen
  \bibfield  {author} {\bibinfo {author} {\bibfnamefont {F.}~\bibnamefont
  {{Hoyle}}}\ and\ \bibinfo {author} {\bibfnamefont {R.~A.}\ \bibnamefont
  {{Lyttleton}}},\ }\bibfield  {title} {\bibinfo {title} {{The effect of
  interstellar matter on climatic variation}},\ }\href
  {https://doi.org/10.1017/S0305004100021150} {\bibfield  {journal} {\bibinfo
  {journal} {Proceedings of the Cambridge Philosophical Society}\ }\textbf
  {\bibinfo {volume} {35}},\ \bibinfo {pages} {405} (\bibinfo {year}
  {1939})}\BibitemShut {NoStop}%
\bibitem [{\citenamefont {Bondi}\ and\ \citenamefont
  {Hoyle}(1944)}]{10.1093/mnras/104.5.273}%
  \BibitemOpen
  \bibfield  {author} {\bibinfo {author} {\bibfnamefont {H.}~\bibnamefont
  {Bondi}}\ and\ \bibinfo {author} {\bibfnamefont {F.}~\bibnamefont {Hoyle}},\
  }\bibfield  {title} {\bibinfo {title} {{On the Mechanism of Accretion by
  Stars}},\ }\href {https://doi.org/10.1093/mnras/104.5.273} {\bibfield
  {journal} {\bibinfo  {journal} {Monthly Notices of the Royal Astronomical
  Society}\ }\textbf {\bibinfo {volume} {104}},\ \bibinfo {pages} {273}
  (\bibinfo {year} {1944})},\ \Eprint
  {https://arxiv.org/abs/https://academic.oup.com/mnras/article-pdf/104/5/273/8072203/mnras104-0273.pdf}
  {https://academic.oup.com/mnras/article-pdf/104/5/273/8072203/mnras104-0273.pdf}
  \BibitemShut {NoStop}%
\bibitem [{\citenamefont {{Bondi}}(1952)}]{1952MNRAS.112..195B}%
  \BibitemOpen
  \bibfield  {author} {\bibinfo {author} {\bibfnamefont {H.}~\bibnamefont
  {{Bondi}}},\ }\bibfield  {title} {\bibinfo {title} {{On spherically
  symmetrical accretion}},\ }\href {https://doi.org/10.1093/mnras/112.2.195}
  {\bibfield  {journal} {\bibinfo  {journal} {Monthly Notices of the Royal
  Astronomical Society}\ }\textbf {\bibinfo {volume} {112}},\ \bibinfo {pages}
  {195} (\bibinfo {year} {1952})}\BibitemShut {NoStop}%
\bibitem [{\citenamefont {{Lewin}}\ \emph {et~al.}(1995)\citenamefont
  {{Lewin}}, \citenamefont {{van Paradijs}},\ and\ \citenamefont {{van den
  Heuvel}}}]{1997CAS....26.....L}%
  \BibitemOpen
  \bibfield  {author} {\bibinfo {author} {\bibfnamefont {W.~H.~G.}\
  \bibnamefont {{Lewin}}}, \bibinfo {author} {\bibfnamefont {J.}~\bibnamefont
  {{van Paradijs}}},\ and\ \bibinfo {author} {\bibfnamefont {E.~P.~J.}\
  \bibnamefont {{van den Heuvel}}},\ }\href@noop {} {\emph {\bibinfo {title}
  {X-ray binaries}}}\ (\bibinfo  {publisher} {Cambridge University Press},\
  \bibinfo {address} {Cambridge, UK},\ \bibinfo {year} {1995})\BibitemShut
  {NoStop}%
\bibitem [{\citenamefont {{Popham}}\ \emph {et~al.}(1999)\citenamefont
  {{Popham}}, \citenamefont {{Woosley}},\ and\ \citenamefont
  {{Fryer}}}]{1999ApJ...518..356P}%
  \BibitemOpen
  \bibfield  {author} {\bibinfo {author} {\bibfnamefont {R.}~\bibnamefont
  {{Popham}}}, \bibinfo {author} {\bibfnamefont {S.~E.}\ \bibnamefont
  {{Woosley}}},\ and\ \bibinfo {author} {\bibfnamefont {C.}~\bibnamefont
  {{Fryer}}},\ }\bibfield  {title} {\bibinfo {title} {{Hyperaccreting Black
  Holes and Gamma-Ray Bursts}},\ }\href {https://doi.org/10.1086/307259}
  {\bibfield  {journal} {\bibinfo  {journal} {\apj}\ }\textbf {\bibinfo
  {volume} {518}},\ \bibinfo {pages} {356} (\bibinfo {year} {1999})},\ \Eprint
  {https://arxiv.org/abs/astro-ph/9807028} {arXiv:astro-ph/9807028 [astro-ph]}
  \BibitemShut {NoStop}%
\bibitem [{\citenamefont {{Williams}}\ and\ \citenamefont
  {{Cieza}}(2011)}]{2011ARA&A..49...67W}%
  \BibitemOpen
  \bibfield  {author} {\bibinfo {author} {\bibfnamefont {J.~P.}\ \bibnamefont
  {{Williams}}}\ and\ \bibinfo {author} {\bibfnamefont {L.~A.}\ \bibnamefont
  {{Cieza}}},\ }\bibfield  {title} {\bibinfo {title} {{Protoplanetary Disks and
  Their Evolution}},\ }\href
  {https://doi.org/10.1146/annurev-astro-081710-102548} {\bibfield  {journal}
  {\bibinfo  {journal} {\araa}\ }\textbf {\bibinfo {volume} {49}},\ \bibinfo
  {pages} {67} (\bibinfo {year} {2011})},\ \Eprint
  {https://arxiv.org/abs/1103.0556} {arXiv:1103.0556 [astro-ph.GA]}
  \BibitemShut {NoStop}%
\bibitem [{\citenamefont {{Krolik}}(1999)}]{1999agnc.book.....K}%
  \BibitemOpen
  \bibfield  {author} {\bibinfo {author} {\bibfnamefont {J.~H.}\ \bibnamefont
  {{Krolik}}},\ }\href@noop {} {\emph {\bibinfo {title} {{Active Galactic
  Nuclei: From the Central Black Hole to the Galactic Environment}}}}\
  (\bibinfo  {publisher} {Princeton University Press},\ \bibinfo {address}
  {Princeton, NJ, USA},\ \bibinfo {year} {1999})\BibitemShut {NoStop}%
\bibitem [{\citenamefont {{Penrose}}(1965)}]{1965PhRvL..14...57P}%
  \BibitemOpen
  \bibfield  {author} {\bibinfo {author} {\bibfnamefont {R.}~\bibnamefont
  {{Penrose}}},\ }\bibfield  {title} {\bibinfo {title} {{Gravitational Collapse
  and Space-Time Singularities}},\ }\href
  {https://doi.org/10.1103/PhysRevLett.14.57} {\bibfield  {journal} {\bibinfo
  {journal} {\prl}\ }\textbf {\bibinfo {volume} {14}},\ \bibinfo {pages} {57}
  (\bibinfo {year} {1965})}\BibitemShut {NoStop}%
\bibitem [{\citenamefont {{Celotti}}\ \emph {et~al.}(1999)\citenamefont
  {{Celotti}}, \citenamefont {{Miller}},\ and\ \citenamefont
  {{Sciama}}}]{1999CQGra..16A...3C}%
  \BibitemOpen
  \bibfield  {author} {\bibinfo {author} {\bibfnamefont {A.}~\bibnamefont
  {{Celotti}}}, \bibinfo {author} {\bibfnamefont {J.~C.}\ \bibnamefont
  {{Miller}}},\ and\ \bibinfo {author} {\bibfnamefont {D.~W.}\ \bibnamefont
  {{Sciama}}},\ }\bibfield  {title} {\bibinfo {title} {{Astrophysical evidence
  for the existence of black holes}},\ }\href@noop {} {\bibfield  {journal}
  {\bibinfo  {journal} {Classical and Quantum Gravity}\ }\textbf {\bibinfo
  {volume} {16}},\ \bibinfo {pages} {A3} (\bibinfo {year} {1999})},\ \Eprint
  {https://arxiv.org/abs/astro-ph/9912186} {arXiv:astro-ph/9912186 [astro-ph]}
  \BibitemShut {NoStop}%
\bibitem [{\citenamefont {{Kormendy}}\ and\ \citenamefont
  {{Richstone}}(1995)}]{1995ARA&A..33..581K}%
  \BibitemOpen
  \bibfield  {author} {\bibinfo {author} {\bibfnamefont {J.}~\bibnamefont
  {{Kormendy}}}\ and\ \bibinfo {author} {\bibfnamefont {D.}~\bibnamefont
  {{Richstone}}},\ }\bibfield  {title} {\bibinfo {title} {{Inward Bound---The
  Search For Supermassive Black Holes In Galactic Nuclei}},\ }\href
  {https://doi.org/10.1146/annurev.aa.33.090195.003053} {\bibfield  {journal}
  {\bibinfo  {journal} {\araa}\ }\textbf {\bibinfo {volume} {33}},\ \bibinfo
  {pages} {581} (\bibinfo {year} {1995})}\BibitemShut {NoStop}%
\bibitem [{\citenamefont {Kormendy}\ and\ \citenamefont
  {Ho}(2013)}]{Kormendy:2013dxa}%
  \BibitemOpen
  \bibfield  {author} {\bibinfo {author} {\bibfnamefont {J.}~\bibnamefont
  {Kormendy}}\ and\ \bibinfo {author} {\bibfnamefont {L.~C.}\ \bibnamefont
  {Ho}},\ }\bibfield  {title} {\bibinfo {title} {{Coevolution (Or Not) of
  Supermassive Black Holes and Host Galaxies}},\ }\href
  {https://doi.org/10.1146/annurev-astro-082708-101811} {\bibfield  {journal}
  {\bibinfo  {journal} {Ann. Rev. Astron. Astrophys.}\ }\textbf {\bibinfo
  {volume} {51}},\ \bibinfo {pages} {511} (\bibinfo {year} {2013})},\ \Eprint
  {https://arxiv.org/abs/1304.7762} {arXiv:1304.7762 [astro-ph.CO]}
  \BibitemShut {NoStop}%
\bibitem [{\citenamefont {{Michel}}(1972)}]{1972Ap&SS..15..153M}%
  \BibitemOpen
  \bibfield  {author} {\bibinfo {author} {\bibfnamefont {F.~C.}\ \bibnamefont
  {{Michel}}},\ }\bibfield  {title} {\bibinfo {title} {{Accretion of Matter by
  Condensed Objects}},\ }\href {https://doi.org/10.1007/BF00649949} {\bibfield
  {journal} {\bibinfo  {journal} {\apss}\ }\textbf {\bibinfo {volume} {15}},\
  \bibinfo {pages} {153} (\bibinfo {year} {1972})}\BibitemShut {NoStop}%
\bibitem [{\citenamefont {{Aguayo-Ortiz}}\ \emph {et~al.}(2021)\citenamefont
  {{Aguayo-Ortiz}}, \citenamefont {{Tejeda}}, \citenamefont {{Sarbach}},\ and\
  \citenamefont {{L{\'o}pez-C{\'a}mara}}}]{Aguayo-Ortiz:2021jzv}%
  \BibitemOpen
  \bibfield  {author} {\bibinfo {author} {\bibfnamefont {A.}~\bibnamefont
  {{Aguayo-Ortiz}}}, \bibinfo {author} {\bibfnamefont {E.}~\bibnamefont
  {{Tejeda}}}, \bibinfo {author} {\bibfnamefont {O.}~\bibnamefont
  {{Sarbach}}},\ and\ \bibinfo {author} {\bibfnamefont {D.}~\bibnamefont
  {{L{\'o}pez-C{\'a}mara}}},\ }\bibfield  {title} {\bibinfo {title} {{Spherical
  accretion: Bondi, Michel, and rotating black holes}},\ }\bibfield  {journal}
  {\bibinfo  {journal} {\mnras}\ }\href
  {https://doi.org/10.1093/mnras/stab1127} {10.1093/mnras/stab1127} (\bibinfo
  {year} {2021}),\ \Eprint {https://arxiv.org/abs/2102.12529} {arXiv:2102.12529
  [astro-ph.HE]} \BibitemShut {NoStop}%
\bibitem [{\citenamefont {Porth}\ \emph {et~al.}(2017)\citenamefont {Porth},
  \citenamefont {Olivares}, \citenamefont {Mizuno}, \citenamefont {Younsi},
  \citenamefont {Rezzolla}, \citenamefont {Moscibrodzka}, \citenamefont
  {Falcke},\ and\ \citenamefont {Kramer}}]{Porth_2017}%
  \BibitemOpen
  \bibfield  {author} {\bibinfo {author} {\bibfnamefont {O.}~\bibnamefont
  {Porth}}, \bibinfo {author} {\bibfnamefont {H.}~\bibnamefont {Olivares}},
  \bibinfo {author} {\bibfnamefont {Y.}~\bibnamefont {Mizuno}}, \bibinfo
  {author} {\bibfnamefont {Z.}~\bibnamefont {Younsi}}, \bibinfo {author}
  {\bibfnamefont {L.}~\bibnamefont {Rezzolla}}, \bibinfo {author}
  {\bibfnamefont {M.}~\bibnamefont {Moscibrodzka}}, \bibinfo {author}
  {\bibfnamefont {H.}~\bibnamefont {Falcke}},\ and\ \bibinfo {author}
  {\bibfnamefont {M.}~\bibnamefont {Kramer}},\ }\bibfield  {title} {\bibinfo
  {title} {The black hole accretion code},\ }\bibfield  {journal} {\bibinfo
  {journal} {Computational Astrophysics and Cosmology}\ }\textbf {\bibinfo
  {volume} {4}},\ \href {https://doi.org/10.1186/s40668-017-0020-2}
  {10.1186/s40668-017-0020-2} (\bibinfo {year} {2017})\BibitemShut {NoStop}%
\bibitem [{\citenamefont {{Ghez}}\ \emph {et~al.}(2003)\citenamefont {{Ghez}}
  \emph {et~al.}}]{2003ApJ...586L.127G}%
  \BibitemOpen
  \bibfield  {author} {\bibinfo {author} {\bibfnamefont {A.~M.}\ \bibnamefont
  {{Ghez}}} \emph {et~al.},\ }\bibfield  {title} {\bibinfo {title} {{The First
  Measurement of Spectral Lines in a Short-Period Star Bound to the Galaxy's
  Central Black Hole: A Paradox of Youth}},\ }\href
  {https://doi.org/10.1086/374804} {\bibfield  {journal} {\bibinfo  {journal}
  {\apj}\ }\textbf {\bibinfo {volume} {586}},\ \bibinfo {pages} {L127}
  (\bibinfo {year} {2003})},\ \Eprint {https://arxiv.org/abs/astro-ph/0302299}
  {arXiv:astro-ph/0302299 [astro-ph]} \BibitemShut {NoStop}%
\bibitem [{\citenamefont {{Gillessen}}\ \emph {et~al.}(2009)\citenamefont
  {{Gillessen}}, \citenamefont {{Eisenhauer}}, \citenamefont {{Trippe}},
  \citenamefont {{Alexander}}, \citenamefont {{Genzel}}, \citenamefont
  {{Martins}},\ and\ \citenamefont {{Ott}}}]{2009ApJ...692.1075G}%
  \BibitemOpen
  \bibfield  {author} {\bibinfo {author} {\bibfnamefont {S.}~\bibnamefont
  {{Gillessen}}}, \bibinfo {author} {\bibfnamefont {F.}~\bibnamefont
  {{Eisenhauer}}}, \bibinfo {author} {\bibfnamefont {S.}~\bibnamefont
  {{Trippe}}}, \bibinfo {author} {\bibfnamefont {T.}~\bibnamefont
  {{Alexander}}}, \bibinfo {author} {\bibfnamefont {R.}~\bibnamefont
  {{Genzel}}}, \bibinfo {author} {\bibfnamefont {F.}~\bibnamefont
  {{Martins}}},\ and\ \bibinfo {author} {\bibfnamefont {T.}~\bibnamefont
  {{Ott}}},\ }\bibfield  {title} {\bibinfo {title} {{Monitoring Stellar Orbits
  Around the Massive Black Hole in the Galactic Center}},\ }\href
  {https://doi.org/10.1088/0004-637X/692/2/1075} {\bibfield  {journal}
  {\bibinfo  {journal} {\apj}\ }\textbf {\bibinfo {volume} {692}},\ \bibinfo
  {pages} {1075} (\bibinfo {year} {2009})},\ \Eprint
  {https://arxiv.org/abs/0810.4674} {arXiv:0810.4674 [astro-ph]} \BibitemShut
  {NoStop}%
\bibitem [{\citenamefont {Falcke}\ and\ \citenamefont
  {Markoff}(2013)}]{Falcke_2013}%
  \BibitemOpen
  \bibfield  {author} {\bibinfo {author} {\bibfnamefont {H.}~\bibnamefont
  {Falcke}}\ and\ \bibinfo {author} {\bibfnamefont {S.~B.}\ \bibnamefont
  {Markoff}},\ }\bibfield  {title} {\bibinfo {title} {Toward the event
  horizon—the supermassive black hole in the galactic center},\ }\href
  {https://doi.org/10.1088/0264-9381/30/24/244003} {\bibfield  {journal}
  {\bibinfo  {journal} {Classical and Quantum Gravity}\ }\textbf {\bibinfo
  {volume} {30}},\ \bibinfo {pages} {244003} (\bibinfo {year}
  {2013})}\BibitemShut {NoStop}%
\bibitem [{\citenamefont {{Event Horizon Telescope
  Collaboration}}(2019{\natexlab{a}})}]{2019ApJ...875L...1E}%
  \BibitemOpen
  \bibfield  {author} {\bibinfo {author} {\bibnamefont {{Event Horizon
  Telescope Collaboration}}},\ }\bibfield  {title} {\bibinfo {title} {{First
  M87 Event Horizon Telescope Results. I. The Shadow of the Supermassive Black
  Hole}},\ }\href {https://doi.org/10.3847/2041-8213/ab0ec7} {\bibfield
  {journal} {\bibinfo  {journal} {\apjl}\ }\textbf {\bibinfo {volume} {875}},\
  \bibinfo {eid} {L1} (\bibinfo {year} {2019}{\natexlab{a}})},\ \Eprint
  {https://arxiv.org/abs/1906.11238} {arXiv:1906.11238 [astro-ph.GA]}
  \BibitemShut {NoStop}%
\bibitem [{\citenamefont {{Event Horizon Telescope
  Collaboration}}(2019{\natexlab{b}})}]{EHTCV}%
  \BibitemOpen
  \bibfield  {author} {\bibinfo {author} {\bibnamefont {{Event Horizon
  Telescope Collaboration}}},\ }\bibfield  {title} {\bibinfo {title} {{First
  M87 Event Horizon Telescope Results. V. Physical Origin of the Asymmetric
  Ring}},\ }\href {https://doi.org/10.3847/2041-8213/ab0f43} {\bibfield
  {journal} {\bibinfo  {journal} {\apjl}\ }\textbf {\bibinfo {volume} {875}},\
  \bibinfo {eid} {L5} (\bibinfo {year} {2019}{\natexlab{b}})},\ \Eprint
  {https://arxiv.org/abs/1906.11242} {arXiv:1906.11242 [astro-ph.GA]}
  \BibitemShut {NoStop}%
\bibitem [{\citenamefont {Mahadevan}\ and\ \citenamefont
  {Quataert}(1997)}]{Mahadevan_1997}%
  \BibitemOpen
  \bibfield  {author} {\bibinfo {author} {\bibfnamefont {R.}~\bibnamefont
  {Mahadevan}}\ and\ \bibinfo {author} {\bibfnamefont {E.}~\bibnamefont
  {Quataert}},\ }\bibfield  {title} {\bibinfo {title} {Are particles in
  advection‐dominated accretion flows thermal?},\ }\href
  {https://doi.org/10.1086/304908} {\bibfield  {journal} {\bibinfo  {journal}
  {The Astrophysical Journal}\ }\textbf {\bibinfo {volume} {490}},\ \bibinfo
  {pages} {605–618} (\bibinfo {year} {1997})}\BibitemShut {NoStop}%
\bibitem [{\citenamefont {{Harris}}\ \emph {et~al.}(1998)\citenamefont
  {{Harris}}, \citenamefont {{Harris}},\ and\ \citenamefont
  {{McLaughlin}}}]{1998AJ....115.1801H}%
  \BibitemOpen
  \bibfield  {author} {\bibinfo {author} {\bibfnamefont {W.~E.}\ \bibnamefont
  {{Harris}}}, \bibinfo {author} {\bibfnamefont {G.~L.~H.}\ \bibnamefont
  {{Harris}}},\ and\ \bibinfo {author} {\bibfnamefont {D.~E.}\ \bibnamefont
  {{McLaughlin}}},\ }\bibfield  {title} {\bibinfo {title} {{{M}87, Globular
  Clusters, and Galactic Winds: Issues in Giant Galaxy Formation}},\ }\href
  {https://doi.org/10.1086/300322} {\bibfield  {journal} {\bibinfo  {journal}
  {\aj}\ }\textbf {\bibinfo {volume} {115}},\ \bibinfo {pages} {1801} (\bibinfo
  {year} {1998})},\ \Eprint {https://arxiv.org/abs/astro-ph/9801214}
  {arXiv:astro-ph/9801214 [astro-ph]} \BibitemShut {NoStop}%
\bibitem [{\citenamefont {{Baganoff}}\ \emph {et~al.}(2003)\citenamefont
  {{Baganoff}} \emph {et~al.}}]{2003ApJ...591..891B}%
  \BibitemOpen
  \bibfield  {author} {\bibinfo {author} {\bibfnamefont {F.~K.}\ \bibnamefont
  {{Baganoff}}} \emph {et~al.},\ }\bibfield  {title} {\bibinfo {title}
  {{Chandra X-Ray Spectroscopic Imaging of Sagittarius A* and the Central
  Parsec of the Galaxy}},\ }\href {https://doi.org/10.1086/375145} {\bibfield
  {journal} {\bibinfo  {journal} {\apjl}\ }\textbf {\bibinfo {volume} {591}},\
  \bibinfo {pages} {891} (\bibinfo {year} {2003})},\ \Eprint
  {https://arxiv.org/abs/astro-ph/0102151} {arXiv:astro-ph/0102151 [astro-ph]}
  \BibitemShut {NoStop}%
\bibitem [{\citenamefont {Read}\ and\ \citenamefont
  {Gilmore}(2003)}]{Read:2002wb}%
  \BibitemOpen
  \bibfield  {author} {\bibinfo {author} {\bibfnamefont {J.~I.}\ \bibnamefont
  {Read}}\ and\ \bibinfo {author} {\bibfnamefont {G.}~\bibnamefont {Gilmore}},\
  }\bibfield  {title} {\bibinfo {title} {{Can supermassive black holes alter
  cold dark matter cusps through accretion?}},\ }\href
  {https://doi.org/10.1046/j.1365-8711.2003.06232.x} {\bibfield  {journal}
  {\bibinfo  {journal} {Mon. Not. Roy. Astron. Soc.}\ }\textbf {\bibinfo
  {volume} {339}},\ \bibinfo {pages} {949} (\bibinfo {year} {2003})},\ \Eprint
  {https://arxiv.org/abs/astro-ph/0210658} {arXiv:astro-ph/0210658}
  \BibitemShut {NoStop}%
\bibitem [{\citenamefont {{Choquette}}\ \emph {et~al.}(2019)\citenamefont
  {{Choquette}}, \citenamefont {{Cline}},\ and\ \citenamefont
  {{Cornell}}}]{Choquette:2018lvq}%
  \BibitemOpen
  \bibfield  {author} {\bibinfo {author} {\bibfnamefont {J.}~\bibnamefont
  {{Choquette}}}, \bibinfo {author} {\bibfnamefont {J.~M.}\ \bibnamefont
  {{Cline}}},\ and\ \bibinfo {author} {\bibfnamefont {J.~M.}\ \bibnamefont
  {{Cornell}}},\ }\bibfield  {title} {\bibinfo {title} {{Early formation of
  supermassive black holes via dark matter self-interactions}},\ }\href
  {https://doi.org/10.1088/1475-7516/2019/07/036} {\bibfield  {journal}
  {\bibinfo  {journal} {\jcap}\ }\textbf {\bibinfo {volume} {2019}},\ \bibinfo
  {eid} {036} (\bibinfo {year} {2019})},\ \Eprint
  {https://arxiv.org/abs/1812.05088} {arXiv:1812.05088 [astro-ph.CO]}
  \BibitemShut {NoStop}%
\bibitem [{\citenamefont {{Arg{\"u}elles}}\ \emph {et~al.}(2021)\citenamefont
  {{Arg{\"u}elles}}, \citenamefont {{D{\'\i}az}}, \citenamefont {{Krut}},\ and\
  \citenamefont {{Yunis}}}]{2021MNRAS.502.4227A}%
  \BibitemOpen
  \bibfield  {author} {\bibinfo {author} {\bibfnamefont {C.~R.}\ \bibnamefont
  {{Arg{\"u}elles}}}, \bibinfo {author} {\bibfnamefont {M.~I.}\ \bibnamefont
  {{D{\'\i}az}}}, \bibinfo {author} {\bibfnamefont {A.}~\bibnamefont
  {{Krut}}},\ and\ \bibinfo {author} {\bibfnamefont {R.}~\bibnamefont
  {{Yunis}}},\ }\bibfield  {title} {\bibinfo {title} {{On the formation and
  stability of fermionic dark matter haloes in a cosmological framework}},\
  }\href {https://doi.org/10.1093/mnras/staa3986} {\bibfield  {journal}
  {\bibinfo  {journal} {\mnras}\ }\textbf {\bibinfo {volume} {502}},\ \bibinfo
  {pages} {4227} (\bibinfo {year} {2021})},\ \Eprint
  {https://arxiv.org/abs/2012.11709} {arXiv:2012.11709 [astro-ph.GA]}
  \BibitemShut {NoStop}%
\bibitem [{\citenamefont {Cercignani}\ and\ \citenamefont
  {Kremer}(2002)}]{CercignaniKremer-Book}%
  \BibitemOpen
  \bibfield  {author} {\bibinfo {author} {\bibfnamefont {C.}~\bibnamefont
  {Cercignani}}\ and\ \bibinfo {author} {\bibfnamefont {G.}~\bibnamefont
  {Kremer}},\ }\href@noop {} {\emph {\bibinfo {title} {The Relativistic
  Boltzmann Equation: Theory and Applications}}}\ (\bibinfo  {publisher}
  {{Birkh\"auser}},\ \bibinfo {address} {Basel},\ \bibinfo {year}
  {2002})\BibitemShut {NoStop}%
\bibitem [{\citenamefont {{Zeldovich}}\ and\ \citenamefont
  {{Novikov}}(1971)}]{1971reas.book.....Z}%
  \BibitemOpen
  \bibfield  {author} {\bibinfo {author} {\bibfnamefont {Y.~B.}\ \bibnamefont
  {{Zeldovich}}}\ and\ \bibinfo {author} {\bibfnamefont {I.~D.}\ \bibnamefont
  {{Novikov}}},\ }\href@noop {} {\emph {\bibinfo {title} {{Relativistic
  Astrophysics. Vol.1: Stars and Relativity}}}}\ (\bibinfo  {publisher}
  {University of Chicago Press},\ \bibinfo {address} {Chicago, USA},\ \bibinfo
  {year} {1971})\BibitemShut {NoStop}%
\bibitem [{\citenamefont {{Shapiro}}\ and\ \citenamefont
  {{Teukolsky}}(1983)}]{1983bhwd.book.....S}%
  \BibitemOpen
  \bibfield  {author} {\bibinfo {author} {\bibfnamefont {S.~L.}\ \bibnamefont
  {{Shapiro}}}\ and\ \bibinfo {author} {\bibfnamefont {S.~A.}\ \bibnamefont
  {{Teukolsky}}},\ }\href@noop {} {\emph {\bibinfo {title} {{Black Holes, White
  Dwarfs, and Neutron Stars: The Physics of Compact Objects}}}}\ (\bibinfo
  {publisher} {Wiley},\ \bibinfo {address} {New York, USA},\ \bibinfo {year}
  {1983})\BibitemShut {NoStop}%
\bibitem [{\citenamefont {Rioseco}\ and\ \citenamefont
  {Sarbach}(2017{\natexlab{a}})}]{pRoS17a}%
  \BibitemOpen
  \bibfield  {author} {\bibinfo {author} {\bibfnamefont {P.}~\bibnamefont
  {Rioseco}}\ and\ \bibinfo {author} {\bibfnamefont {O.}~\bibnamefont
  {Sarbach}},\ }\bibfield  {title} {\bibinfo {title} {Accretion of a
  relativistic, collisionless kinetic gas into a {S}chwarzschild black hole},\
  }\href {https://doi.org/10.1088/1361-6382/aa65fa} {\bibfield  {journal}
  {\bibinfo  {journal} {Class. Quantum Grav.}\ }\textbf {\bibinfo {volume}
  {34}},\ \bibinfo {pages} {095007} (\bibinfo {year}
  {2017}{\natexlab{a}})}\BibitemShut {NoStop}%
\bibitem [{\citenamefont {Rioseco}\ and\ \citenamefont
  {Sarbach}(2017{\natexlab{b}})}]{pRoS17b}%
  \BibitemOpen
  \bibfield  {author} {\bibinfo {author} {\bibfnamefont {P.}~\bibnamefont
  {Rioseco}}\ and\ \bibinfo {author} {\bibfnamefont {O.}~\bibnamefont
  {Sarbach}},\ }\bibfield  {title} {\bibinfo {title} {Spherical steady-state
  accretion of a relativistic collisionless gas into a {S}chwarzschild black
  hole},\ }\bibfield  {booktitle} {\emph {\bibinfo {booktitle} {{70\&70
  Classical and Quantum Gravitation Party: Meeting with Two Latin American
  Masters on Theoretical Physics Cartagena, Colombia, September 28-30,
  2016}}},\ }\href {https://doi.org/10.1088/1742-6596/831/1/012009} {\bibfield
  {journal} {\bibinfo  {journal} {J. Phys. Conf. Ser.}\ }\textbf {\bibinfo
  {volume} {831}},\ \bibinfo {pages} {012009} (\bibinfo {year}
  {2017}{\natexlab{b}})}\BibitemShut {NoStop}%
\bibitem [{\citenamefont {Mach}\ and\ \citenamefont
  {Odrzywo\l{}ek}(2021{\natexlab{a}})}]{pMaO21a}%
  \BibitemOpen
  \bibfield  {author} {\bibinfo {author} {\bibfnamefont {P.}~\bibnamefont
  {Mach}}\ and\ \bibinfo {author} {\bibfnamefont {A.}~\bibnamefont
  {Odrzywo\l{}ek}},\ }\bibfield  {title} {\bibinfo {title} {Accretion of the
  relativistic {V}lasov gas onto a moving {S}chwarzschild black hole: Exact
  solutions},\ }\href {https://doi.org/10.1103/PhysRevD.103.024044} {\bibfield
  {journal} {\bibinfo  {journal} {Phys. Rev. D}\ }\textbf {\bibinfo {volume}
  {103}},\ \bibinfo {pages} {024044} (\bibinfo {year}
  {2021}{\natexlab{a}})}\BibitemShut {NoStop}%
\bibitem [{\citenamefont {Mach}\ and\ \citenamefont
  {Odrzywo\l{}ek}(2021{\natexlab{b}})}]{pMaO21b}%
  \BibitemOpen
  \bibfield  {author} {\bibinfo {author} {\bibfnamefont {P.}~\bibnamefont
  {Mach}}\ and\ \bibinfo {author} {\bibfnamefont {A.}~\bibnamefont
  {Odrzywo\l{}ek}},\ }\bibfield  {title} {\bibinfo {title} {Accretion of dark
  matter onto a moving {S}chwarzschild black hole: An exact solution},\ }\href
  {https://doi.org/10.1103/PhysRevLett.126.101104} {\bibfield  {journal}
  {\bibinfo  {journal} {Phys. Rev. Lett.}\ }\textbf {\bibinfo {volume} {126}},\
  \bibinfo {pages} {101104} (\bibinfo {year} {2021}{\natexlab{b}})}\BibitemShut
  {NoStop}%
\bibitem [{\citenamefont {Sharma}\ \emph {et~al.}(2006)\citenamefont {Sharma},
  \citenamefont {Hammett}, \citenamefont {Quataert},\ and\ \citenamefont
  {Stone}}]{Sharma:2005ez}%
  \BibitemOpen
  \bibfield  {author} {\bibinfo {author} {\bibfnamefont {P.}~\bibnamefont
  {Sharma}}, \bibinfo {author} {\bibfnamefont {G.~W.}\ \bibnamefont {Hammett}},
  \bibinfo {author} {\bibfnamefont {E.}~\bibnamefont {Quataert}},\ and\
  \bibinfo {author} {\bibfnamefont {J.~M.}\ \bibnamefont {Stone}},\ }\bibfield
  {title} {\bibinfo {title} {{Shearing box simulations of the mri in a
  collisionless plasma}},\ }\href {https://doi.org/10.1086/498405} {\bibfield
  {journal} {\bibinfo  {journal} {Astrophys. J.}\ }\textbf {\bibinfo {volume}
  {637}},\ \bibinfo {pages} {952} (\bibinfo {year} {2006})},\ \Eprint
  {https://arxiv.org/abs/astro-ph/0508502} {arXiv:astro-ph/0508502}
  \BibitemShut {NoStop}%
\bibitem [{\citenamefont {Chandra}\ \emph {et~al.}(2015)\citenamefont
  {Chandra}, \citenamefont {Gammie}, \citenamefont {Foucart},\ and\
  \citenamefont {Quataert}}]{Chandra:2015iza}%
  \BibitemOpen
  \bibfield  {author} {\bibinfo {author} {\bibfnamefont {M.}~\bibnamefont
  {Chandra}}, \bibinfo {author} {\bibfnamefont {C.~F.}\ \bibnamefont {Gammie}},
  \bibinfo {author} {\bibfnamefont {F.}~\bibnamefont {Foucart}},\ and\ \bibinfo
  {author} {\bibfnamefont {E.}~\bibnamefont {Quataert}},\ }\bibfield  {title}
  {\bibinfo {title} {{An Extended Magnetohydrodynamics Model for Relativistic
  Weakly Collisional Plasmas}},\ }\href
  {https://doi.org/10.1088/0004-637X/810/2/162} {\bibfield  {journal} {\bibinfo
   {journal} {Astrophys. J.}\ }\textbf {\bibinfo {volume} {810}},\ \bibinfo
  {pages} {162} (\bibinfo {year} {2015})},\ \Eprint
  {https://arxiv.org/abs/1508.00878} {arXiv:1508.00878 [astro-ph.HE]}
  \BibitemShut {NoStop}%
\bibitem [{\citenamefont {Foucart}\ \emph {et~al.}(2017)\citenamefont
  {Foucart}, \citenamefont {Chandra}, \citenamefont {Gammie}, \citenamefont
  {Quataert},\ and\ \citenamefont {Tchekhovskoy}}]{Foucart:2017axc}%
  \BibitemOpen
  \bibfield  {author} {\bibinfo {author} {\bibfnamefont {F.}~\bibnamefont
  {Foucart}}, \bibinfo {author} {\bibfnamefont {M.}~\bibnamefont {Chandra}},
  \bibinfo {author} {\bibfnamefont {C.~F.}\ \bibnamefont {Gammie}}, \bibinfo
  {author} {\bibfnamefont {E.}~\bibnamefont {Quataert}},\ and\ \bibinfo
  {author} {\bibfnamefont {A.}~\bibnamefont {Tchekhovskoy}},\ }\bibfield
  {title} {\bibinfo {title} {{How important is non-ideal physics in simulations
  of sub-Eddington accretion on to spinning black holes?}},\ }\href
  {https://doi.org/10.1093/mnras/stx1368} {\bibfield  {journal} {\bibinfo
  {journal} {Mon. Not. Roy. Astron. Soc.}\ }\textbf {\bibinfo {volume} {470}},\
  \bibinfo {pages} {2240} (\bibinfo {year} {2017})},\ \Eprint
  {https://arxiv.org/abs/1706.01533} {arXiv:1706.01533 [astro-ph.HE]}
  \BibitemShut {NoStop}%
\bibitem [{\citenamefont {Kunz}\ \emph {et~al.}(2016)\citenamefont {Kunz},
  \citenamefont {Stone},\ and\ \citenamefont {Quataert}}]{Kunz_2016}%
  \BibitemOpen
  \bibfield  {author} {\bibinfo {author} {\bibfnamefont {M.~W.}\ \bibnamefont
  {Kunz}}, \bibinfo {author} {\bibfnamefont {J.~M.}\ \bibnamefont {Stone}},\
  and\ \bibinfo {author} {\bibfnamefont {E.}~\bibnamefont {Quataert}},\
  }\bibfield  {title} {\bibinfo {title} {Magnetorotational turbulence and
  dynamo in a collisionless plasma},\ }\bibfield  {journal} {\bibinfo
  {journal} {Physical Review Letters}\ }\textbf {\bibinfo {volume} {117}},\
  \href {https://doi.org/10.1103/physrevlett.117.235101}
  {10.1103/physrevlett.117.235101} (\bibinfo {year} {2016})\BibitemShut
  {NoStop}%
\bibitem [{\citenamefont {Rioseco}\ and\ \citenamefont
  {Sarbach}(2018)}]{pRoS18}%
  \BibitemOpen
  \bibfield  {author} {\bibinfo {author} {\bibfnamefont {P.}~\bibnamefont
  {Rioseco}}\ and\ \bibinfo {author} {\bibfnamefont {O.}~\bibnamefont
  {Sarbach}},\ }\bibfield  {title} {\bibinfo {title} {Phase space mixing in the
  equatorial plane of a {K}err black hole},\ }\href
  {https://doi.org/10.1103/PhysRevD.98.124024} {\bibfield  {journal} {\bibinfo
  {journal} {Phys. Rev. D}\ }\textbf {\bibinfo {volume} {98}},\ \bibinfo
  {pages} {124024} (\bibinfo {year} {2018})}\BibitemShut {NoStop}%
\bibitem [{\citenamefont {Aguayo-Ortiz}\ \emph {et~al.}(2019)\citenamefont
  {Aguayo-Ortiz}, \citenamefont {Tejeda},\ and\ \citenamefont
  {Hernandez}}]{Aguayo-Ortiz:2019fap}%
  \BibitemOpen
  \bibfield  {author} {\bibinfo {author} {\bibfnamefont {A.}~\bibnamefont
  {Aguayo-Ortiz}}, \bibinfo {author} {\bibfnamefont {E.}~\bibnamefont
  {Tejeda}},\ and\ \bibinfo {author} {\bibfnamefont {X.}~\bibnamefont
  {Hernandez}},\ }\bibfield  {title} {\bibinfo {title} {{Choked accretion: from
  radial infall to bipolar outflows by breaking spherical symmetry}},\ }\href
  {https://doi.org/10.1093/mnras/stz2989} {\bibfield  {journal} {\bibinfo
  {journal} {Mon. Not. Roy. Astron. Soc.}\ }\textbf {\bibinfo {volume} {490}},\
  \bibinfo {pages} {5078} (\bibinfo {year} {2019})},\ \Eprint
  {https://arxiv.org/abs/1909.00884} {arXiv:1909.00884 [astro-ph.HE]}
  \BibitemShut {NoStop}%
\bibitem [{\citenamefont {Tejeda}\ \emph {et~al.}(2020)\citenamefont {Tejeda},
  \citenamefont {Aguayo-Ortiz},\ and\ \citenamefont
  {Hernandez}}]{Tejeda:2019fwr}%
  \BibitemOpen
  \bibfield  {author} {\bibinfo {author} {\bibfnamefont {E.}~\bibnamefont
  {Tejeda}}, \bibinfo {author} {\bibfnamefont {A.}~\bibnamefont
  {Aguayo-Ortiz}},\ and\ \bibinfo {author} {\bibfnamefont {X.}~\bibnamefont
  {Hernandez}},\ }\bibfield  {title} {\bibinfo {title} {{Choked accretion onto
  a Schwarzschild black hole: A hydrodynamical jet-launching mechanism}},\
  }\href {https://doi.org/10.3847/1538-4357/ab7ffe} {\bibfield  {journal}
  {\bibinfo  {journal} {Astrophys. J.}\ }\textbf {\bibinfo {volume} {893}},\
  \bibinfo {pages} {81} (\bibinfo {year} {2020})},\ \Eprint
  {https://arxiv.org/abs/1909.01527} {arXiv:1909.01527 [astro-ph.HE]}
  \BibitemShut {NoStop}%
\bibitem [{\citenamefont {Aguayo-Ortiz}\ \emph {et~al.}(2021)\citenamefont
  {Aguayo-Ortiz}, \citenamefont {Sarbach},\ and\ \citenamefont
  {Tejeda}}]{Aguayo-Ortiz:2020qro}%
  \BibitemOpen
  \bibfield  {author} {\bibinfo {author} {\bibfnamefont {A.}~\bibnamefont
  {Aguayo-Ortiz}}, \bibinfo {author} {\bibfnamefont {O.}~\bibnamefont
  {Sarbach}},\ and\ \bibinfo {author} {\bibfnamefont {E.}~\bibnamefont
  {Tejeda}},\ }\bibfield  {title} {\bibinfo {title} {{Choked accretion onto a
  Kerr black hole}},\ }\href {https://doi.org/10.1103/PhysRevD.103.023003}
  {\bibfield  {journal} {\bibinfo  {journal} {Phys. Rev. D}\ }\textbf {\bibinfo
  {volume} {103}},\ \bibinfo {pages} {023003} (\bibinfo {year} {2021})},\
  \Eprint {https://arxiv.org/abs/2009.06653} {arXiv:2009.06653 [astro-ph.HE]}
  \BibitemShut {NoStop}%
\bibitem [{\citenamefont {{Di Matteo}}\ \emph {et~al.}(2003)\citenamefont {{Di
  Matteo}}, \citenamefont {{Allen}}, \citenamefont {{Fabian}}, \citenamefont
  {{Wilson}},\ and\ \citenamefont {{Young}}}]{2003ApJ...582..133D}%
  \BibitemOpen
  \bibfield  {author} {\bibinfo {author} {\bibfnamefont {T.}~\bibnamefont {{Di
  Matteo}}}, \bibinfo {author} {\bibfnamefont {S.~W.}\ \bibnamefont {{Allen}}},
  \bibinfo {author} {\bibfnamefont {A.~C.}\ \bibnamefont {{Fabian}}}, \bibinfo
  {author} {\bibfnamefont {A.~S.}\ \bibnamefont {{Wilson}}},\ and\ \bibinfo
  {author} {\bibfnamefont {A.~J.}\ \bibnamefont {{Young}}},\ }\bibfield
  {title} {\bibinfo {title} {{Accretion onto the Supermassive Black Hole in
  M87}},\ }\href {https://doi.org/10.1086/344504} {\bibfield  {journal}
  {\bibinfo  {journal} {\apj}\ }\textbf {\bibinfo {volume} {582}},\ \bibinfo
  {pages} {133} (\bibinfo {year} {2003})},\ \Eprint
  {https://arxiv.org/abs/astro-ph/0202238} {arXiv:astro-ph/0202238 [astro-ph]}
  \BibitemShut {NoStop}%
\bibitem [{\citenamefont {{Aitken}}\ \emph {et~al.}(2000)\citenamefont
  {{Aitken}}, \citenamefont {{Greaves}}, \citenamefont {{Chrysostomou}},
  \citenamefont {{Jenness}}, \citenamefont {{Holland}}, \citenamefont
  {{Hough}}, \citenamefont {{Pierce-Price}},\ and\ \citenamefont
  {{Richer}}}]{2000ApJ...534L.173A}%
  \BibitemOpen
  \bibfield  {author} {\bibinfo {author} {\bibfnamefont {D.~K.}\ \bibnamefont
  {{Aitken}}}, \bibinfo {author} {\bibfnamefont {J.}~\bibnamefont {{Greaves}}},
  \bibinfo {author} {\bibfnamefont {A.}~\bibnamefont {{Chrysostomou}}},
  \bibinfo {author} {\bibfnamefont {T.}~\bibnamefont {{Jenness}}}, \bibinfo
  {author} {\bibfnamefont {W.}~\bibnamefont {{Holland}}}, \bibinfo {author}
  {\bibfnamefont {J.~H.}\ \bibnamefont {{Hough}}}, \bibinfo {author}
  {\bibfnamefont {D.}~\bibnamefont {{Pierce-Price}}},\ and\ \bibinfo {author}
  {\bibfnamefont {J.}~\bibnamefont {{Richer}}},\ }\bibfield  {title} {\bibinfo
  {title} {{Detection of Polarized Millimeter and Submillimeter Emission from
  Sagittarius A*}},\ }\href {https://doi.org/10.1086/312685} {\bibfield
  {journal} {\bibinfo  {journal} {\apjl}\ }\textbf {\bibinfo {volume} {534}},\
  \bibinfo {pages} {L173} (\bibinfo {year} {2000})}\BibitemShut {NoStop}%
\bibitem [{\citenamefont {{Kuo}}\ \emph {et~al.}(2014)\citenamefont {{Kuo}}
  \emph {et~al.}}]{2014ApJ...783L..33K}%
  \BibitemOpen
  \bibfield  {author} {\bibinfo {author} {\bibfnamefont {C.~Y.}\ \bibnamefont
  {{Kuo}}} \emph {et~al.},\ }\bibfield  {title} {\bibinfo {title} {{Measuring
  Mass Accretion Rate onto the Supermassive Black Hole in M87 Using Faraday
  Rotation Measure with the Submillimeter Array}},\ }\href
  {https://doi.org/10.1088/2041-8205/783/2/L33} {\bibfield  {journal} {\bibinfo
   {journal} {\apjl}\ }\textbf {\bibinfo {volume} {783}},\ \bibinfo {eid} {L33}
  (\bibinfo {year} {2014})},\ \Eprint {https://arxiv.org/abs/1402.5238}
  {arXiv:1402.5238 [astro-ph.GA]} \BibitemShut {NoStop}%
\bibitem [{\citenamefont {{Event Horizon Telescope
  Collaboration}}(2021{\natexlab{a}})}]{ehtc7}%
  \BibitemOpen
  \bibfield  {author} {\bibinfo {author} {\bibnamefont {{Event Horizon
  Telescope Collaboration}}},\ }\bibfield  {title} {\bibinfo {title} {{First
  M87 Event Horizon Telescope Results. VII. Polarization of the Ring}},\ }\href
  {https://doi.org/10.3847/2041-8213/abe71d} {\bibfield  {journal} {\bibinfo
  {journal} {\apjl}\ }\textbf {\bibinfo {volume} {910}},\ \bibinfo {eid} {L12}
  (\bibinfo {year} {2021}{\natexlab{a}})},\ \Eprint
  {https://arxiv.org/abs/2105.01169} {arXiv:2105.01169 [astro-ph.HE]}
  \BibitemShut {NoStop}%
\bibitem [{\citenamefont {{Event Horizon Telescope
  Collaboration}}(2021{\natexlab{b}})}]{EHTCVIII}%
  \BibitemOpen
  \bibfield  {author} {\bibinfo {author} {\bibnamefont {{Event Horizon
  Telescope Collaboration}}},\ }\bibfield  {title} {\bibinfo {title} {{First
  M87 Event Horizon Telescope Results. VIII. Magnetic Field Structure near The
  Event Horizon}},\ }\href {https://doi.org/10.3847/2041-8213/abe4de}
  {\bibfield  {journal} {\bibinfo  {journal} {\apjl}\ }\textbf {\bibinfo
  {volume} {910}},\ \bibinfo {eid} {L13} (\bibinfo {year}
  {2021}{\natexlab{b}})},\ \Eprint {https://arxiv.org/abs/2105.01173}
  {arXiv:2105.01173 [astro-ph.HE]} \BibitemShut {NoStop}%
\bibitem [{\citenamefont {{Quataert}}\ and\ \citenamefont
  {{Gruzinov}}(2000)}]{2000ApJ...545..842Q}%
  \BibitemOpen
  \bibfield  {author} {\bibinfo {author} {\bibfnamefont {E.}~\bibnamefont
  {{Quataert}}}\ and\ \bibinfo {author} {\bibfnamefont {A.}~\bibnamefont
  {{Gruzinov}}},\ }\bibfield  {title} {\bibinfo {title} {{Constraining the
  Accretion Rate onto Sagittarius A* Using Linear Polarization}},\ }\href
  {https://doi.org/10.1086/317845} {\bibfield  {journal} {\bibinfo  {journal}
  {\apj}\ }\textbf {\bibinfo {volume} {545}},\ \bibinfo {pages} {842} (\bibinfo
  {year} {2000})},\ \Eprint {https://arxiv.org/abs/astro-ph/0004286}
  {arXiv:astro-ph/0004286 [astro-ph]} \BibitemShut {NoStop}%
\bibitem [{\citenamefont {{Jim{\'e}nez-Rosales}}\ and\ \citenamefont
  {{Dexter}}(2018)}]{2018MNRAS.478.1875J}%
  \BibitemOpen
  \bibfield  {author} {\bibinfo {author} {\bibfnamefont {A.}~\bibnamefont
  {{Jim{\'e}nez-Rosales}}}\ and\ \bibinfo {author} {\bibfnamefont
  {J.}~\bibnamefont {{Dexter}}},\ }\bibfield  {title} {\bibinfo {title} {{The
  impact of Faraday effects on polarized black hole images of Sagittarius
  A*}},\ }\href {https://doi.org/10.1093/mnras/sty1210} {\bibfield  {journal}
  {\bibinfo  {journal} {\mnras}\ }\textbf {\bibinfo {volume} {478}},\ \bibinfo
  {pages} {1875} (\bibinfo {year} {2018})},\ \Eprint
  {https://arxiv.org/abs/1805.02652} {arXiv:1805.02652 [astro-ph.HE]}
  \BibitemShut {NoStop}%
\bibitem [{\citenamefont {{Narayan}}\ and\ \citenamefont
  {{Yi}}(1995)}]{1995ApJ...452..710N}%
  \BibitemOpen
  \bibfield  {author} {\bibinfo {author} {\bibfnamefont {R.}~\bibnamefont
  {{Narayan}}}\ and\ \bibinfo {author} {\bibfnamefont {I.}~\bibnamefont
  {{Yi}}},\ }\bibfield  {title} {\bibinfo {title} {{Advection-dominated
  Accretion: Underfed Black Holes and Neutron Stars}},\ }\href
  {https://doi.org/10.1086/176343} {\bibfield  {journal} {\bibinfo  {journal}
  {\apj}\ }\textbf {\bibinfo {volume} {452}},\ \bibinfo {pages} {710} (\bibinfo
  {year} {1995})},\ \Eprint {https://arxiv.org/abs/astro-ph/9411059}
  {arXiv:astro-ph/9411059 [astro-ph]} \BibitemShut {NoStop}%
\bibitem [{\citenamefont {{Quataert}}\ and\ \citenamefont
  {{Narayan}}(1999)}]{1999ApJ...516..399Q}%
  \BibitemOpen
  \bibfield  {author} {\bibinfo {author} {\bibfnamefont {E.}~\bibnamefont
  {{Quataert}}}\ and\ \bibinfo {author} {\bibfnamefont {R.}~\bibnamefont
  {{Narayan}}},\ }\bibfield  {title} {\bibinfo {title} {{On the Energetics of
  Advection-dominated Accretion Flows}},\ }\href
  {https://doi.org/10.1086/307097} {\bibfield  {journal} {\bibinfo  {journal}
  {\apj}\ }\textbf {\bibinfo {volume} {516}},\ \bibinfo {pages} {399} (\bibinfo
  {year} {1999})},\ \Eprint {https://arxiv.org/abs/astro-ph/9810117}
  {arXiv:astro-ph/9810117 [astro-ph]} \BibitemShut {NoStop}%
\bibitem [{\citenamefont {{Yuan}}\ \emph {et~al.}(2003)\citenamefont {{Yuan}},
  \citenamefont {{Quataert}},\ and\ \citenamefont
  {{Narayan}}}]{2003ApJ...598..301Y}%
  \BibitemOpen
  \bibfield  {author} {\bibinfo {author} {\bibfnamefont {F.}~\bibnamefont
  {{Yuan}}}, \bibinfo {author} {\bibfnamefont {E.}~\bibnamefont {{Quataert}}},\
  and\ \bibinfo {author} {\bibfnamefont {R.}~\bibnamefont {{Narayan}}},\
  }\bibfield  {title} {\bibinfo {title} {{Nonthermal Electrons in Radiatively
  Inefficient Accretion Flow Models of Sagittarius A*}},\ }\href
  {https://doi.org/10.1086/378716} {\bibfield  {journal} {\bibinfo  {journal}
  {\apj}\ }\textbf {\bibinfo {volume} {598}},\ \bibinfo {pages} {301} (\bibinfo
  {year} {2003})},\ \Eprint {https://arxiv.org/abs/astro-ph/0304125}
  {arXiv:astro-ph/0304125 [astro-ph]} \BibitemShut {NoStop}%
\bibitem [{\citenamefont {{Yuan}}\ and\ \citenamefont
  {{Narayan}}(2014)}]{2014ARA&A..52..529Y}%
  \BibitemOpen
  \bibfield  {author} {\bibinfo {author} {\bibfnamefont {F.}~\bibnamefont
  {{Yuan}}}\ and\ \bibinfo {author} {\bibfnamefont {R.}~\bibnamefont
  {{Narayan}}},\ }\bibfield  {title} {\bibinfo {title} {{Hot Accretion Flows
  Around Black Holes}},\ }\href
  {https://doi.org/10.1146/annurev-astro-082812-141003} {\bibfield  {journal}
  {\bibinfo  {journal} {\araa}\ }\textbf {\bibinfo {volume} {52}},\ \bibinfo
  {pages} {529} (\bibinfo {year} {2014})},\ \Eprint
  {https://arxiv.org/abs/1401.0586} {arXiv:1401.0586 [astro-ph.HE]}
  \BibitemShut {NoStop}%
\bibitem [{\citenamefont {{Debbasch}}\ and\ \citenamefont {{van
  Leeuwen}}(2009)}]{2009PhyA..388.1079D}%
  \BibitemOpen
  \bibfield  {author} {\bibinfo {author} {\bibfnamefont {F.}~\bibnamefont
  {{Debbasch}}}\ and\ \bibinfo {author} {\bibfnamefont {W.~A.}\ \bibnamefont
  {{van Leeuwen}}},\ }\bibfield  {title} {\bibinfo {title} {{General
  relativistic Boltzmann equation, I: Covariant treatment}},\ }\href
  {https://doi.org/10.1016/j.physa.2008.12.023} {\bibfield  {journal} {\bibinfo
   {journal} {Physica A Statistical Mechanics and its Applications}\ }\textbf
  {\bibinfo {volume} {388}},\ \bibinfo {pages} {1079} (\bibinfo {year}
  {2009})}\BibitemShut {NoStop}%
\bibitem [{\citenamefont {Sarbach}\ and\ \citenamefont
  {Zannias}(2014)}]{Sarbach:2013uba}%
  \BibitemOpen
  \bibfield  {author} {\bibinfo {author} {\bibfnamefont {O.}~\bibnamefont
  {Sarbach}}\ and\ \bibinfo {author} {\bibfnamefont {T.}~\bibnamefont
  {Zannias}},\ }\bibfield  {title} {\bibinfo {title} {{The geometry of the
  tangent bundle and the relativistic kinetic theory of gases}},\ }\href
  {https://doi.org/10.1088/0264-9381/31/8/085013} {\bibfield  {journal}
  {\bibinfo  {journal} {Class. Quant. Grav.}\ }\textbf {\bibinfo {volume}
  {31}},\ \bibinfo {pages} {085013} (\bibinfo {year} {2014})},\ \Eprint
  {https://arxiv.org/abs/1309.2036} {arXiv:1309.2036 [gr-qc]} \BibitemShut
  {NoStop}%
\bibitem [{\citenamefont {Acu{\~n}a{-}C{\'a}rdenas}\ \emph
  {et~al.}(2021)\citenamefont {Acu{\~n}a{-}C{\'a}rdenas}, \citenamefont
  {Gabarrete},\ and\ \citenamefont {Sarbach}}]{rAcGoS2021}%
  \BibitemOpen
  \bibfield  {author} {\bibinfo {author} {\bibfnamefont {R.}~\bibnamefont
  {Acu{\~n}a{-}C{\'a}rdenas}}, \bibinfo {author} {\bibfnamefont
  {C.}~\bibnamefont {Gabarrete}},\ and\ \bibinfo {author} {\bibfnamefont
  {O.}~\bibnamefont {Sarbach}},\ }\bibfield  {title} {\bibinfo {title} {An
  introduction to the relativistic kinetic theory on curved spacetimes},\
  }\href@noop {} {\bibfield  {journal} {\bibinfo  {journal} {arXiv:2106.09235}\
  } (\bibinfo {year} {2021})}\BibitemShut {NoStop}%
\bibitem [{\citenamefont {Rioseco}\ and\ \citenamefont
  {Sarbach}(2020)}]{pRoS20}%
  \BibitemOpen
  \bibfield  {author} {\bibinfo {author} {\bibfnamefont {P.}~\bibnamefont
  {Rioseco}}\ and\ \bibinfo {author} {\bibfnamefont {O.}~\bibnamefont
  {Sarbach}},\ }\bibfield  {title} {\bibinfo {title} {{Phase space mixing in an
  external gravitational central potential}},\ }\href
  {https://doi.org/10.1088/1361-6382/ababb3} {\bibfield  {journal} {\bibinfo
  {journal} {Class. Quant. Grav.}\ }\textbf {\bibinfo {volume} {37}},\ \bibinfo
  {pages} {195027} (\bibinfo {year} {2020})},\ \Eprint
  {https://arxiv.org/abs/2005.05988} {arXiv:2005.05988 [gr-qc]} \BibitemShut
  {NoStop}%
\bibitem [{\citenamefont {{Binney}}\ and\ \citenamefont
  {{Tremaine}}(2008)}]{binney2011galactic}%
  \BibitemOpen
  \bibfield  {author} {\bibinfo {author} {\bibfnamefont {J.}~\bibnamefont
  {{Binney}}}\ and\ \bibinfo {author} {\bibfnamefont {S.}~\bibnamefont
  {{Tremaine}}},\ }\href@noop {} {\emph {\bibinfo {title} {Galactic
  Dynamics}}},\ \bibinfo {edition} {2nd}\ ed.\ (\bibinfo  {publisher}
  {Princeton University Press},\ \bibinfo {address} {Princeton, NJ, USA},\
  \bibinfo {year} {2008})\BibitemShut {NoStop}%
\bibitem [{\citenamefont {{Crawford}}\ and\ \citenamefont
  {{Tereno}}(2002)}]{2002GReGr..34.2075C}%
  \BibitemOpen
  \bibfield  {author} {\bibinfo {author} {\bibfnamefont {P.}~\bibnamefont
  {{Crawford}}}\ and\ \bibinfo {author} {\bibfnamefont {I.}~\bibnamefont
  {{Tereno}}},\ }\bibfield  {title} {\bibinfo {title} {{Generalized Observers
  and Velocity Measurements in General Relativity}},\ }\href
  {https://doi.org/10.1023/A:1021131401034} {\bibfield  {journal} {\bibinfo
  {journal} {General Relativity and Gravitation}\ }\textbf {\bibinfo {volume}
  {34}},\ \bibinfo {pages} {2075} (\bibinfo {year} {2002})},\ \Eprint
  {https://arxiv.org/abs/gr-qc/0111073} {arXiv:gr-qc/0111073 [gr-qc]}
  \BibitemShut {NoStop}%
\bibitem [{\citenamefont
  {Jüttner}(1911)}]{https://doi.org/10.1002/andp.19113390503}%
  \BibitemOpen
  \bibfield  {author} {\bibinfo {author} {\bibfnamefont {F.}~\bibnamefont
  {Jüttner}},\ }\bibfield  {title} {\bibinfo {title} {Das maxwellsche gesetz
  der geschwindigkeitsverteilung in der relativtheorie},\ }\href
  {https://doi.org/https://doi.org/10.1002/andp.19113390503} {\bibfield
  {journal} {\bibinfo  {journal} {Annalen der Physik}\ }\textbf {\bibinfo
  {volume} {339}},\ \bibinfo {pages} {856} (\bibinfo {year}
  {1911})}\BibitemShut {NoStop}%
\bibitem [{\citenamefont {Abramowitz}\ and\ \citenamefont
  {Stegun}(1964)}]{abramowitz2012handbook}%
  \BibitemOpen
  \bibfield  {author} {\bibinfo {author} {\bibfnamefont {M.}~\bibnamefont
  {Abramowitz}}\ and\ \bibinfo {author} {\bibfnamefont {I.~A.}\ \bibnamefont
  {Stegun}},\ }\href@noop {} {\emph {\bibinfo {title} {Handbook of Mathematical
  Functions with Formulas, Graphs, and Mathematical Tables}}},\ \bibinfo
  {edition} {ninth dover printing, tenth gpo printing}\ ed.\ (\bibinfo
  {publisher} {Dover},\ \bibinfo {address} {New York},\ \bibinfo {year}
  {1964})\BibitemShut {NoStop}%
\bibitem [{\citenamefont {Misner}\ \emph {et~al.}(1973)\citenamefont {Misner},
  \citenamefont {Thorne},\ and\ \citenamefont {Wheeler}}]{MTW-Book}%
  \BibitemOpen
  \bibfield  {author} {\bibinfo {author} {\bibfnamefont {C.}~\bibnamefont
  {Misner}}, \bibinfo {author} {\bibfnamefont {K.}~\bibnamefont {Thorne}},\
  and\ \bibinfo {author} {\bibfnamefont {J.}~\bibnamefont {Wheeler}},\
  }\href@noop {} {\emph {\bibinfo {title} {Gravitation}}}\ (\bibinfo
  {publisher} {W. H. Freeman},\ \bibinfo {year} {1973})\BibitemShut {NoStop}%
\bibitem [{\citenamefont {Straumann}(2013)}]{Straumann-Book}%
  \BibitemOpen
  \bibfield  {author} {\bibinfo {author} {\bibfnamefont {N.}~\bibnamefont
  {Straumann}},\ }\href@noop {} {\emph {\bibinfo {title} {General
  Relativity}}}\ (\bibinfo  {publisher} {Springer-Verlag},\ \bibinfo {address}
  {Berlin},\ \bibinfo {year} {2013})\BibitemShut {NoStop}%
\bibitem [{\citenamefont {Korol}\ \emph {et~al.}(2016)\citenamefont {Korol},
  \citenamefont {Ciotti},\ and\ \citenamefont {Pellegrini}}]{Korol_2016}%
  \BibitemOpen
  \bibfield  {author} {\bibinfo {author} {\bibfnamefont {V.}~\bibnamefont
  {Korol}}, \bibinfo {author} {\bibfnamefont {L.}~\bibnamefont {Ciotti}},\ and\
  \bibinfo {author} {\bibfnamefont {S.}~\bibnamefont {Pellegrini}},\ }\bibfield
   {title} {\bibinfo {title} {Bondi accretion in early-type galaxies},\ }\href
  {https://doi.org/10.1093/mnras/stw1029} {\bibfield  {journal} {\bibinfo
  {journal} {Monthly Notices of the Royal Astronomical Society}\ }\textbf
  {\bibinfo {volume} {460}},\ \bibinfo {pages} {1188–1200} (\bibinfo {year}
  {2016})}\BibitemShut {NoStop}%
\bibitem [{\citenamefont {{Eatough}}\ \emph {et~al.}(2013)\citenamefont
  {{Eatough}} \emph {et~al.}}]{2013Natur.501..391E}%
  \BibitemOpen
  \bibfield  {author} {\bibinfo {author} {\bibfnamefont {R.~P.}\ \bibnamefont
  {{Eatough}}} \emph {et~al.},\ }\bibfield  {title} {\bibinfo {title} {{A
  strong magnetic field around the supermassive black hole at the centre of the
  Galaxy}},\ }\href {https://doi.org/10.1038/nature12499} {\bibfield  {journal}
  {\bibinfo  {journal} {\nat}\ }\textbf {\bibinfo {volume} {501}},\ \bibinfo
  {pages} {391} (\bibinfo {year} {2013})},\ \Eprint
  {https://arxiv.org/abs/1308.3147} {arXiv:1308.3147 [astro-ph.GA]}
  \BibitemShut {NoStop}%
\bibitem [{\citenamefont {{Cuadra}}\ and\ \citenamefont
  {{Nayakshin}}(2006)}]{2006JPhCS..54..436C}%
  \BibitemOpen
  \bibfield  {author} {\bibinfo {author} {\bibfnamefont {J.}~\bibnamefont
  {{Cuadra}}}\ and\ \bibinfo {author} {\bibfnamefont {S.}~\bibnamefont
  {{Nayakshin}}},\ }\bibfield  {title} {\bibinfo {title} {{Variable accretion
  of stellar winds onto Sgr A*}},\ }in\ \href
  {https://doi.org/10.1088/1742-6596/54/1/068} {\emph {\bibinfo {booktitle}
  {Journal of Physics Conference Series}}},\ \bibinfo {series} {Journal of
  Physics Conference Series}, Vol.~\bibinfo {volume} {54}\ (\bibinfo {year}
  {2006})\ pp.\ \bibinfo {pages} {436--442}\BibitemShut {NoStop}%
\bibitem [{\citenamefont {{Cuadra}}\ \emph {et~al.}(2008)\citenamefont
  {{Cuadra}}, \citenamefont {{Nayakshin}},\ and\ \citenamefont
  {{Martins}}}]{2008MNRAS.383..458C}%
  \BibitemOpen
  \bibfield  {author} {\bibinfo {author} {\bibfnamefont {J.}~\bibnamefont
  {{Cuadra}}}, \bibinfo {author} {\bibfnamefont {S.}~\bibnamefont
  {{Nayakshin}}},\ and\ \bibinfo {author} {\bibfnamefont {F.}~\bibnamefont
  {{Martins}}},\ }\bibfield  {title} {\bibinfo {title} {{Variable accretion and
  emission from the stellar winds in the Galactic Centre}},\ }\href
  {https://doi.org/10.1111/j.1365-2966.2007.12573.x} {\bibfield  {journal}
  {\bibinfo  {journal} {\mnras}\ }\textbf {\bibinfo {volume} {383}},\ \bibinfo
  {pages} {458} (\bibinfo {year} {2008})},\ \Eprint
  {https://arxiv.org/abs/0705.0769} {arXiv:0705.0769 [astro-ph]} \BibitemShut
  {NoStop}%
\bibitem [{\citenamefont {Cuadra}\ \emph {et~al.}(2015)\citenamefont {Cuadra},
  \citenamefont {Nayakshin},\ and\ \citenamefont {Wang}}]{cuadra2015}%
  \BibitemOpen
  \bibfield  {author} {\bibinfo {author} {\bibfnamefont {J.}~\bibnamefont
  {Cuadra}}, \bibinfo {author} {\bibfnamefont {S.}~\bibnamefont {Nayakshin}},\
  and\ \bibinfo {author} {\bibfnamefont {Q.~D.}\ \bibnamefont {Wang}},\
  }\bibfield  {title} {\bibinfo {title} {{The role of feedback in accretion on
  low-luminosity AGN: Sgr A* case study}},\ }\href
  {https://doi.org/10.1093/mnras/stv584} {\bibfield  {journal} {\bibinfo
  {journal} {Monthly Notices of the Royal Astronomical Society}\ }\textbf
  {\bibinfo {volume} {450}},\ \bibinfo {pages} {277} (\bibinfo {year}
  {2015})},\ \Eprint
  {https://arxiv.org/abs/https://academic.oup.com/mnras/article-pdf/450/1/277/18502570/stv584.pdf}
  {https://academic.oup.com/mnras/article-pdf/450/1/277/18502570/stv584.pdf}
  \BibitemShut {NoStop}%
\bibitem [{\citenamefont {{Bower}}\ \emph {et~al.}(2003)\citenamefont
  {{Bower}}, \citenamefont {{Wright}}, \citenamefont {{Falcke}},\ and\
  \citenamefont {{Backer}}}]{2003ApJ...588..331B}%
  \BibitemOpen
  \bibfield  {author} {\bibinfo {author} {\bibfnamefont {G.~C.}\ \bibnamefont
  {{Bower}}}, \bibinfo {author} {\bibfnamefont {M.~C.~H.}\ \bibnamefont
  {{Wright}}}, \bibinfo {author} {\bibfnamefont {H.}~\bibnamefont {{Falcke}}},\
  and\ \bibinfo {author} {\bibfnamefont {D.~C.}\ \bibnamefont {{Backer}}},\
  }\bibfield  {title} {\bibinfo {title} {{Interferometric Detection of Linear
  Polarization from Sagittarius A* at 230 GHz}},\ }\href
  {https://doi.org/10.1086/373989} {\bibfield  {journal} {\bibinfo  {journal}
  {\apj}\ }\textbf {\bibinfo {volume} {588}},\ \bibinfo {pages} {331} (\bibinfo
  {year} {2003})},\ \Eprint {https://arxiv.org/abs/astro-ph/0302227}
  {arXiv:astro-ph/0302227 [astro-ph]} \BibitemShut {NoStop}%
\bibitem [{\citenamefont {{Marrone}}\ \emph {et~al.}(2006)\citenamefont
  {{Marrone}}, \citenamefont {{Moran}}, \citenamefont {{Zhao}},\ and\
  \citenamefont {{Rao}}}]{2006ApJ...640..308M}%
  \BibitemOpen
  \bibfield  {author} {\bibinfo {author} {\bibfnamefont {D.~P.}\ \bibnamefont
  {{Marrone}}}, \bibinfo {author} {\bibfnamefont {J.~M.}\ \bibnamefont
  {{Moran}}}, \bibinfo {author} {\bibfnamefont {J.-H.}\ \bibnamefont
  {{Zhao}}},\ and\ \bibinfo {author} {\bibfnamefont {R.}~\bibnamefont
  {{Rao}}},\ }\bibfield  {title} {\bibinfo {title} {{Interferometric
  Measurements of Variable 340 GHz Linear Polarization in Sagittarius A*}},\
  }\href {https://doi.org/10.1086/500106} {\bibfield  {journal} {\bibinfo
  {journal} {\apj}\ }\textbf {\bibinfo {volume} {640}},\ \bibinfo {pages} {308}
  (\bibinfo {year} {2006})},\ \Eprint {https://arxiv.org/abs/astro-ph/0511653}
  {arXiv:astro-ph/0511653 [astro-ph]} \BibitemShut {NoStop}%
\bibitem [{\citenamefont {{Marrone}}\ \emph {et~al.}(2007)\citenamefont
  {{Marrone}}, \citenamefont {{Moran}}, \citenamefont {{Zhao}},\ and\
  \citenamefont {{Rao}}}]{2007ApJ...654L..57M}%
  \BibitemOpen
  \bibfield  {author} {\bibinfo {author} {\bibfnamefont {D.~P.}\ \bibnamefont
  {{Marrone}}}, \bibinfo {author} {\bibfnamefont {J.~M.}\ \bibnamefont
  {{Moran}}}, \bibinfo {author} {\bibfnamefont {J.-H.}\ \bibnamefont
  {{Zhao}}},\ and\ \bibinfo {author} {\bibfnamefont {R.}~\bibnamefont
  {{Rao}}},\ }\bibfield  {title} {\bibinfo {title} {{An Unambiguous Detection
  of Faraday Rotation in Sagittarius A*}},\ }\href
  {https://doi.org/10.1086/510850} {\bibfield  {journal} {\bibinfo  {journal}
  {\apjl}\ }\textbf {\bibinfo {volume} {654}},\ \bibinfo {pages} {L57}
  (\bibinfo {year} {2007})},\ \Eprint {https://arxiv.org/abs/astro-ph/0611791}
  {arXiv:astro-ph/0611791 [astro-ph]} \BibitemShut {NoStop}%
\bibitem [{\citenamefont {{Bower}}\ \emph {et~al.}(2018)\citenamefont {{Bower}}
  \emph {et~al.}}]{2018ApJ...868..101B}%
  \BibitemOpen
  \bibfield  {author} {\bibinfo {author} {\bibfnamefont {G.~C.}\ \bibnamefont
  {{Bower}}} \emph {et~al.},\ }\bibfield  {title} {\bibinfo {title} {{ALMA
  Polarimetry of Sgr A*: Probing the Accretion Flow from the Event Horizon to
  the Bondi Radius}},\ }\href {https://doi.org/10.3847/1538-4357/aae983}
  {\bibfield  {journal} {\bibinfo  {journal} {\apj}\ }\textbf {\bibinfo
  {volume} {868}},\ \bibinfo {eid} {101} (\bibinfo {year} {2018})},\ \Eprint
  {https://arxiv.org/abs/1810.07317} {arXiv:1810.07317 [astro-ph.HE]}
  \BibitemShut {NoStop}%
\bibitem [{\citenamefont {{Allen}}\ \emph {et~al.}(1990)\citenamefont
  {{Allen}}, \citenamefont {{Hyland}},\ and\ \citenamefont
  {{Hillier}}}]{1990MNRAS.244..706A}%
  \BibitemOpen
  \bibfield  {author} {\bibinfo {author} {\bibfnamefont {D.~A.}\ \bibnamefont
  {{Allen}}}, \bibinfo {author} {\bibfnamefont {A.~R.}\ \bibnamefont
  {{Hyland}}},\ and\ \bibinfo {author} {\bibfnamefont {D.~J.}\ \bibnamefont
  {{Hillier}}},\ }\bibfield  {title} {\bibinfo {title} {{The source of
  luminosity at the Galactic Centre.}},\ }\href@noop {} {\bibfield  {journal}
  {\bibinfo  {journal} {\mnras}\ }\textbf {\bibinfo {volume} {244}},\ \bibinfo
  {pages} {706} (\bibinfo {year} {1990})}\BibitemShut {NoStop}%
\bibitem [{\citenamefont {{Puls}}\ \emph {et~al.}(1996)\citenamefont {{Puls}}
  \emph {et~al.}}]{1996A&A...305..171P}%
  \BibitemOpen
  \bibfield  {author} {\bibinfo {author} {\bibfnamefont {J.}~\bibnamefont
  {{Puls}}} \emph {et~al.},\ }\bibfield  {title} {\bibinfo {title} {{O-star
  mass-loss and wind momentum rates in the Galaxy and the Magellanic Clouds
  Observations and theoretical predictions.}},\ }\href@noop {} {\bibfield
  {journal} {\bibinfo  {journal} {\aap}\ }\textbf {\bibinfo {volume} {305}},\
  \bibinfo {pages} {171} (\bibinfo {year} {1996})}\BibitemShut {NoStop}%
\bibitem [{\citenamefont {{Repolust}}\ \emph {et~al.}(2004)\citenamefont
  {{Repolust}}, \citenamefont {{Puls}},\ and\ \citenamefont
  {{Herrero}}}]{2004A&A...415..349R}%
  \BibitemOpen
  \bibfield  {author} {\bibinfo {author} {\bibfnamefont {T.}~\bibnamefont
  {{Repolust}}}, \bibinfo {author} {\bibfnamefont {J.}~\bibnamefont {{Puls}}},\
  and\ \bibinfo {author} {\bibfnamefont {A.}~\bibnamefont {{Herrero}}},\
  }\bibfield  {title} {\bibinfo {title} {{Stellar and wind parameters of
  Galactic O-stars. The influence of line-blocking/blanketing}},\ }\href
  {https://doi.org/10.1051/0004-6361:20034594} {\bibfield  {journal} {\bibinfo
  {journal} {\aap}\ }\textbf {\bibinfo {volume} {415}},\ \bibinfo {pages} {349}
  (\bibinfo {year} {2004})}\BibitemShut {NoStop}%
\bibitem [{\citenamefont {{Paumard}}\ \emph {et~al.}(2006)\citenamefont
  {{Paumard}} \emph {et~al.}}]{2006ApJ...643.1011P}%
  \BibitemOpen
  \bibfield  {author} {\bibinfo {author} {\bibfnamefont {T.}~\bibnamefont
  {{Paumard}}} \emph {et~al.},\ }\bibfield  {title} {\bibinfo {title} {{The Two
  Young Star Disks in the Central Parsec of the Galaxy: Properties, Dynamics,
  and Formation}},\ }\href {https://doi.org/10.1086/503273} {\bibfield
  {journal} {\bibinfo  {journal} {\apj}\ }\textbf {\bibinfo {volume} {643}},\
  \bibinfo {pages} {1011} (\bibinfo {year} {2006})},\ \Eprint
  {https://arxiv.org/abs/astro-ph/0601268} {arXiv:astro-ph/0601268 [astro-ph]}
  \BibitemShut {NoStop}%
\bibitem [{\citenamefont {{Martins}}\ \emph {et~al.}(2007)\citenamefont
  {{Martins}}, \citenamefont {{Genzel}}, \citenamefont {{Hillier}},
  \citenamefont {{Eisenhauer}}, \citenamefont {{Paumard}}, \citenamefont
  {{Gillessen}}, \citenamefont {{Ott}},\ and\ \citenamefont
  {{Trippe}}}]{2007A&A...468..233M}%
  \BibitemOpen
  \bibfield  {author} {\bibinfo {author} {\bibfnamefont {F.}~\bibnamefont
  {{Martins}}}, \bibinfo {author} {\bibfnamefont {R.}~\bibnamefont {{Genzel}}},
  \bibinfo {author} {\bibfnamefont {D.~J.}\ \bibnamefont {{Hillier}}}, \bibinfo
  {author} {\bibfnamefont {F.}~\bibnamefont {{Eisenhauer}}}, \bibinfo {author}
  {\bibfnamefont {T.}~\bibnamefont {{Paumard}}}, \bibinfo {author}
  {\bibfnamefont {S.}~\bibnamefont {{Gillessen}}}, \bibinfo {author}
  {\bibfnamefont {T.}~\bibnamefont {{Ott}}},\ and\ \bibinfo {author}
  {\bibfnamefont {S.}~\bibnamefont {{Trippe}}},\ }\bibfield  {title} {\bibinfo
  {title} {{Stellar and wind properties of massive stars in the central parsec
  of the Galaxy}},\ }\href {https://doi.org/10.1051/0004-6361:20066688}
  {\bibfield  {journal} {\bibinfo  {journal} {\aap}\ }\textbf {\bibinfo
  {volume} {468}},\ \bibinfo {pages} {233} (\bibinfo {year} {2007})},\ \Eprint
  {https://arxiv.org/abs/astro-ph/0703211} {arXiv:astro-ph/0703211 [astro-ph]}
  \BibitemShut {NoStop}%
\bibitem [{\citenamefont {{Ressler}}\ \emph {et~al.}(2018)\citenamefont
  {{Ressler}}, \citenamefont {{Quataert}},\ and\ \citenamefont
  {{Stone}}}]{2018MNRAS.478.3544R}%
  \BibitemOpen
  \bibfield  {author} {\bibinfo {author} {\bibfnamefont {S.~M.}\ \bibnamefont
  {{Ressler}}}, \bibinfo {author} {\bibfnamefont {E.}~\bibnamefont
  {{Quataert}}},\ and\ \bibinfo {author} {\bibfnamefont {J.~M.}\ \bibnamefont
  {{Stone}}},\ }\bibfield  {title} {\bibinfo {title} {{Hydrodynamic simulations
  of the inner accretion flow of Sagittarius A* fuelled by stellar winds}},\
  }\href {https://doi.org/10.1093/mnras/sty1146} {\bibfield  {journal}
  {\bibinfo  {journal} {\mnras}\ }\textbf {\bibinfo {volume} {478}},\ \bibinfo
  {pages} {3544} (\bibinfo {year} {2018})},\ \Eprint
  {https://arxiv.org/abs/1805.00474} {arXiv:1805.00474 [astro-ph.HE]}
  \BibitemShut {NoStop}%
\bibitem [{\citenamefont {{Ressler}}\ \emph {et~al.}(2020)\citenamefont
  {{Ressler}}, \citenamefont {{White}}, \citenamefont {{Quataert}},\ and\
  \citenamefont {{Stone}}}]{2020ApJ...896L...6R}%
  \BibitemOpen
  \bibfield  {author} {\bibinfo {author} {\bibfnamefont {S.~M.}\ \bibnamefont
  {{Ressler}}}, \bibinfo {author} {\bibfnamefont {C.~J.}\ \bibnamefont
  {{White}}}, \bibinfo {author} {\bibfnamefont {E.}~\bibnamefont
  {{Quataert}}},\ and\ \bibinfo {author} {\bibfnamefont {J.~M.}\ \bibnamefont
  {{Stone}}},\ }\bibfield  {title} {\bibinfo {title} {{Ab Initio Horizon-scale
  Simulations of Magnetically Arrested Accretion in Sagittarius A* Fed by
  Stellar Winds}},\ }\href {https://doi.org/10.3847/2041-8213/ab9532}
  {\bibfield  {journal} {\bibinfo  {journal} {\apjl}\ }\textbf {\bibinfo
  {volume} {896}},\ \bibinfo {eid} {L6} (\bibinfo {year} {2020})},\ \Eprint
  {https://arxiv.org/abs/2006.00005} {arXiv:2006.00005 [astro-ph.HE]}
  \BibitemShut {NoStop}%
\bibitem [{\citenamefont {{Calder{\'o}n}}\ \emph {et~al.}(2020)\citenamefont
  {{Calder{\'o}n}}, \citenamefont {{Cuadra}}, \citenamefont {{Schartmann}},
  \citenamefont {{Burkert}},\ and\ \citenamefont
  {{Russell}}}]{2020ApJ...888L...2C}%
  \BibitemOpen
  \bibfield  {author} {\bibinfo {author} {\bibfnamefont {D.}~\bibnamefont
  {{Calder{\'o}n}}}, \bibinfo {author} {\bibfnamefont {J.}~\bibnamefont
  {{Cuadra}}}, \bibinfo {author} {\bibfnamefont {M.}~\bibnamefont
  {{Schartmann}}}, \bibinfo {author} {\bibfnamefont {A.}~\bibnamefont
  {{Burkert}}},\ and\ \bibinfo {author} {\bibfnamefont {C.~M.~P.}\ \bibnamefont
  {{Russell}}},\ }\bibfield  {title} {\bibinfo {title} {{Stellar Winds Pump the
  Heart of the Milky Way}},\ }\href {https://doi.org/10.3847/2041-8213/ab5e81}
  {\bibfield  {journal} {\bibinfo  {journal} {\apjl}\ }\textbf {\bibinfo
  {volume} {888}},\ \bibinfo {eid} {L2} (\bibinfo {year} {2020})},\ \Eprint
  {https://arxiv.org/abs/1910.06976} {arXiv:1910.06976 [astro-ph.GA]}
  \BibitemShut {NoStop}%
\bibitem [{\citenamefont {{Mo{\'s}cibrodzka}}\ \emph
  {et~al.}(2009)\citenamefont {{Mo{\'s}cibrodzka}}, \citenamefont {{Gammie}},
  \citenamefont {{Dolence}}, \citenamefont {{Shiokawa}},\ and\ \citenamefont
  {{Leung}}}]{2009ApJ...706..497M}%
  \BibitemOpen
  \bibfield  {author} {\bibinfo {author} {\bibfnamefont {M.}~\bibnamefont
  {{Mo{\'s}cibrodzka}}}, \bibinfo {author} {\bibfnamefont {C.~F.}\ \bibnamefont
  {{Gammie}}}, \bibinfo {author} {\bibfnamefont {J.~C.}\ \bibnamefont
  {{Dolence}}}, \bibinfo {author} {\bibfnamefont {H.}~\bibnamefont
  {{Shiokawa}}},\ and\ \bibinfo {author} {\bibfnamefont {P.~K.}\ \bibnamefont
  {{Leung}}},\ }\bibfield  {title} {\bibinfo {title} {{Radiative Models of SGR
  A* from GRMHD Simulations}},\ }\href
  {https://doi.org/10.1088/0004-637X/706/1/497} {\bibfield  {journal} {\bibinfo
   {journal} {\apj}\ }\textbf {\bibinfo {volume} {706}},\ \bibinfo {pages}
  {497} (\bibinfo {year} {2009})},\ \Eprint {https://arxiv.org/abs/0909.5431}
  {arXiv:0909.5431 [astro-ph.HE]} \BibitemShut {NoStop}%
\bibitem [{\citenamefont {{Mo{\'s}cibrodzka}}\ and\ \citenamefont
  {{Falcke}}(2013)}]{2013A&A...559L...3M}%
  \BibitemOpen
  \bibfield  {author} {\bibinfo {author} {\bibfnamefont {M.}~\bibnamefont
  {{Mo{\'s}cibrodzka}}}\ and\ \bibinfo {author} {\bibfnamefont
  {H.}~\bibnamefont {{Falcke}}},\ }\bibfield  {title} {\bibinfo {title}
  {{Coupled jet-disk model for Sagittarius A*: explaining the flat-spectrum
  radio core with GRMHD simulations of jets}},\ }\href
  {https://doi.org/10.1051/0004-6361/201322692} {\bibfield  {journal} {\bibinfo
   {journal} {\aap}\ }\textbf {\bibinfo {volume} {559}},\ \bibinfo {eid} {L3}
  (\bibinfo {year} {2013})},\ \Eprint {https://arxiv.org/abs/1310.4951}
  {arXiv:1310.4951 [astro-ph.HE]} \BibitemShut {NoStop}%
\bibitem [{\citenamefont {{Shcherbakov}}\ and\ \citenamefont
  {{Baganoff}}(2010)}]{2010ApJ...716..504S}%
  \BibitemOpen
  \bibfield  {author} {\bibinfo {author} {\bibfnamefont {R.~V.}\ \bibnamefont
  {{Shcherbakov}}}\ and\ \bibinfo {author} {\bibfnamefont {F.~K.}\ \bibnamefont
  {{Baganoff}}},\ }\bibfield  {title} {\bibinfo {title} {{Inflow-Outflow Model
  with Conduction and Self-consistent Feeding for Sgr A*}},\ }\href
  {https://doi.org/10.1088/0004-637X/716/1/504} {\bibfield  {journal} {\bibinfo
   {journal} {\apj}\ }\textbf {\bibinfo {volume} {716}},\ \bibinfo {pages}
  {504} (\bibinfo {year} {2010})},\ \Eprint {https://arxiv.org/abs/1004.0702}
  {arXiv:1004.0702 [astro-ph.HE]} \BibitemShut {NoStop}%
\bibitem [{\citenamefont {Chandrasekhar}(1983)}]{Chandrasekhar83}%
  \BibitemOpen
  \bibfield  {author} {\bibinfo {author} {\bibfnamefont {S.}~\bibnamefont
  {Chandrasekhar}},\ }\href@noop {} {\emph {\bibinfo {title} {The Mathematical
  Theory of Black Holes}}}\ (\bibinfo  {publisher} {Oxford University Press},\
  \bibinfo {address} {Oxford, England},\ \bibinfo {year} {1983})\BibitemShut
  {NoStop}%
\bibitem [{\citenamefont {N${\rm\acute{u}\tilde{n}}$ez}\ and\ \citenamefont
  {Degollado}(2005)}]{Nunez05}%
  \BibitemOpen
  \bibfield  {author} {\bibinfo {author} {\bibfnamefont {D.}~\bibnamefont
  {N${\rm\acute{u}\tilde{n}}$ez}}\ and\ \bibinfo {author} {\bibfnamefont
  {J.~C.}\ \bibnamefont {Degollado}},\ }\href@noop {} {\emph {\bibinfo {title}
  {Relatividad General}}}\ (\bibinfo  {publisher} {online},\ \bibinfo {address}
  {Facultad de Ciencias, UNAM, Mexico City, Mexico},\ \bibinfo {year}
  {2005})\BibitemShut {NoStop}%
\bibitem [{\citenamefont {{Event Horizon Telescope
  Collaboration}}(2019{\natexlab{c}})}]{EHTCVI}%
  \BibitemOpen
  \bibfield  {author} {\bibinfo {author} {\bibnamefont {{Event Horizon
  Telescope Collaboration}}},\ }\bibfield  {title} {\bibinfo {title} {{First
  M87 Event Horizon Telescope Results. VI. The Shadow and Mass of the Central
  Black Hole}},\ }\href {https://doi.org/10.3847/2041-8213/ab1141} {\bibfield
  {journal} {\bibinfo  {journal} {\apjl}\ }\textbf {\bibinfo {volume} {875}},\
  \bibinfo {eid} {L6} (\bibinfo {year} {2019}{\natexlab{c}})},\ \Eprint
  {https://arxiv.org/abs/1906.11243} {arXiv:1906.11243 [astro-ph.GA]}
  \BibitemShut {NoStop}%
\bibitem [{\citenamefont {Chaverra}\ and\ \citenamefont
  {Sarbach}(2015)}]{Chaverra:2015bya}%
  \BibitemOpen
  \bibfield  {author} {\bibinfo {author} {\bibfnamefont {E.}~\bibnamefont
  {Chaverra}}\ and\ \bibinfo {author} {\bibfnamefont {O.}~\bibnamefont
  {Sarbach}},\ }\bibfield  {title} {\bibinfo {title} {{Radial accretion flows
  on static spherically symmetric black holes}},\ }\href
  {https://doi.org/10.1088/0264-9381/32/15/155006} {\bibfield  {journal}
  {\bibinfo  {journal} {Class. Quant. Grav.}\ }\textbf {\bibinfo {volume}
  {32}},\ \bibinfo {pages} {155006} (\bibinfo {year} {2015})},\ \Eprint
  {https://arxiv.org/abs/1501.01641} {arXiv:1501.01641 [gr-qc]} \BibitemShut
  {NoStop}%
\end{thebibliography}%


\end{document}